# Development of Rotating Target with Ferrofluid Seal for ILC Electron-Driven Positron Source

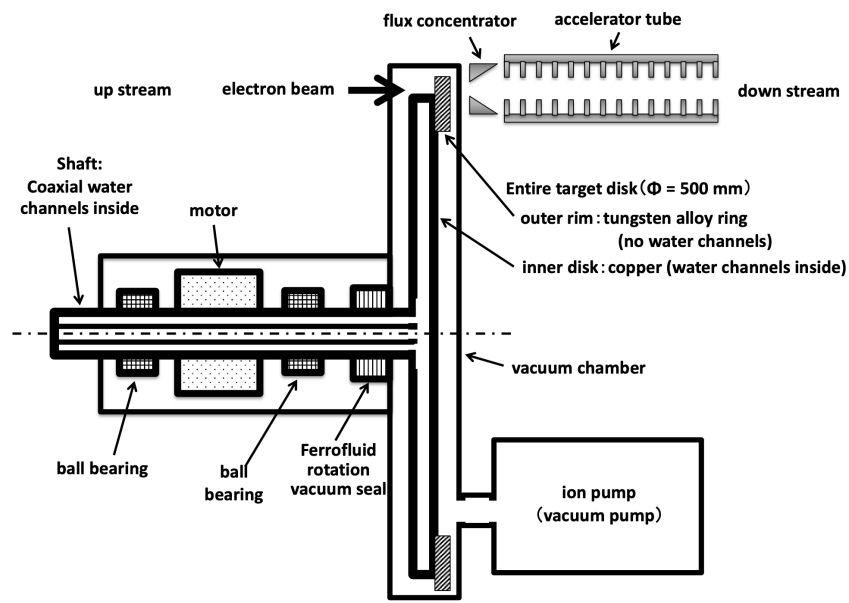

Tsunehiko OMORI*, Kaoru YOKOYA, Junji URAKAWA
KEK, High Energy Accelerator Research Organization

Tohru TAKAHASHI, Masao KURIKI
Hiroshima University

Masakazu WASHIO
Waseda University

Masaru KURIBAYASHI, Masaaki YAMAKATA, Chie NAKAISHI
Rigaku Corporation

---

* e-mail : tsunehiko.omori@biglobe.jp

**Note**

This paper is an English translation of KEK Report 2023-4 (language: Japanese).
The translation was done by the authors themselves.

**Title in original language (Japanese)**

ILC 電子駆動陽電子源のための磁性流体シールを用いた回転ターゲットの開発

**Author names and affiliations in original language (Japanese)**

大森恒彦、横谷馨、浦川順治
高エネルギー加速器研究機構

高橋徹、栗木雅夫
広島大学

鷲尾方一
早稲田大学

栗林勝、山片正明、中石千恵
株式会社リガク

# Contents

# List of Figures

## Table List of Tables

# Development of Rotating Target with Ferrofluid Seal for ILC Electron-Driven Positron Source

Tsunehiko OMORI, Kaoru YOKOYA, Junji URAKAWA
KEK, High Energy Accelerator Research Organization

Tohru TAKAHASHI, Masao KURIKI
Hiroshima University

Masakazu WASHIO
Waseda University

Masaru KURIBAYASHI, Masaaki YAMAKATA, Chie NAKAISHI
Rigaku Corporation

## Abstract

In the ILC electron-driven positron source, acceleration tubes are located immediately downstream of the positron generation target to capture the generated positrons, necessitating high vacuum for their operation. We designed a rotating target using a ferrofluid rotating vacuum seal that can stably maintain a vacuum of $10^{-6}$ Pa and a tangential velocity of 5 m/sec, which meets the requirements of the ILC positron source. We fabricated a full-scale prototype and confirmed the compatibility of rotation and stable vacuum maintenance. Various tests on the radiation resistance of the ferrofluid demonstrated its ability to withstand six years of radiation exposure in normal ILC operation without degradation and to continue exhibiting vacuum sealing performance. Radiation exposure tests on a small-scale target with a structure and material similar to the full-scale target confirmed its operability after two years of radiation exposure in normal ILC operation. The design, including thermal load, stress analysis, and cooling, led to the conclusion of an estimated lifespan of the target of over one year, based on comparison with the SLC target. These results demonstrate and establish the technology of positron generation targets in the ILC electron-driven positron source.

## Introduction

The International Linear Collider (ILC) is an electron-positron linear collider designed for experimental exploration in particle physics at center-of-mass energies ranging from 250 GeV to 1 TeV. The ILC Technical Design Report (TDR) was issued in 2013 [1]. A notable feature of the ILC is the adoption of superconducting RF acceleration technology in its main linac, enabling the acceleration of high-current beams with pulses lasting 0.7 milliseconds from approximately 1300 positron (electron) bunches, thus achieving high luminosity.

The ILC requires an unprecedentedly large number of positrons, approximately 30 times that of the SLC and about 80 times that of SuperKEKB, in terms of production rate per unit time. Consequently, the positron source poses one of the greatest challenges for the realization of the ILC project.

In the ILC Technical Design Report (TDR), an undulator-based positron source is adopted as the baseline positron source. In this undulator-based positron source, high-energy beams (>125 GeV) from the main linac (with a beam pulse length of approximately 0.7 milliseconds) are passed through a helical undulator to generate gamma rays, from which positrons are produced. This type of positron source has never been used in an accelerator before, so an electron-driven positron source was planned as a technical backup to reduce the risks.

The ILC electron-driven positron source generates positrons by directing an electron beam onto a target, where positrons are produced through electromagnetic showers [2][3]. The fundamental concept can be summarized as follows. Given that the energy of the electron beam is not very high, on the order of several GeV, a separate and independent system is prepared apart from the main linac (Figure 1). In the first stage of the ILC, the energy of this electron beam is 3 GeV. Since the ILC operates at a low repetition rate of 5 Hz (200 milliseconds), advantage is taken of the strength of an independent particle source to slowly create positrons over approximately 60 milliseconds. This approach serves to protect the target from unprecedentedly large thermal loads. This paper summarizes the status of target development for the electron-driven positron source up to the end of the fiscal year 2022, based on the design at the time of compiling proposals for R&D by the International Development Team of the ILC at the Pre ILC laboratory [4].

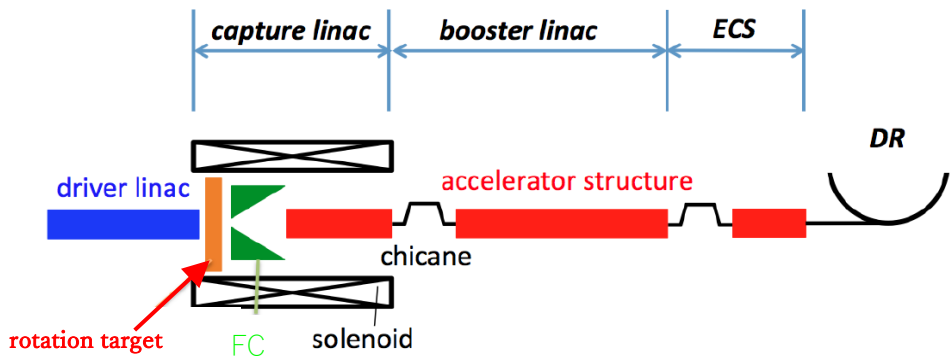

Fig. 1.    Conceptual diagram of ILC electron-driven positron source.
(Note: This figure is adapted from Reference [3].)

The undulator-based positron source offers the advantage of generating polarized positrons; however, it necessitates the creation of 1312 bunches within a short pulse length of 0.7 ms from the main linac. Due to this constraint, the undulator-based positron source requires an extremely high-speed rotating target with a tangential speed of 100 m/s for positron generation. Additionally, it cannot be prepared as an independent system from the main linac. In contrast, the electron-driven positron source provides freedom from these constraints. As a result, a positron source with lower risks and greater operational flexibility can be realized by leveraging this freedom. This flexibility is expected to be advantageous during commissioning after construction and during machine tunings following long-term shutdowns.

The main linac of the ILC operates at 5 Hz, with each RF pulse containing 1312 bunches. Before acceleration by the main linac, all bunches must be prepared in the damping ring (DR). Therefore, positrons need to be generated such that 1312 bunches are produced within 200 milliseconds corresponding to 5 Hz. Since the damping ring is designed to radiatively damp the emittance of the beam within approximately 100 milliseconds to achieve low emittance, the remaining approximately 100 milliseconds can be utilized for positron generation. In our proposed electron-driven positron source, positrons for one pulse (1312 bunches) are generated as 20 sub-pulses at a repetition frequency of 300 Hz (Figure 2). This generation utilizes an electron beam from a 300 Hz normal conducting electron linac.

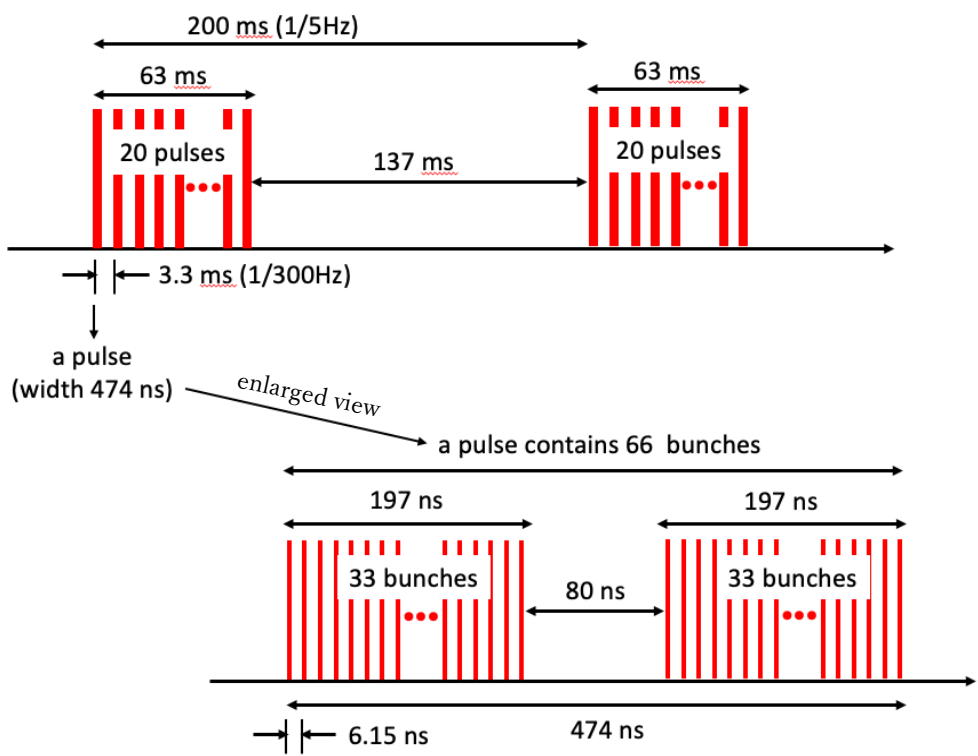

Fig. 2.    Time structure of the beam at the positron source

The 20 sub-pulses consist of the first 12 sub-pulses composed of two 33-bunch mini-trains, and the last 8 pulses composed of a 33-bunch mini-train and a 32-bunch mini-train. In this way, 1312 bunches are generated in 63 milliseconds. Please refer to the following equation.

$$12 \times (33+33) + 8 \times (33+32) = 1312$$

Figure 2 illustrates the overall time structure of the 200-millisecond interval repetition and the time structure of one sub-pulse within it (which includes two 33-bunch mini-trains). Adopting this time structure, when the target is moved at an appropriate speed, thermal accumulation associated with only 66 (or 65) bunches accumulates at a specific point on the target. This results in a significant reduction in thermal load on the target compared to generating 1312 bunches simultaneously.

The spacing between bunches in the damping ring is 6.15 nanoseconds. Therefore, the 300 Hz electron linac of the positron source adopts a bunch spacing of 6.15 ns to match this. To avoid instability caused by electron clouds in the positron damping ring, the positron beam is structured such that approximately 33 bunches form a mini-train. There is approximately an 80 nanosecond gap between adjacent mini-trains, which is key to preventing the instability. The beam time structure of the 300 Hz linac in the positron source follows a similar pattern. Each sub-pulse of this positron source is directly injected into the damping ring using a kicker (Figure 3). The kicker has a pulse length of 1 microsecond and a rise time of about 80 nanoseconds. The generated positrons are accumulated in the damping ring for a sufficient duration of 137 milliseconds for radiative damping, then sent to the main linac for acceleration before being directed to the collision point.

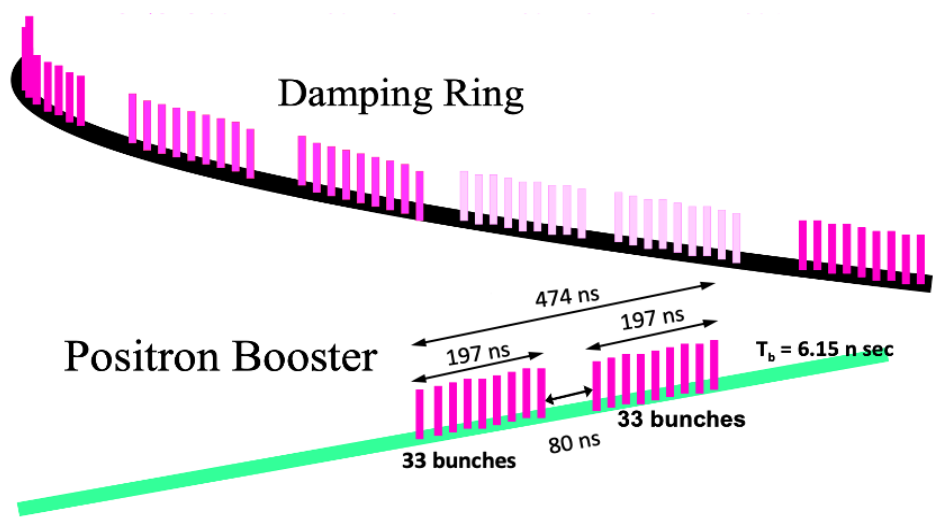

Fig. 3. The sub-pulses generated in the positron source (typically composed of 66 bunches) are accelerated up to 5 GeV in the booster before collectively being injected into the damping ring, all 66 bunches at once.

In this positron source, a rotating target with a tangential speed of 5 m/s is utilized. At immediate downstream of the target, high-gradient acceleration tubes are installed for positron capture. To ensure the operation of these acceleration tubes, it is necessary to simultaneously maintain target rotation and ultra-high vacuum. Therefore, ferrofluid is employed in the vacuum seal of the rotation shaft. The functionality, structure, and characteristics of this rotating target are similar to those of a rotating-anode X-ray generator, necessitating efficient dissipation of heat generated by the beam to keep the target temperature low and to mitigate exposure to generated radiation. Leveraging the technology of small rotating targets utilizing ferrofluid vacuum seals, commonly used in high-intensity X-ray generators with accumulated technological expertise across various practical and research fields, we have been advancing the development of this rotating target. Progress has been made in confirming the performance of ferrofluid rotary seals, verifying radiation resistance characteristics, designing the target, manufacturing two types of prototypes, and conducting long-term continuous rotation vacuum tests on prototypes. This paper discusses the overall development of this rotating target, along with its history and progression.

## (I) Determination of Basic Technology

The ILC operates with a standard beam consisting of $2 \times 10^{10}$ particles per bunch and approximately 1300 bunches per RF pulse. However, for future high luminosity upgrades, operation with 2600 bunches per RF pulse is also planned. Therefore, the basic technology of the target should be selected to withstand operation with 2600 bunches.

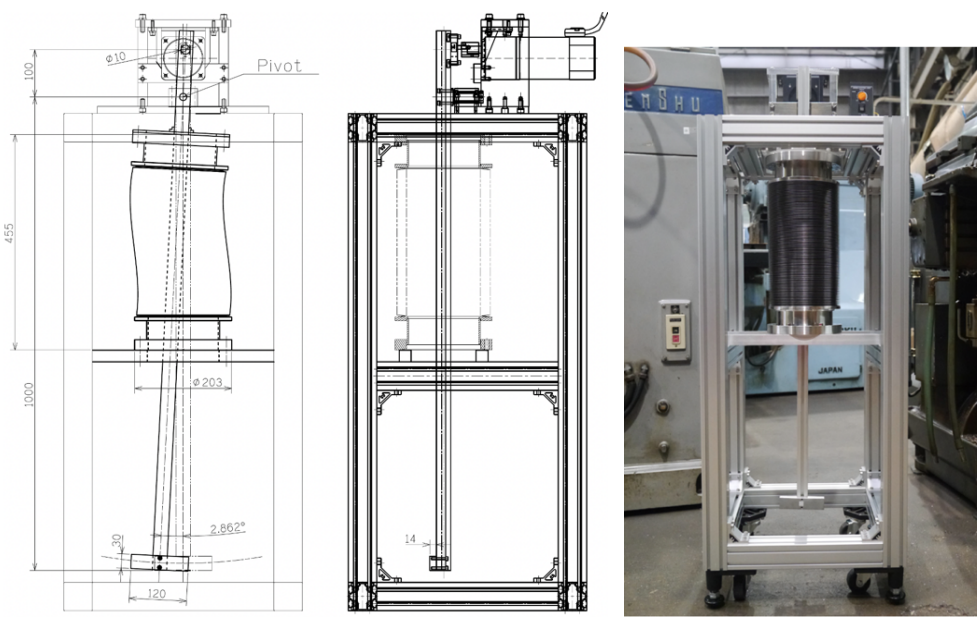

Fig. 4. Prototype of the pendulum target. Left and center: schematic drawings, Right: photograph.

Upon initial consideration, two candidate target designs were compared: a rotating target using a rotating vacuum seal and a pendulum target utilizing bellows to perform pendulum motion. It was believed that the rotating target could generate a higher intensity beam due to its ability to achieve higher target speed. Conversely, in the case of the bellows-based design, which uses bellows to separate vacuum and atmosphere, there was no concern regarding vacuum maintenance, but achieving high speeds was challenging due to the necessity of reciprocating motion. We attached a weight resembling the target to the end of a pendulum with a length of about 1 meter and introduced it into a vacuum container using bellows, allowing it to undergo pendulum motion in the container (Figure 4, made in 2013). The tangential velocity at the tip of the pendulum was set to 1 m/s. This prototype underwent several rounds of operation and improvement, and in 2014, the bellows and target vacuum container were actually evacuated, and the bellows were operated under atmospheric pressure stress for several months without damage. As a result of observing the vibration during the operation of this target, it was recognized that the speed limit of the pendulum method was approximately 1 m/s.

Concurrently, an investigation was conducted into the relation between heat load and target speed. Given that the time interval between sub-pulses is 3.3 milliseconds, with a speed of about 1 m/s, the centers of the two high-temperature spots created by two sub-pulses would be approximately 3.3 mm apart. At the outset of the study around 2012-2013, there was an optimistic outlook that such a separation of the two spots would be sufficient. Figure 5 is a citation from an initial paper [2] on the overall design of the electron-driven positron source, dating back to 2012. Anticipating a future luminosity upgrade (2600 bunch operation in the main linear accelerator), a simple simulation of target temperature rise was performed, assuming a drive beam energy of 6 GeV, beam diameter of $\sigma$ = 4mm, 2600 bunches, charge per bunch of 3.2 nC, and 132 sub-pulses (with the same 20 sub-pulses as in the standard mode). Cooling was neglected for simplicity. Additionally, the operation was not continuous over a long period; it involved only three sub-pulses being fired at the target. The motion was linear rather than rotational for simplicity. The figures (a), (b), (c), and (d) in Figure 5 represent tangential speeds of target motion at 5 m/s, 2 m/s, 1 m/s, and 0.5 m/s, respectively. In the case of 5 m/s, the distribution of heat is clearly separated, indicating a noticeable reduction in heat load due to the target's movement. However, at velocities below this, the distribution of heat is not separated. Nonetheless, even at a velocity of 0.5 m/s, the temperature is sufficiently lower than the melting point of the assumed target material, which is a tungsten alloy. Based on this, the authors concluded in reference [2] that even 0.5 m/s would be acceptable. However, researchers from DESY in Germany, along with those from the University of Hamburg and the University of Applied Sciences in Wildau, conducted simulations from 2013 to 2014, considering not only heat but also stress and PEDD (Peak Energy Deposit Density) [5][6]. They concluded that velocities such as 1 m/s were inadequate and that 5 m/s was necessary. The findings of this German research group were compelling, leading to a consensus that a tangential velocity of about 5 m/s was required. Since the issue of target stress is crucial, it will be discussed in detail later in Chapter (V).

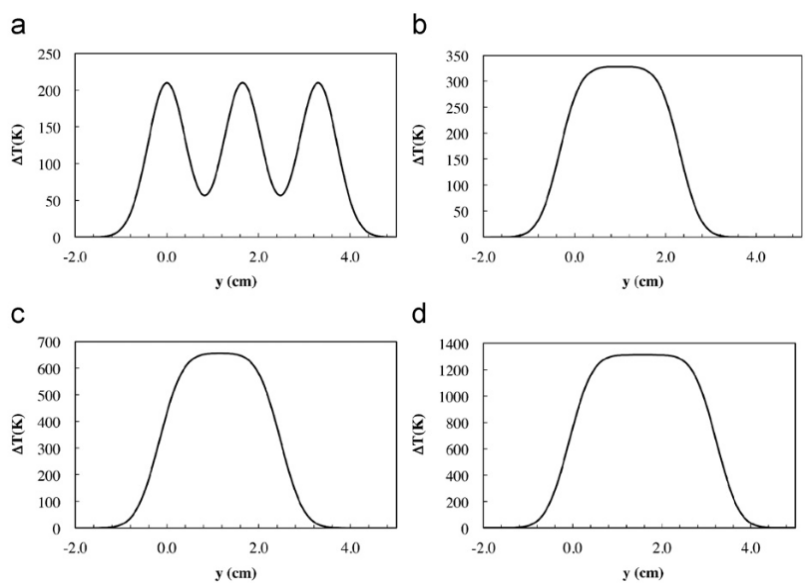

Fig. 5. Tangential Speed and Target Temperature. Results of simplified simulations. Cooling not considered. Only three sub-pulses are injected. (a) 5 m/s, (b) 2 m/s, (c) 1 m/s, (d) 0.5 m/s (Note: This figure is quoted from reference [2].)

Following these simulation results and considering test results of the pendulum targe, the development of the pendulum target was halted, as it was found that a tangential speed of around 1 m/s was the upper limit. Instead, efforts were directed towards developing a rotating target aiming for a tangential speed of about 5 m/s.

In the development of the rotating target, we decided to base the development of the ILC positron target on the technology of a small rotating target that uses ferrofluid vacuum seals, which are used in high-intensity X-ray generation devices. Figure 6 illustrates an example of such a rotating target. Ferrofluid is a liquid consisting of synthetic polymers with low vapor pressure (referred to as base oil) dispersed with particles of ferromagnetic material such as magnetite. It is inserted into the tiny gap between the rotating shaft and the pole pieces at the stator, where it is trapped and held in place by the magnetic flux of the permanent magnet, forming a rotating vacuum seal. This is the ferrofluid rotary seal (Figure 7). The rotation of the shaft is supported by ball-bearings, independent of the ferrofluid rotary vacuum seal. Since high vacuum conditions are required for our application, two sets of these ball-bearings are installed to support the shaft, with both sets positioned on the atmospheric side outside the ferrofluid seal, with a driving motor sandwiched in between.

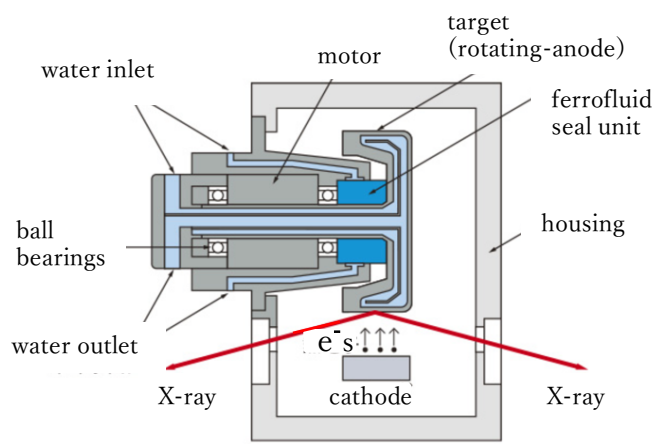

Fig. 6.    Example of a commercially available small rotating target utilizing ferrofluid vacuum seals for X-ray generation. Note: This figure is cited from the Rigaku Corporation website.

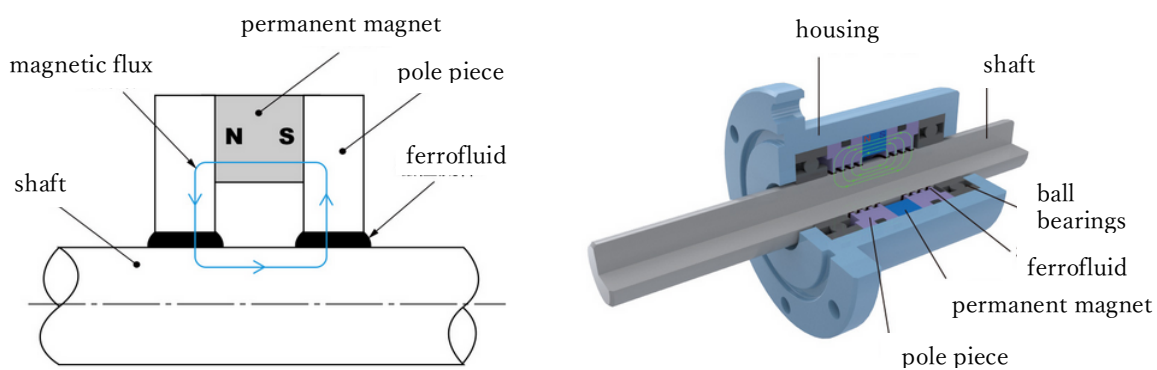

Fig. 7.    Diagram of the principle of a ferrofluid rotary vacuum seal. Left: Schematic of the magnetic circuit consisting of a rotating shaft, pole pieces, permanent magnet, and ferrofluid. Right: Overview diagram of the entire assembly (drive motor omitted). Note: Quoted from Rigaku Corporation's web page.

## (II) Preliminary Tests of Rotating Target

### (1) Confirmation of Attainable Vacuum Level with Commercial Small Rotating Targets

As mentioned in the previous chapter (I), commercially available small rotating targets exist in this format. This fact was a significant advantage in advancing the development. Initially, we confirmed the attainable vacuum level with commercially available small rotating targets and estimated the leak rate from there (in 2013-2014, Figure 8).

Tests were conducted on three types of ferrofluids used by Rigaku, confirming that each of them possesses sufficient performance for our purposes (Figure 9). Specifically, from these results, we confirmed that for ILC targets, reasonable-sized pumps could maintain the required vacuum level with an acceptable leak rate.

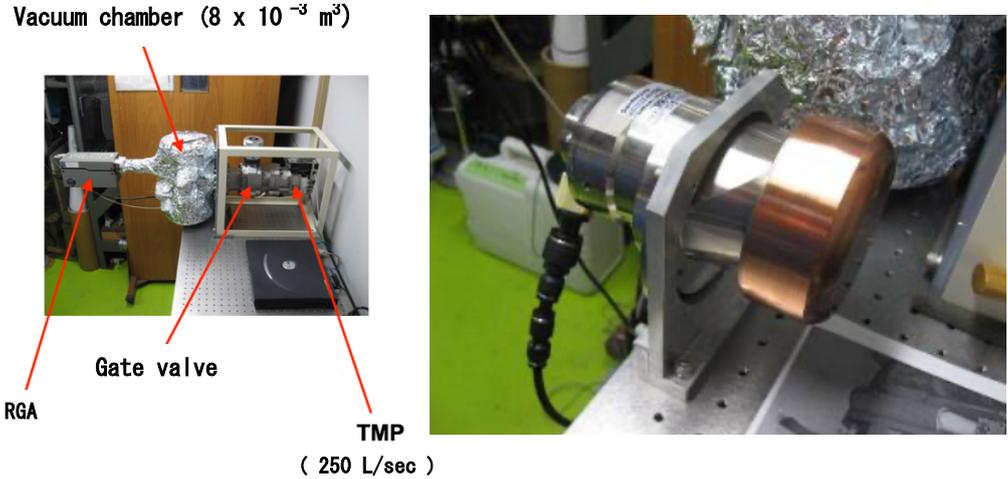

Fig. 8. Confirmation of the achieved vacuum level using a commercially available small rotating target. Left: Testing process, Right: Appearance of the small rotating target used.

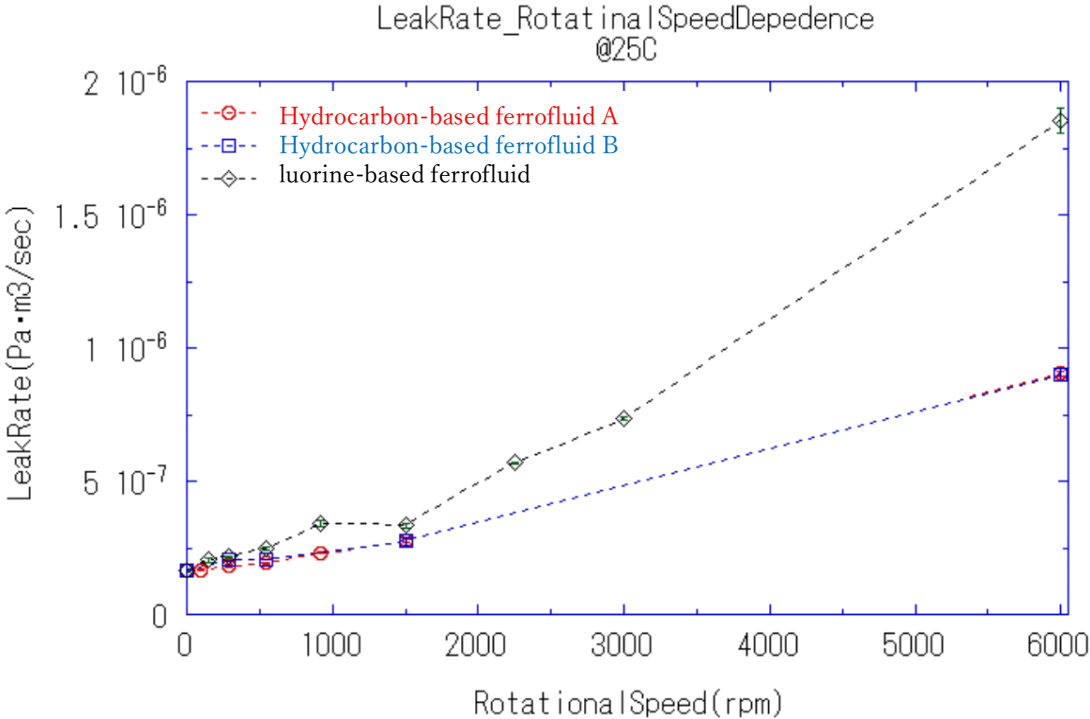

Fig. 9. Results of vacuum tests using a commercially available small rotating target. Horizontal axis: Rotation speed, Vertical axis: Leak rate. The three lines represent the types of ferrofluids used.

(2) Confirmation of Radiation Resistance of Ferrofluids: First Stage

Our target development is based on the rotating target technology used in high-intensity X-ray generation devices. Therefore, materials used there, including ferrofluids, are naturally possess appropriate radiation resistance. However, the radiation environment of the ILC positron source is extremely harsh, necessitating thorough radiation resistance testing. In particular, ferrofluids, being based on polymer oils, raised concerns about their ability to withstand the radiation environment of the ILC.

(2-1) Radiation Irradiation Tests on Ferrofluids

Two types of ferrofluids, "fluorine-based" and "hydrocarbon-based," were irradiated at the QST Takasaki Laboratory's $^{60}$Co irradiation facility (high-intensity gamma-ray source) with doses of 0.25 MGy and 3.2 MGy (in 2013-2014, Figure 10). The fluorine-based ferrofluid produced a strong odor at 0.25 MGy and completely decomposed with a strong odor at 3.2 MGy, likely generating fluorine gas. In contrast, the hydrocarbon-based ferrofluid changed into a gel-like substance at 3.2 MGy (increasing in viscosity) but did not decompose. There were also no observable changes in color or appearance.

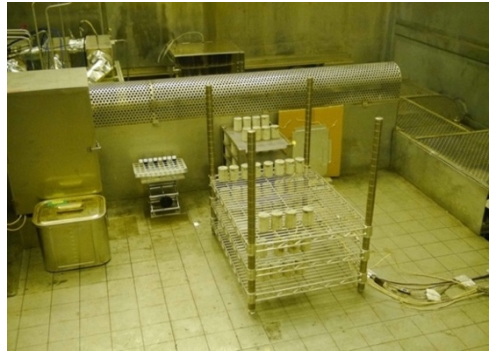

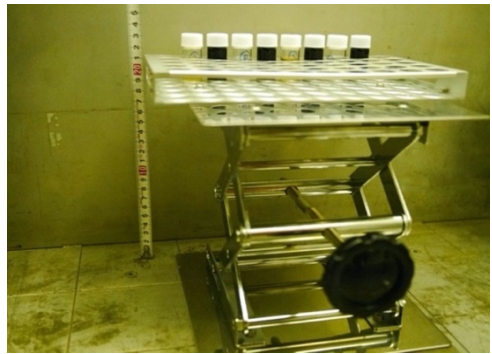

Fig. 10. Scenes from the radiation exposure test at QST Takasaki Research Institute. Top: Inside the irradiation chamber. Bottom: Samples of ferrofluid in small vials.

(2-2) Estimation of Radiation Dose Received by Ferrofluid in ILC Target

To compare with the results of the tests in the previous section (2-1), a simple simulation of radiation dose estimation was conducted using GEANT4, assuming two geometries of the ILC target. The parameters were based on the assumption of future luminosity upgrade (Main linac 2600 bunch operation). Figure 11 illustrates the schematics of the two geometries. The first (a) assumes a target diameter of 250 mm with no material between the shaft containing the ferrofluid and the outer ring where the beam is irradiated. The second (b) assumes a target diameter of 400 mm with spatial allowance between the shaft containing the ferrofluid and the outer ring where the beam is irradiated, accommodating a tungsten shield with a thickness of approximately 70 mm. The results showed that the irradiation dose for one year was 280 MGy in case (a) and 1.5 MGy in case (b). Considering the results of the radiation tests conducted at QST Takasaki Laboratory in the previous section (2-1), case (a) was deemed impractical. Therefore, from this point forward, case (b) with a diameter of 400 mm or larger was assumed as the base. Additionally, based on the results of the irradiation tests in section (2-1), only hydrocarbon-based ferrofluid would be considered for further testing.

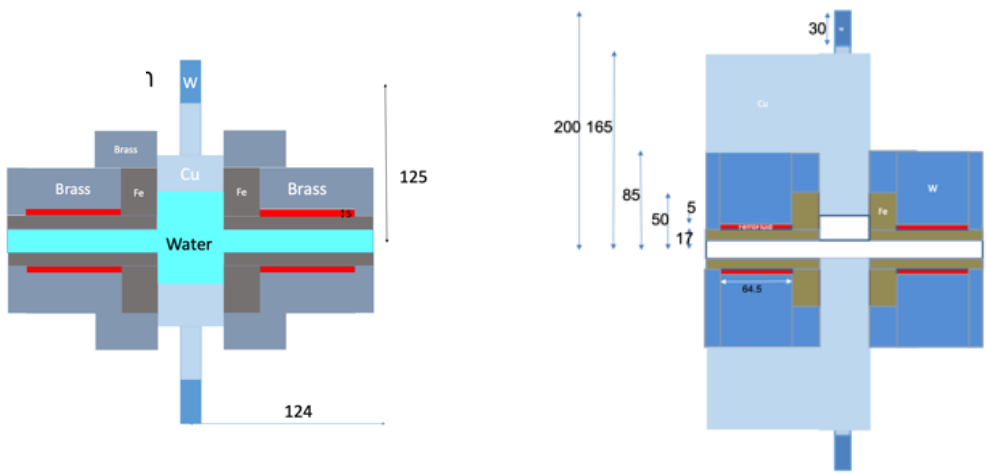

Geometry (a)
   Diameter: 250 mm, No tungsten shield
   Annual irradiation dose: 280 MGy

Geometry (b)
   Diameter: 400 mm, Tungsten shield installed
   Annual irradiation dose: 1.5 MGy

Fig. 11. Simulation of radiation dose estimation assuming two geometries.
Parameters correspond to future luminosity upgrade (Main linac 2600 bunch operation).

(3) Confirmation of Radiation Resistance of Ferrofluid: Second Stage

(3-1) Measurement of Viscosity

Stepwise radiation exposure up to 4.7 MGy was conducted on hydrocarbon-based ferrofluid in 2014. The dose of 4.7 MGy roughly corresponds to the radiation dose received over three years of operation after the luminosity upgrade of the ILC (Main linac 2600 bunch operation),

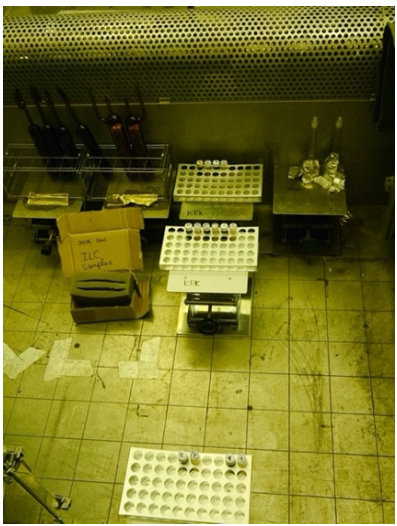

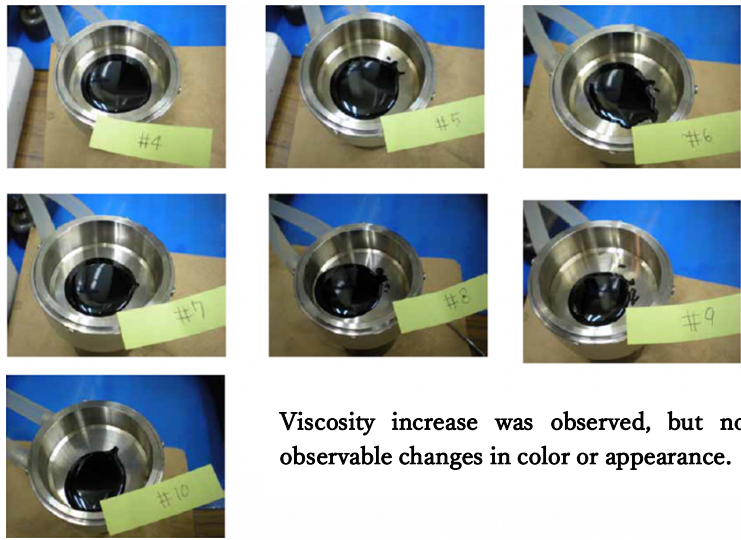

Viscosity increase was observed, but no observable changes in color or appearance.

Fig. 12. Samples arranged in the gamma-ray irradiation chamber.

Fig. 13. Samples of ferrofluid after irradiation. From top left: unirradiated, 0.5 MGy, 1 MGy, 1.5 MGy, 2.1 MGy, 3.1 MGy, 4.7 MGy.

assuming the design that protects the ferrofluid section with a radiation shield as adopted in the previous section (2-2) with geometry (b). The facility used was the same high-intensity gamma-ray facility at QST Takasaki Laboratory as in section (2-1). See Figures 12 and 13 for reference. Multiple samples were placed at different distances from the gamma-ray source, and the exposure dose was varied stepwise by adjusting the combination of distance and exposure time. The irradiated ferrofluids showed no visible abnormalities, and none decomposed up to 4.7 MGy. (Figure 13.)

The viscosity of the irradiated ferrofluid was measured using a cone-plate viscometer. It was found that the viscosity increased roughly proportionally with the exposure dose (Figure 14). The viscosity measured after 4.7 MGy irradiation was approximately 9000 mPa·s. This represents a significant increase compared to the viscosity of the unirradiated ferrofluid, which was about 1400 mPa·s. While the increased viscosity is not directly problematic for vacuum maintenance, the generation of frictional heat due to rotation increases with viscosity, especially at higher rotation speeds. Excessive temperature rise of the ferrofluid can increase its vapor pressure, necessitating attention to vacuum maintenance. Particularly at high rotation speeds, heat generation due to friction can become an issue. Therefore, determining the desired rotation speed is crucial. Based on the research conducted so far, we have decided to set the tangential velocity to approximately 5 m/s (chapter (I)) to disperse the thermal load from the electromagnetic shower on the target, and to ensure a target diameter of at least 400 mm to accommodate a shield protecting the ferrofluid from radiation (chapter(II), section (2-2)), We have also assumed a target diameter of 500 mm, which corresponds to a rotational speed of 200 rpm to achieve a tangential speed of 5 m/s. A rotational speed of 200 rpm falls within the range of low-speed rotation based on past experiences with rotating targets used in high-intensity X-ray generation devices, and the extent of viscosity increase observed in this irradiation test was deemed acceptable.

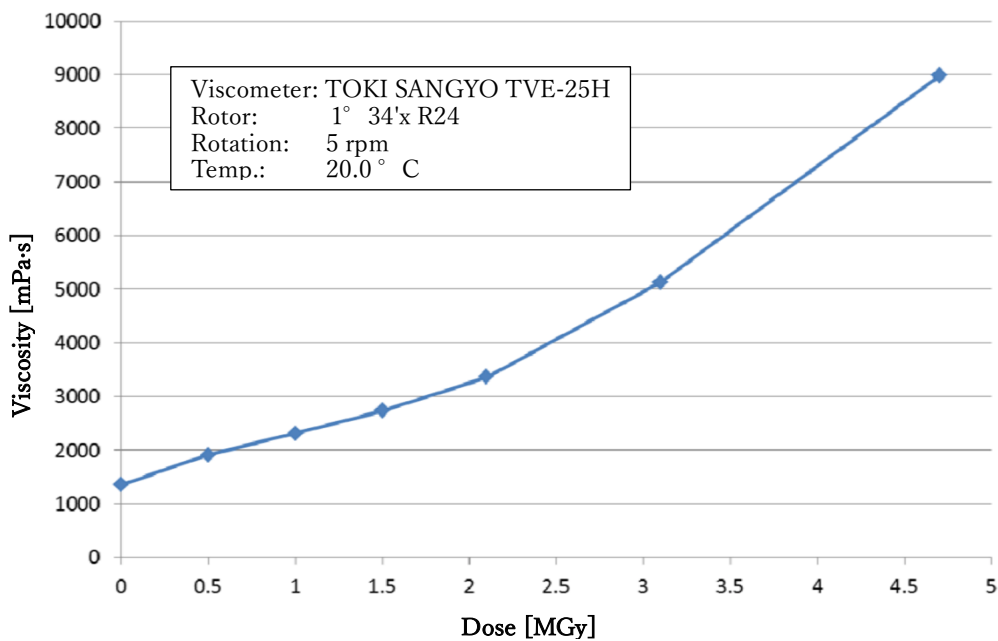

Fig. 14. Changes in viscosity as a function of irradiation dose.

(3-2) Applying Irradiated Ferrofluid in Commercial Small Rotation Targets for Test

Incorporation of the 4.7 MGy irradiated ferrofluid from the previous section (3-1) into commercial small rotating targets and conducting vacuum maintenance tests while rotating was performed (in 2014). The tests were conducted at a rotation speed of 600 rpm, approximately three times higher than the anticipated rated rotation speed of about 200 rpm. It was confirmed that vacuum maintenance was achieved without any issues and that there was no temperature rise in the sealing part (Figure 15). Although a sudden short-term pressure increase (spike) was observed once during the approximately two-week test run.

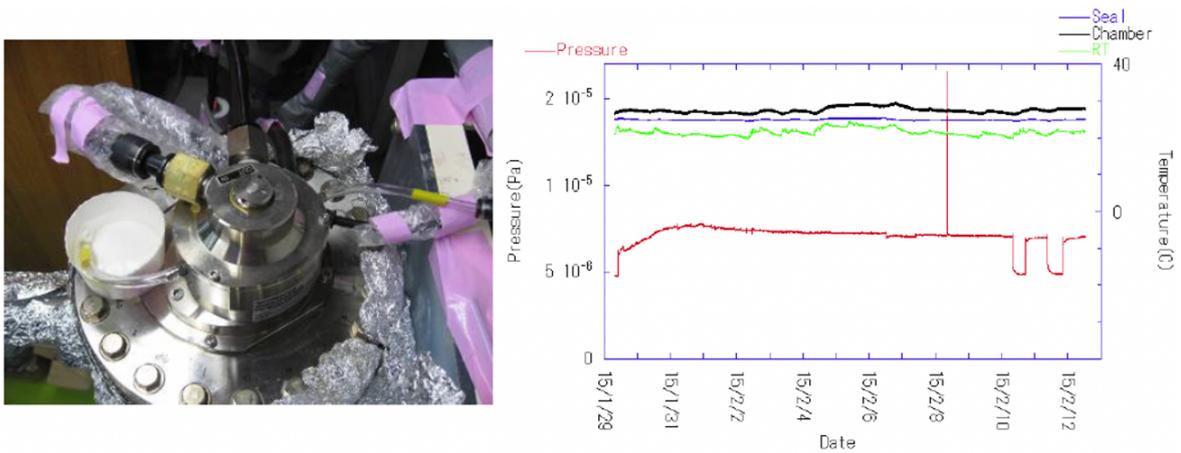

Fig. 15. Testing of commercially available small rotating target with ferrofluid irradiated at 4.7 MGy. Left: Test setup. Right: X-axis represents dates from January 29, 2019, to February 12, Y-axis represents pressure (red line, left scale), and temperatures of various components (right scale).

## (4) Confirmation of Radiation Resistance of the Entire Target Structure

In addition to ferrofluid, various materials such as permanent magnets, various resins, ball bearings, and various metals are used in the target. To examine the radiation resistance of all these materials, radiation irradiation was performed on the entire commercially available small rotating target in 2014 (Figure 16). The rotating target for the ILC positron source that we are producing differs significantly in size, weight, incoming beam energy, power received from the beam, etc., from the commercially available small rotating target. However, the basic structure and materials used are the same. Therefore, sufficient evaluation can be performed by irradiation testing on commercially available small rotating targets. The irradiation was conducted using a high-intensity cobalt gamma-ray source at the QST Takasaki Laboratory. The irradiation dose was set at 0.6 MGy at the motor location. This corresponds to two years' worth of irradiation at the motor position during ILC 1300 bunch operation. The appearance of the target after irradiation was darkened, which was attributed to the ozoneization of the air in the irradiation chamber during radiation exposure and subsequent oxidation by the ozone. Rotation and vacuum maintenance tests were conducted using the target after irradiation. The rotation speed during the test was set at 600 rpm. No abnormalities were observed in the motor, ball bearings, etc., and the rotation proceeded smoothly without any issues. Furthermore, vacuum testing while rotating confirmed that vacuum maintenance was achieved without any problems.

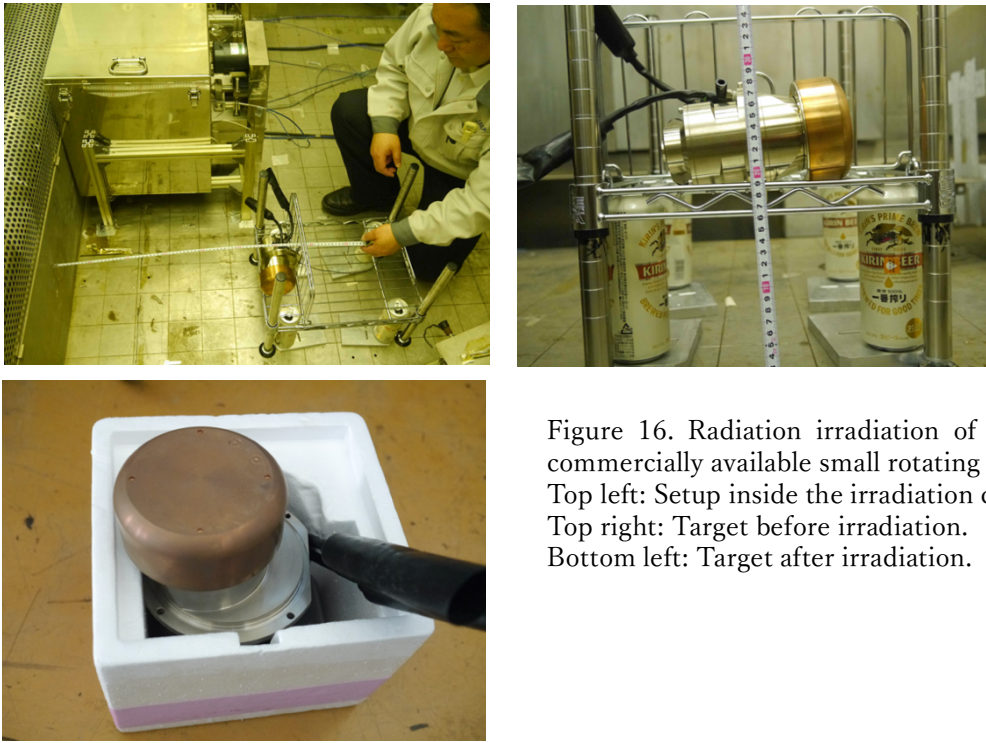

Figure 16. Radiation irradiation of the entire commercially available small rotating target.
Top left: Setup inside the irradiation chamber.
Top right: Target before irradiation.
Bottom left: Target after irradiation.

## (III) Production and Testing of Positron Rotating Target Prototype

Following the preliminary tests and evaluations described in (II), it has been determined that a rotating target using ferrofluids is feasible for use in the ILC. With the basic parameters such as target diameter determined, the overall configuration has been finalized. Figure 17 illustrates a conceptual diagram of the entire rotating target. In the following explanation, upstream and downstream are defined based on the direction of the electron beam incident for positron generation. The rotating mechanism assembly, including ball bearings and motors, is housed in the casing along with the ferrofluid rotary seal to maintain vacuum and is positioned upstream of the target disk. To achieve high vacuum, the ball bearings are not placed inside the vacuum. Instead, two ball bearing units are positioned on the atmosphere side, and the shaft is supported in a cantilevered manner. The target disk supported by the shaft is placed inside the downstream vacuum chamber. The entire target disk consists of a tungsten alloy ring attached to a copper disk, with the tungsten alloy section being irradiated by the electron beam to generate electromagnetic showers. Cooling water channels are provided within the copper disk, with water entering and exiting through the rotating shaft. No cooling water channels are incorporated into the tungsten alloy ring. By placing the rotating mechanism casing upstream, the amount of radiation hitting the ferrofluid and motors due to electromagnetic showers can be minimized. Although not depicted in the figure, a shield made of tungsten or similar material is installed between the electromagnetic shower generation section and the target rotation mechanism casing to protect the rotating mechanism including the ferrofluid seal from radiation. A flux concentrator generating pulsed high magnetic fields is placed downstream of the target disk, and further downstream, an acceleration tube is positioned, allowing for the capture of positrons by these components. Based on these configurations, we aimed to proceed with the detailed design and fabrication of the prototype.

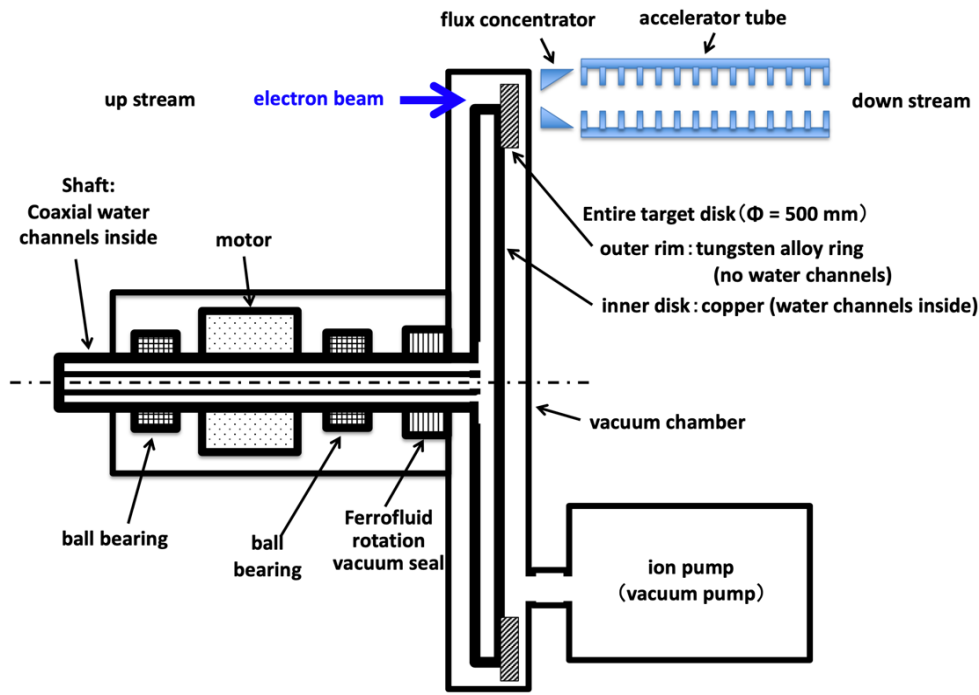

Figure 17. Conceptual diagram of the entire positron rotating target (cross-section including the central axis). Size is not scaled.

## (1) Design, Fabrication, and Testing of Prototype 1

The design and fabrication of a full-scale prototype were carried out in 2015. This prototype is referred to as Prototype 1 to distinguish it from later improved versions. At this stage, the positron-production target for the International Linear Collider (ILC) had been determined to have a diameter of approximately 500 mm and a rotation speed of approximately 200 rpm. Therefore, the prototype was fabricated accordingly. Figure 18 shows a schematic diagram combining Prototype 1 with its vacuum testing apparatus. Due to budget constraints, Prototype 1 does not include the target disk; however, its mechanical structure is designed to support the actual disk (diameter: 500 mm, mass: approximately 52 kg). Prototype 1 consists of components such as a motor, rotating shaft, ferrofluid seal, and ball bearings. Inside the rotating shaft, there are water channels for the water cooling of the target plate. As mentioned earlier, Prototype 1 does not have a target disk; therefore, it is designed to bend back and return to its original position at the tip of the rotating shaft. The water channels are arranged coaxially in a double-layered structure, with the outer layer serving as the inlet and the inner layer serving as the outlet.

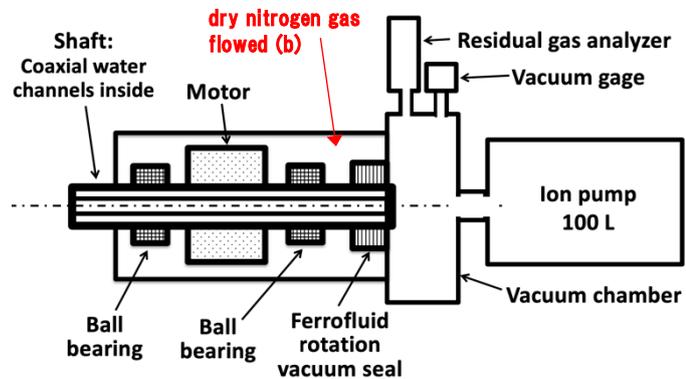

Fig. 18. Schematic diagram combining Prototype 1 with its vacuum testing apparatus.

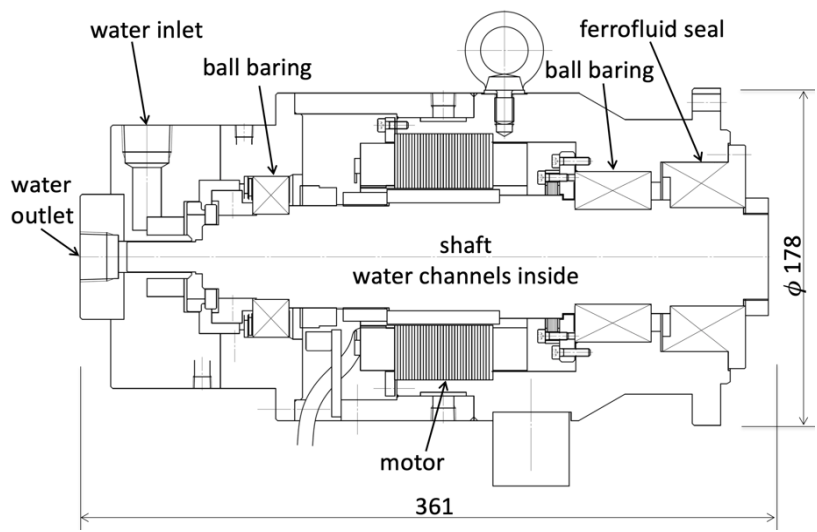

The ferrofluid seal unit is installed to be in contact with the outside of the rotating shaft, and this arrangement of water channels ensures that the ferrofluid seal unit is constantly maintained at the desired temperature by the circulation of cold water, safeguarding it from the heat generated by the beam on the target. In addition to the waterways within the rotating shaft, there are cooling water channels outside the ferrofluid seal unit within the housing, protecting the ferrofluid not only from target heat but also from heat during baking. Furthermore, there are waterways in the motor section for cooling purposes. Two ball bearing units are positioned at both ends of the motor to support the load of the target disk. The ball bearings and the motor are exposed to the atmosphere (Figure 19). On the right side of the figure, the ferrofluid rotary vacuum seal is placed, with the high vacuum region located to the right of the ferrofluid seal. Prototype 1 has a mass of approximately 45 kg.

Fig.19. Schematic of Prototype 1 Structure.
(Note: This figure is adapted from Reference [3].)

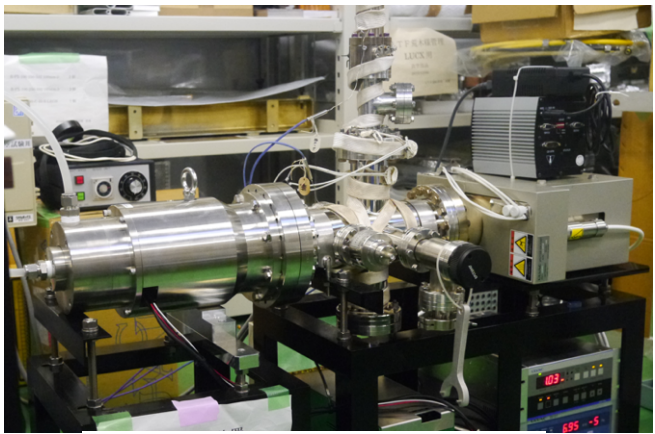

Fig. 20. Vacuum Test of Prototype 1

Using Prototype 1, a one-year-long rotating vacuum test was conducted in 2017. Figure 20 shows the test in progress. An ion pump with a pumping rate of 100 L/s was attached to the vacuum chamber to maintain the vacuum. The rotation speed was set to 225 rpm based on thermal and stress simulations (as mentioned later in V (3)). Vacuum maintenance tests were conducted at the rated speed of 225 rpm. A vacuum level of 5 x $10^{-7}$ Pa was achieved with Prototype 1, significantly surpassing the required vacuum level of $10^{-6}$ Pa for the ILC. Figure 21 shows the vacuum level variations over four days. While the base level of vacuum remained at 5 x $10^{-7}$ Pa, occasional jumps to the 2 x $10^{-5}$ Pa level were observed (We call them spikes). The typical duration of these jumps was a few minutes. From a practical perspective, it was considered that jumps of this magnitude did not pose a significant problem; however, to further improve this aspect, an improved version, as described later, was developed. The achieved vacuum level was measured under the following two conditions with Prototype 1: (a) when the atmospheric side (the part belonging to the atmosphere within the housing, including the motor and ball bearings) was open to the atmosphere, and (b) when the hole in the housing through which the motor cable passes was sealed with putty, and dry nitrogen gas was flowed into the atmosphere side. Among these conditions, the measurement conducted under condition (b) with dry nitrogen gas flow yielded better results. This was attributed to the suppression of moisture from the atmosphere from infiltrating into the ferrofluid. Figure 21 shows the results obtained under condition (b) with dry nitrogen gas flow.

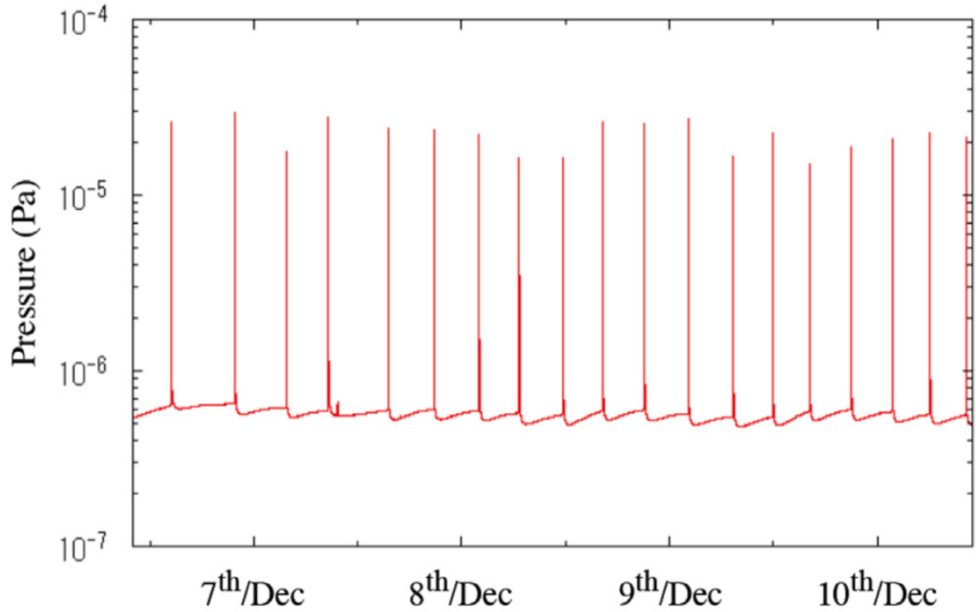

Fig. 21. Results of the vacuum test for Prototype 1. The rotation speed is 225 rpm. The horizontal axis represents the dates from December 6th to December 10th. The vertical axis represents the vacuum level (Pa). The base pressure is approximately $6\times10^{-7}$ Pa. Occasional rapid pressure increases ($6\times10^{-7}$ Pa to the mid-$10^{-5}$ Pa range, approximately 50-fold increase, lasting 1-2 minutes) are observed. Dry nitrogen gas is flowed near the ferrofluid seal on the atmospheric side at a rate of 100 mL/min. Note: The vacuum level was approximately $3.5\times10^{-6}$ Pa before nitrogen flow. (Note: This figure is adapted from Reference [3].)

## (2) Design, Fabrication, and Testing of Prototype 2

In response to the results obtained from Prototype 1 and aiming for higher performance, particularly in suppressing spikes, an improved version was designed and fabricated between 2019 and 2020. This improvement involved a two-stage configuration of ferrofluid seals, with an intermediate point evacuateded by an ion pump, referred to as Prototype 2. Due to budget constraints, Prototype 2 does not have a target disk; however, its mechanical structure is designed to support the actual disk (with a diameter of 500 mm and a mass of approximately 52 kg). Additionally, the shaft end is designed to allow for the addition of a target disk in the future. Prototype 2 consists of a motor, rotating shaft, ferrofluid seals, ball bearings, and other components. Inside the rotating shaft, there are water channels for the water cooling of the target plate. The routing of these channels is the same as in Prototype 1. Drawing from the successful experience of flowing dry nitrogen gas on the atmospheric side in Prototype 1 (refer to the previous section (1)), Prototype 2 anticipates a similar approach. Therefore, the holes in the housing for passing the motor cables were minimized and made easier to seal with putty. Prototype 2 has a mass of approximately 66 kg.

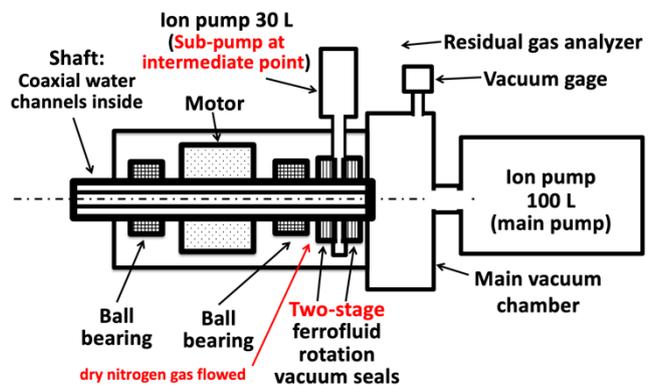

Fig. 22. Schematic diagram combining Prototype 2 with its vacuum testing apparatus.

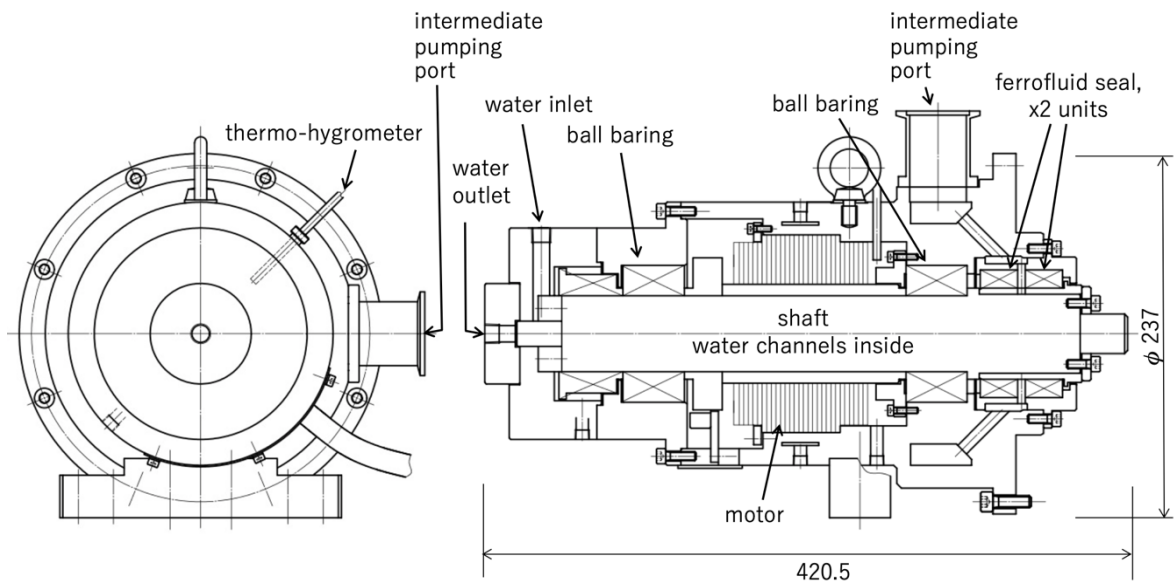

Fig.23. Schematic of Prototype 2 Structure.

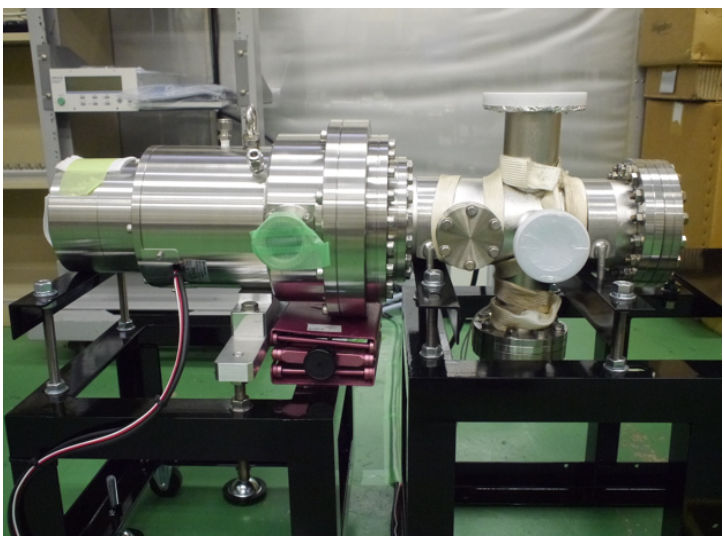

Fig. 24. Assembly of the vacuum test apparatus for Prototype 2. The duct protected by green masking tape in the center of the apparatus is the intermediate pumpig port.

At the vacuum chamber, similar to Prototype 1, a 100 L/s ion pump is installed for vacuum evacuation. Additionally, a 30 L/s ion pump is installed for vacuum evacuation at the intermediate point (Figures 22, 23, 24). Dry nitrogen gas is flowed into the atmospheric side inside the housing.

We conducted a vacuum maintenance test at the rated rotation speed of 225 rpm, similar to the testing of Prototype 1 (refer to the previous section (1)). The test commenced on September 9, 2020. Initially, the intermediate point was left open to the atmosphere, and only the main pump was operated, as shown in Figure 25. In this state, spikes similar to those observed in Prototype 1 occurred. Subsequently, on November 6, we initiated vacuum evacuation of the intermediate point using a turbomolecular pump. At this moment, the occurrence of spikes ceased. The vacuum level in the intermediate point was approximately 5 x 10⁻⁴ Pa. As shown in Figure 25, this led to a rapid improvement in the vacuum level of the main chamber. Furthermore, on December 3, we switched the vacuum evacuation of the intermediate point to an ion pump, achieving the pump configuration as per the design. . At this stage, the vacuum level in the intermediate exhaust section was approximately 3 x 10⁻⁶ Pa.

The results of this two-stage exhaust system in Prototype 2 were satisfactory. Spikes were eliminated, and on January 7, 2021, the main chamber reached a vacuum level of 1.1 x 10⁻⁶ Pa. We concluded the experiment at this point in preparation for the next modification (addition of a dummy disk). As indicated in the graph in Figure 25, the vacuum level was still improving (pressure was decreasing) at this stage. Hence, it is presumed that the vacuum would have further improved if the experiment had continued.

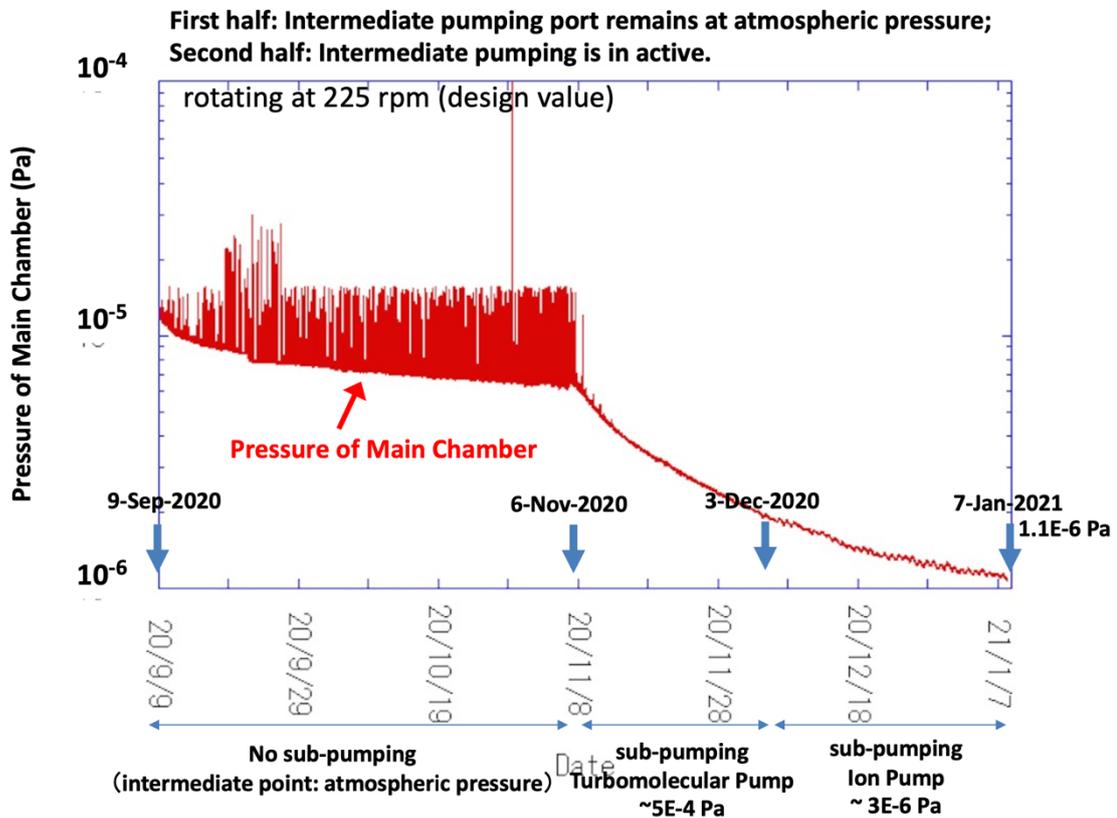

Figure 25. Results of the vacuum test for Prototype 2. The rotation speed is 225 rpm. The horizontal axis represents dates from September 9, 2020, to January 7, 2021. The vertical axis represents vacuum level (Pa). The achieved vacuum level is approximately 1.1 x 10⁻⁷ Pa at January 7th. Transient pressure rises (spikes) which was observed in Prototype 1 test are observed before the start of intermediate pumping but are absent after the start of intermediate pumping. The experiment was terminated on January 7th for the next modification (dummy disk addition). However, as this graph indicates, the vacuum level was still improving (pressure was decreasing) at this point, so it is expected that the vacuum would have further improved if the experiment had continued. Dry nitrogen is flowed near the ferrofluid seal inside the housing (atmosphere side).

(3) Addition of Dummy Disk to Prototype 2

(3-1) Stage 1

As stated in the preceding section (2), Prototype 2 was initially fabricated without a target disk; however, the shaft tip was designed to allow for the future addition of a target disk. Leveraging this design feature, a dummy disk of the same size, weight, and rotational inertia as the actual target disk was added to Prototype 2 for testing purposes (Figures 26, 27).

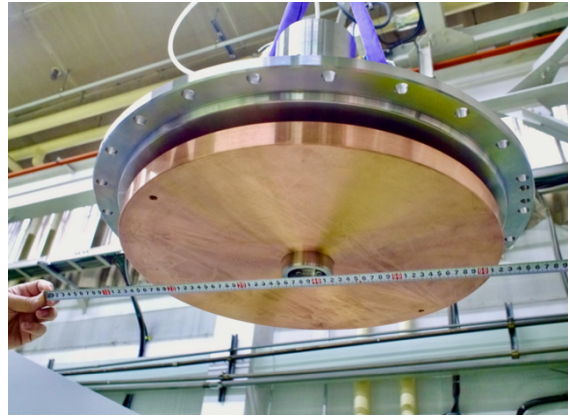

Fig. 26. Attachment of a dummy disk to Prototype 2. Photographed at Rigaku Corporation's factory. The prototype weighs approximately 118 kg with the dummy disk attached, which has a diameter of approximately 500 mm.

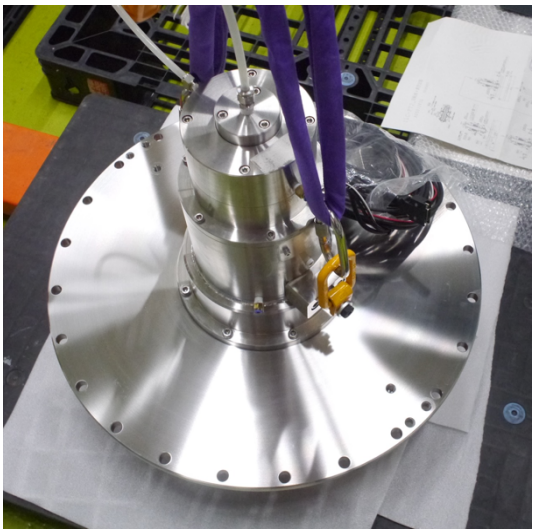 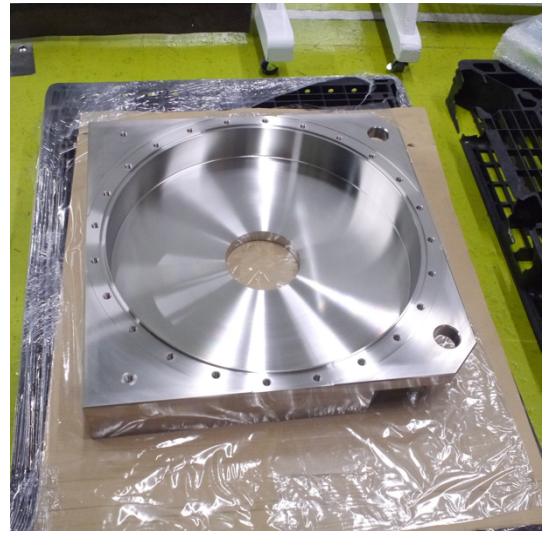

Fig. 27. Installation work in progress for Prototype 2 with the dummy disk mounted into a large vacuum chamber. Photographed at Rigaku Corporation's factory. The prototype weighs approximately 325 kg when integrated into the vacuum chamber.

The actual target disk is predominantly made of copper, with its outer rim consisting of a tungsten alloy ring, which serves as the target. Additionally, within the copper disk of the actual target, cooling channels are incorporated. Due to budget constraints, the dummy disk is constructed solely of copper and does not contain any cooling channels. Consequently, the water channels within the rotating shaft are not connected to the disk, and the cooling water continues to flow as before, returning to its source at the tip of the rotating shaft. The ion pumps in the vacuum chamber and intermediate point maintain the same pumping rates as when the disk is absent, at 100 L/s and 30 L/s, respectively. Dry nitrogen gas is circulated within the atmospheric side of the housing.

Rotation and vacuum maintenance test were conducted on Prototype 2 equipped with the dummy disk (Figure 28). The rotation speed was set to the rated value of 225 rpm. In contrast to the test without the disk, both the main chamber and intermediate point were evacuated using ion pumps from the beginning.

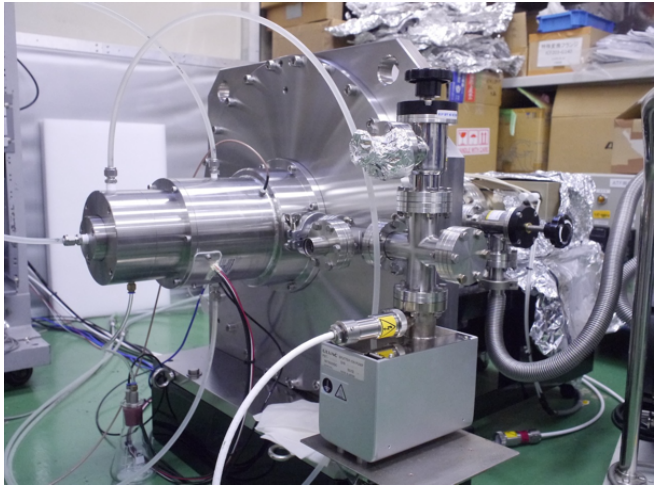

Fig. 28. Prototype 2 with dummy disk during test preparation. The square box-shaped object in the foreground is the ion pump for intermediate pumping, rated at 30 L/s.

Figure 29 illustrates the changes in vacuum levels for both the main chamber and intermediate point. The vacuum level in the main chamber reached 7.4 x 10⁻⁷ Pa approximately three months after the start of the experiment. Spikes, which had disappeared during the experiment without the disk, unfortunately reappeared when the disk was attached. However, the height of the spikes was significantly smaller than during Prototype 1. While spikes during Prototype 1 were approximately 40 times the base pressure, spikes with the dummy disk-equipped Prototype 2 were only about 1.5 times the base pressure. Additionally, the frequency of spikes was reduced. Overall, the performance was deemed satisfactory.

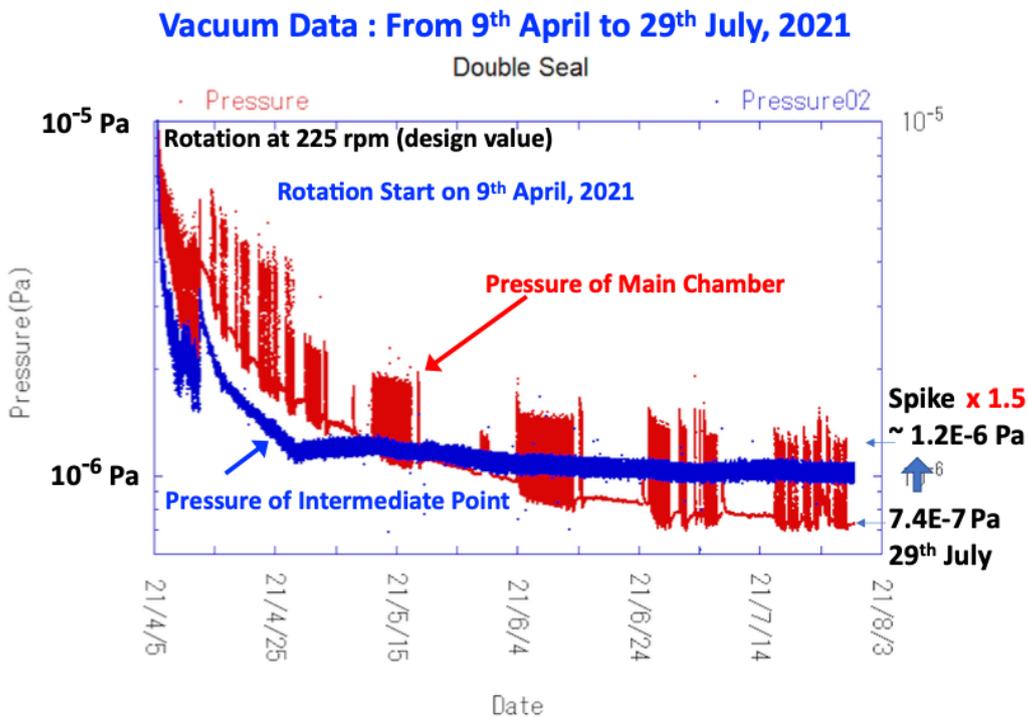

Fig. 29. Results of the vacuum test for Prototype 2 with a dummy disk. Stage 1, intermediate pumping at 30 L/s. The horizontal axis represents dates from April 5, 2021, to August 3, 2021. The vertical axis represents vacuum level (Pa). The achieved ultimate vacuum level (base pressure) is approximately 7.4 x 10⁻⁷ Pa. Intermediate pumping is performed from the initial startup. Transient pressure rises (spikes) are observed. However, the pressure peaks at around 1.2 x 10⁻⁶ Pa, approximately 1.5 times the base pressure.

(3-2) Stage 2

A long-term rotational vacuum test was conducted with the ion pump of the intermediate point change to 50 L/s (2022-2023). The results are shown in Figure 30. As shown in the figure, the results were favorable, indicating significant improvement.

To clearly demonstrate the effect of increasing the capacity of the intermediate point pump from 30 L/s to 50 L/s, graphs before (30 L/s) and after enhancement (50 L/s) are shown side by side in Figure 31.

The improvement effects are as follows:

    Improved ultimate vacuum level (base pressure 7.4 x $10^{-7}$ Pa → 5 x $10^{-7}$ Pa)
    Improved vacuum level during spikes (spike section 1.2 x $10^{-6}$ Pa → 6.5 x $10^{-7}$ Pa)
    Significant reduction in spike frequency.

These three points clearly demonstrate the improvement effects.

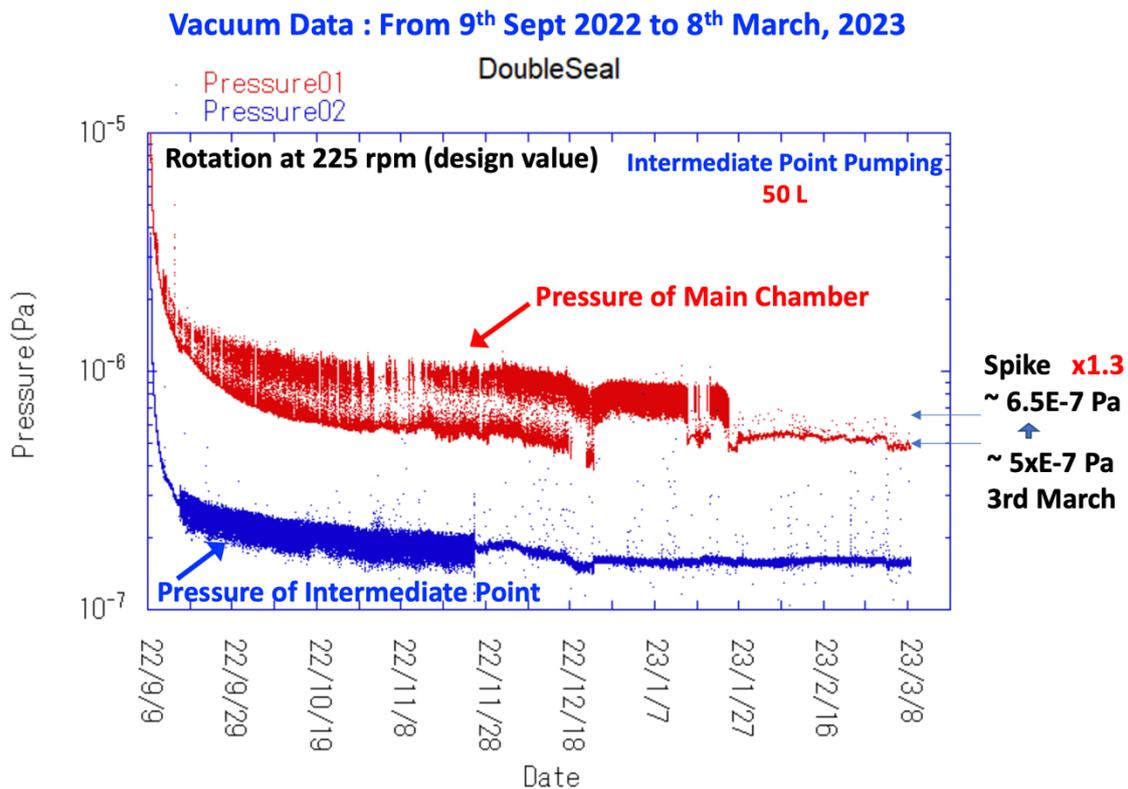

Fig. 30. Results of the vacuum test for Prototype 2 with a dummy disk. Stage 2, intermediate pumping at 50 L/s. The horizontal axis represents dates from September 9, 2022, to March 8, 2023. The vertical axis represents vacuum level (Pa). The achieved ultimate vacuum level (base pressure) is approximately 5x$10^{-7}$ Pa. Intermediate pumping is performed from the initial startup.

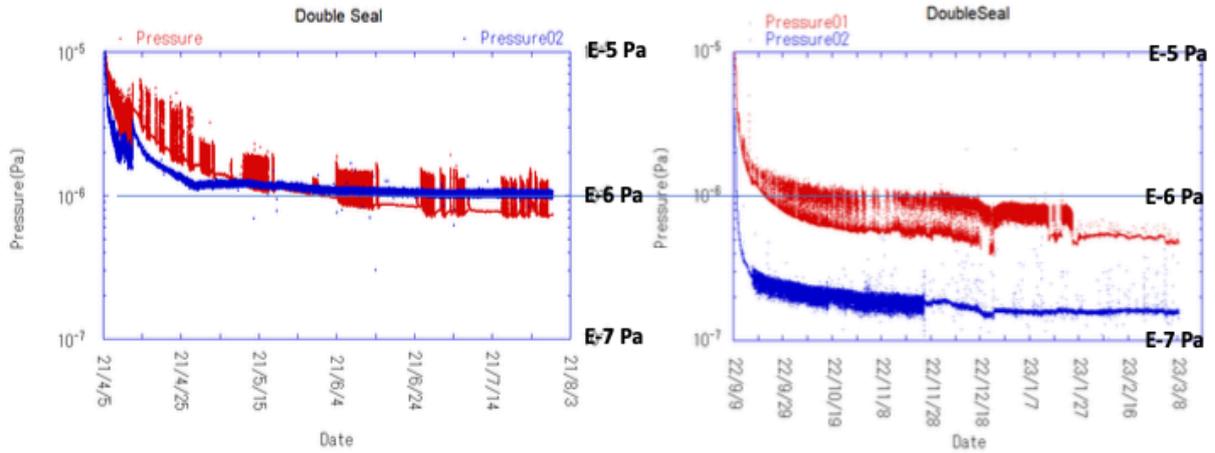

Fig. 31. Results of the vacuum test for Prototype 2 with the dummy disk. Comparison between Stage 1 (intermediate pumping at 30 L/s) and Stage 2 (intermediate pumping at 50 L/s). The left side represents Stage 1, while the right side represents Stage 2. The horizontal axis represents dates, with the left side ranging from April 5, 2021, to August 3, 2021, and the right side ranging from September 9, 2022, to March 8, 2023. The vertical axis represents vacuum level (Pa). The vertical axis scale is consistent on both the right and left sides. Note that the horizontal axis scales are not aligned.

(4) Notes on Prototype Design and Fabrication

(4-1) Size and Mass

It is worth noting that a relatively compact and lightweight design was achieved. This is particularly important for target replacement. In the ILC TDR [1], it is anticipated that targets will be replaced annually. This is primarily to address the degradation of target plates due to heat and stress from electromagnetic showers, based on experience from the SLC. While the scheme for target replacement is beyond the scope of this paper and detailed discussion is omitted, it is anticipated that targets, which are highly activated, will be removed as an integral part with surrounding equipment through remote or semi-remote operations. We believe that the compact and lightweight design contributes to such a scheme. Further details regarding experiences at the SLC will be discussed in Chapters (V), Appendices 2, and 3.

(4-2) Materials of Major Components

Both Prototype 1 and Prototype 2 are made of SUS for the shafts and main parts of the housing.

## (IV) Tests to Understand the Characteristics of Ferrofluids Regarding Radiation

### (1) Confirmation of Radiation Resistance of Ferrofluids: Understanding of the Changes 1

The radiation resistance of ferrofluids was confirmed through tests described in section (II) (2)-(3), indicating that while the viscosity of ferrofluids increases under radiation exposure, there are no problems in practical point of view. However, here we took a further steps to investigate the changes in molecular structure of ferrofluids caused by radiation exposure. (in February 2018). The tests focused on samples both ferrofluid and solely base oil. The objects are the hydrocarbon-based ferrofluid and its base oils which is intended for use in the ILC positron production target (Figure 32).

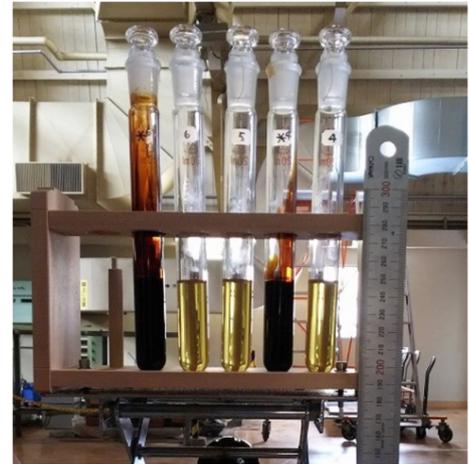

The irradiation doses were set at seven levels: no irradiation, and 0.097, 0.47, 0.94, 2.1, 2.8, and 4.7 MGy. Following irradiation, changes in molecular weight and molecular structure were investigated through viscosity measurements, gel permeation chromatography (GPC), and UV-Vis spectroscopy. While viscosity measurements had been previously conducted, they were limited to examining viscosity changes only in ferrofluids (see II(3-1)). The unique feature of this test is that both the base oil and the ferrofluid were exposed to radiation in the same way and the changes in viscosity were compared.

Fig. 32. Samples under preparation. The transparent samples are the base oil samples, while the black samples are the ferrofluid samples.

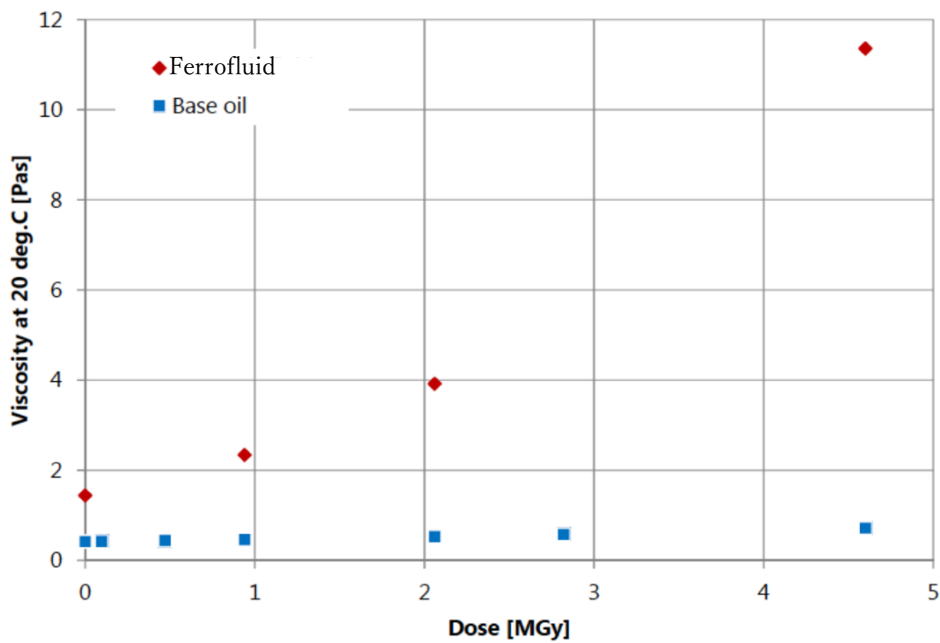

Fig. 33. Changes in viscosity as a function of irradiation dose. Red circles represent the ferrofluid, while blue circles represent the base oil. The viscosity of the ferrofluid significantly increases, whereas no increase in viscosity was observed for the base oil alone.

Analysis Results of Radiation Tests in February 2018

- The results of viscosity measurements indicate a significant increase in the viscosity of the ferrofluid due to gamma-ray irradiation, whereas no noticeable increase in viscosity was observed in the base oil alone (Figure 33).

- Gel permeation chromatography (GPC) reveals the generation of higher-molecular-weight components as a result of gamma-ray irradiation. This is presumed to be due to cross-link of molecular chains induced by radiation. A tendency of increased production of higher-molecular-weight components with increasing irradiation dose was observed (Figure 34).

- However, the majority of molecules remain unchanged even after irradiation at 4.7 MGy. This is considered a contributing factor to the continued functionality of the vacuum seal even after irradiation at 4.7 MGy.

- Molecules with small molecular weights were not extensively generated. In other words, significant decomposition of polymers did not occur. The generation of higher-molecular-weight components due to radiation occurred almost equally in both base oil alone and ferrofluid. Additionally, in both base oil alone and ferrofluid, the absence of extensive generation of molecules with small molecular weights was observed.

- Despite the generation of higher-molecular-weight components in both base oil and ferrofluid as a result of radiation, the increase in viscosity was observed only in the ferrofluid. Therefore, it is considered that the generation of higher-molecular-weight components is not the cause of the increase in viscosity.

- Comparing the molecular weight distributions of base oil and ferrofluid irradiated with the same dose, no significant differences were observed. We think that the presence of magnetite dose not make significant promotion of molecular chain scission or cross-link, by radiation.

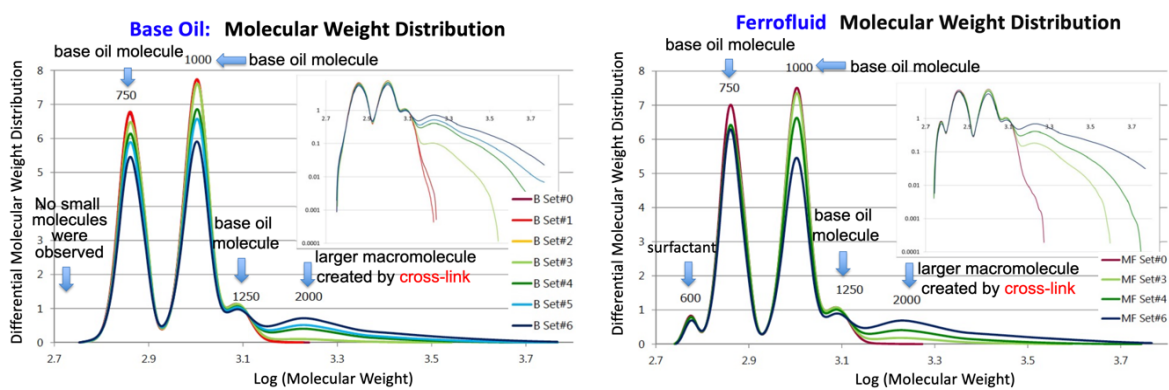

Fig. 34. Changes in molecular weight due to radiation exposure. Left: Base oil, Right: Ferrofluid. There is no significant difference between the two. The x-axis represents Log(molecular weight), with the peak numbers indicating approximate molecular weights. Peaks at around 750, 1000, and 1250 correspond to the base oil. The peak at 600, observed only in the ferrofluid sample, corresponds to the surfactant around the magnetite in the ferrofluid. The broad peak around molecular weight 2000 is believed to be polymer generated by crosslinking, because originally there is no such high molecular weight components in the base oil. The y-axis is proportional to the relative abundance of molecules. Seven stages of irradiation—no irradiation, 0.097, 0.47, 0.94, 2.1, 2.8, 4.7 MGy—are denoted as Set#0, Set#1, Set#2, Set#3, Set#4, Set#5, Set#6, respectively.

- We suspect that the reason for the observed increase in viscosity only in the ferrofluid may be attributed to the potential alteration of surfactant characteristics covering the surface of magnetite, which is dispersed in the base oil, due to radiation.
- Our ferrofluid base oil tends to emit a significant portion of the received radiation energy as UV and visible light outside the system, thereby making it less prone to induce prominent chemical reactions, which we consider a reason for its radiation resistance [7]. This base oil possesses a molecular structure containing aromatic groups, which is believed to facilitate the emission of received radiation energy as UV to visible light outside the system.
- Unfortunately, significant information could not be obtained from UV-Vis (UV-visible spectroscopy) results.

(2) Confirmation of Radiation Resistance of Ferrofluids: Understanding of the Changes 2

In February-March 2019, we continued investigating the structural changes in ferrofluid induced by radiation, following the previous year's work.

Compared to the previous year, we conducted more detailed analyses using not only liquid chromatography but also UV-Vis spectroscopy. Additionally, we collected gases generated during radiation exposure and confirmed their consistency with the data on changes in ferrofluid through gas chromatography.

＊ Stepwise radiation exposure up to 4.7 MGy was performed on hydrocarbon-based fluids (following the same dose setting approach as in 2014 and 2018).

＊ Unlike previous tests where samples were irradiated in the atmosphere, samples were placed in sealed glass containers under vacuum for radiation exposure to facilitate gas chromatography (Figure 35).

＊ Irradiated samples included base oil and ferrofluid.

＊ Gas chromatography (GC) was conducted to analyze the collected gases.

＊ The surface of magnetite in the irradiated ferrofluid was examined using Fourier Transform Infrared Spectroscopy (FT-IR).

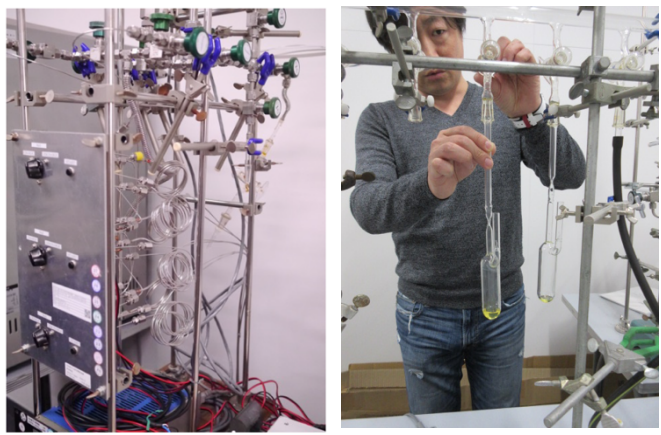

Fig. 35. Left: Gas chromatography (GC) analyzer used (QST Takasaki Research Institute), Right: Sample enclosed in sealable small glass bottle.

＊ Analysis using gel permeation chromatography (GPC) was also carried out (while GPC analysis was conducted also in 2018, the irradiation under vacuum conditions distinguishes this study).

＊ Changes in viscosity were also investigated (while viscosity confirmation was conducted in 2018, the difference exists in the irradiation under vacuum conditions).

<div align="center">

## Produced Gas at 1.49MGy base oil only samples

</div>

| | | Peak Area | Press. [%] | molar [mol] | 発生ガス [mol/g] | G値 |
|---|---|---|---|---|---|---|
| GC1 | H2(1.85) | 13283 | 57.49 | 3.35E-4 | 8.67E-5 | 5.61E-1 |
| | O2(2.94) | 2638 | 0.70 | 4.08E-6 | 1.05E-6 | 6.83E-3 |
| | N2(3.97) | 10658 | 2.85 | 1.66E-5 | 4.30E-6 | 2.79E-2 |
| | CH4(6.70) | 11211 | 1.34 | 7.82E-6 | 2.02E-6 | 1.31E-2 |
| GC2 | CO2(3.84) | 546 | 0.05 | 2.63E-7 | 6.82E-8 | 4.41E-4 |
| | C2H4(6.14) | 17841 | 1.50 | 8.75E-6 | 2.26E-6 | 1.47E-2 |
| | C2H6(8.25) | 15331 | 1.13 | 6.58E-6 | 1.70E-6 | 1.10E-2 |
| unknown | | | 34.94 | 2.03E-4 | 5.27E-5 | 3.41E-1 |
| Total | | | 100.00 | 5.82E-4 | 1.51E-4 | 9.76E-1 |

導入 36.0Torr

<div align="center">

## Produced Gas at 1.49MGy ferrofluid samples

</div>

| | | Peak Area | Press. [%] | molar [mol] | 発生ガス [mol/g] | G値 |
|---|---|---|---|---|---|---|
| GC1 | H2(1.85) | 13052 | 56.80 | 3.33E-4 | 1.01E-4 | 6.57E-1 |
| | O2(2.94) | 4203 | 1.12 | 6.58E-6 | 2.00E-6 | 1.30E-2 |
| | N2(3.97) | 17445 | 4.70 | 2.76E-5 | 8.39E-6 | 5.43E-2 |
| | CH4(6.70) | 9035 | 1.09 | 6.39E-6 | 1.94E-6 | 1.26E-2 |
| GC2 | CO2(3.84) | 20444 | 1.70 | 9.99E-6 | 3.04E-6 | 1.97E-2 |
| | C2H4(6.14) | 12858 | 1.09 | 6.39E-6 | 1.94E-6 | 1.26E-2 |
| | C2H6(8.25) | 10440 | 0.77 | 4.54E-6 | 1.38E-6 | 8.95E-3 |
| unknown | | | 32.73 | 1.92E-4 | 5.85E-5 | 3.78E-1 |
| Total | | | 100.00 | 5.87E-4 | 1.79E-4 | 1.16E+0 |

導入 35.8Torr

Table 1. Analysis Results of Gas Chromatography. Left: Base oil, Right: Ferrofluid. Focus on the G values.

＜Conclusions＞

＊Examination of the gas chromatography (GC) results (Table 1) reveals that the G values for $H_2$ gas production were 0.66 in the ferrofluid and 0.56 in the base oil. These values are significantly smaller compared to the G value of approximately 3 for polyethylene. This small G value is consistent with the conclusion drawn from gel permeation chromatography (GPC) analysis, indicating "no significant decomposition of polymers constituting the base oil." This is the most crucial finding of this study.

＊ Furthermore, a comparison of the GC results shows that the generation rates of gases such as $H_2$, $CH_4$, $C_2H_4$, and $C_2H_6$ were nearly identical between the ferrofluid and base oil. However, the generation rate of $CO_2$ was significantly higher in the ferrofluid compared to the base oil. This is presumed to be attributed to the surfactant covering the surface of magnetite in the ferrofluid.

＊The results of gel permeation chromatography were consistent with those of 2018, regardless of whether the samples were irradiated in the atmosphere or under vacuum. In actual usage environments, the ferrofluid interfaces with one side in a vacuum and the other side in contact with the atmosphere. It was confirmed that decomposition due to radiation is small in both of these conditions.

＊ Unfortunately, significant information could not be obtained from the results of Fourier Transform Infrared Spectroscopy (FT-IR).

### (3) Confirmation of Radiation Resistance of Ferrofluid: Testing with Prototype 1

In the preceding work II (3-2), satisfactory results were obtained by incorporating the irradiated ferrofluids into a commercially available small rotating target for testing. However, these was the small target, and the vacuum level ranged from approximately 5 x 10⁻⁶ Pa to 7 x 10⁻⁶ Pa. For the ILC, a larger target and slightly higher vacuum levels are required. Therefore, ferrofluid irradiated to 4.7 MGy, was incorporated into Prototype 1, the full-scale prototype used for vacuum testing in (III) (1) and a vacuum test was performed. The rotation speed was set at 225 rpm, and dry nitrogen gas was flowed on the atmospheric side. Continuous operation was conducted for approximately one year, during which the vacuum level fluctuated between approximately 3.3 x 10⁻⁶ Pa and 3.7 x 10⁻⁶ Pa (Figure 36). Using ferrofluids irradiated for six years of ILC standard mode (1300 bunch operation mode), the requirements of ILC TDR were met.

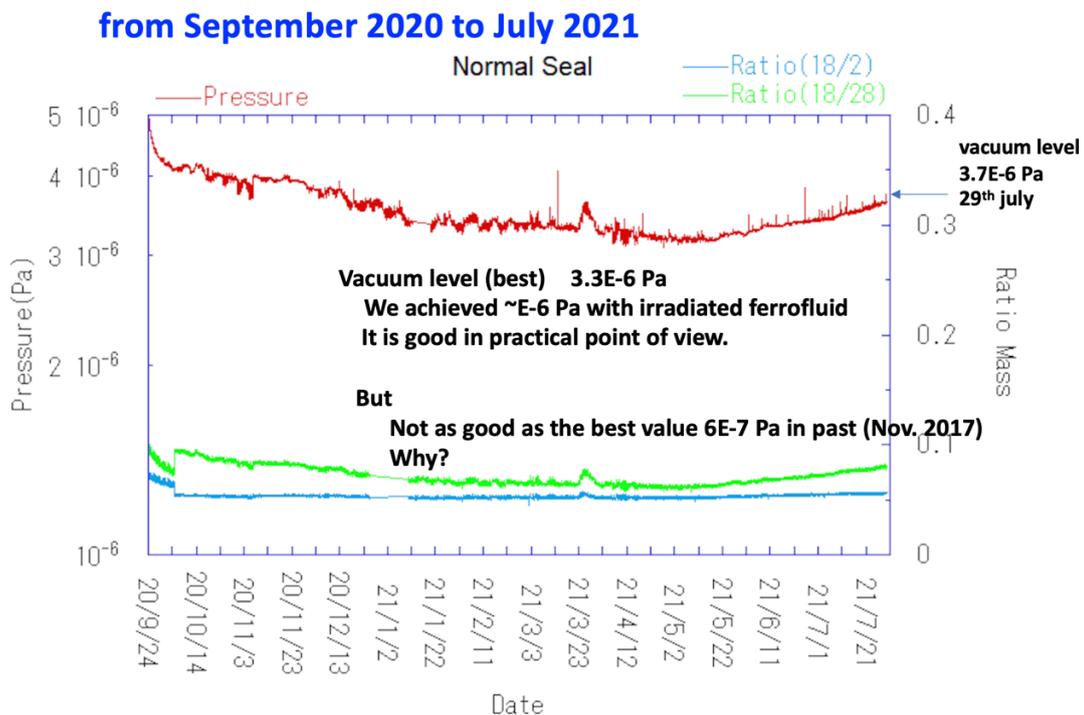

Fig. 36. Vacuum Test with Prototype 1 Incorporating Ferrofluid Irradiated with 4.7 MGy (225 rpm)

However, the vacuum level achieved in this experiment did not reach the 6 x 10⁻⁷ Pa achieved in a previous experiment as shown in Figure 21 of (III) (1). It remained close to the initial vacuum level of Prototype 1 before flowing dry nitrogen gas. Three hypotheses can be considered for this discrepancy:

Degradation of ferrofluids due to radiation exposure.
Deterioration in the performance of the ion pump due to aging.
Humidity in the housing is rather high, ranging from 10% to 20%.

Regarding the aspect of humidity, it should be noted that this test was conducted from September 2020 to July 2021 with dry nitrogen gas flowing; however, the humidity within the housing ranged from 10% to 20%. This humidity was significantly worse than the humidity during operation with dry nitrogen gas from 2018 to 2019, which was at 0% (below the sensitivity limit of the detector). The cause of this elevated humidity remains unclear. The housing of Prototype 1 was not originally designed for the flow of dry nitrogen gas, with a large hole for passing motor cables sealed with rubber sheets and putty to allow gas flow. It is possible that this seal was weakened, or the amount of steam from the rotating water seal for sending cooling water to the rotating shaft increased. Furthermore, the humidity within the housing during the data collection for Figure 21 of (III) (1), when the vacuum level of 6 x 10⁻⁷ Pa was achieved (2017), remains unknown as a humidity sensor was not installed at that time, making a direct comparison of humidity impossible. Despite some lingering questions, the confirmation of meeting the requirements of the TDR led to the conclusion of the experiment at this point.

(4) Direct Observations of Ferrofluids

Long-term continuous operation at vacuum levels in the range of 10⁻⁶ Pa poses a new challenge for the ferrofluid rotary seal system. Therefore, in addition to measuring the vacuum level while rotating, direct observations of the ferrofluid itself was also performed.

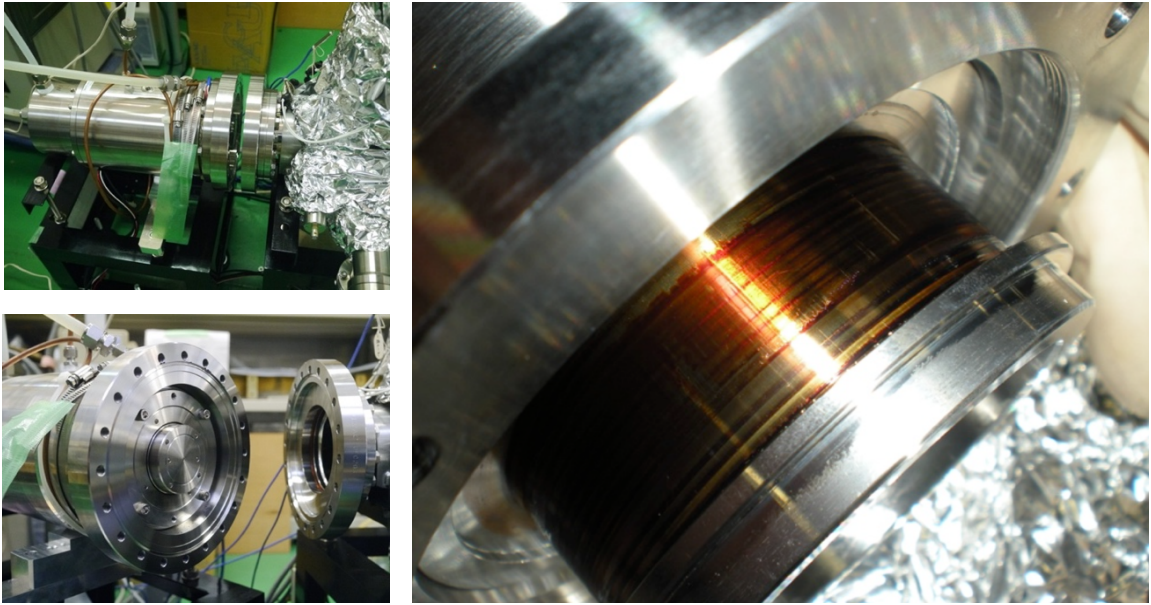

Fig. 37. Visual Inspection of Ferrofluid Rotary Seal after Initial Three Months of Continuous Operation of Prototype 1. Top left: Separation of target and vacuum container. Bottom left: Separation of target and vacuum container. Right: Opening of the ferrofluid rotary seal unit.

(4-1) Visual Inspection of Ferrofluids in Prototype 1 after Operation

Following the commencement of vacuum and rotation tests on Prototype 1, the target was disassembled and the vacuum seal components were inspected visually after the initial approximately three and a half months of continuous vacuum rotation testing (2017). No abnormalities were observed during the visual inspection. There was no noticeable reduction in oil volume as discernible by visual inspection. Additionally, there were no observations of evaporated or dried powdered ferrofluids during the examination (Figure 37).

(4-2) Evaporation Test Using Evaporation Pan

The ferrofluid in the rotary seal section appears as a ring with a diameter of approximately 3 cm and a width of about 100 $\mu$m when viewed from the vacuum container side. In other words, the exposed area is extremely small. In this experiment, the ferrofluid was placed in an evaporation pan with a diameter of 3 cm and the pan was placed in the vacuum chamber to increase the exposed area. The vacuum chamber was then evacuated using an ion pump (Figure 38). A residual gas analyzer (RGA) was installed in the vacuum chamber. The upper limit of hydrocarbon gases was determined after 12 days of operation of the ion pump. The upper limit is the value that is determined by adding unknown components including other hydrocarbons and unknown components (including errors) that arise in the analysis to what was identified as $CH_4$. The purpose of this experiment is to determine the upper limit of the amount of small molecules (CH molecules) that are released as gases, assuming that the polymers in the ferrofluid decompose. Based on the experimental results accumulated through this research, we believe that a significant amount of polymers will not decompose into small molecules and be released as gases into the vacuum chamber. However, this experiment was conducted to directly confirm this assumption.

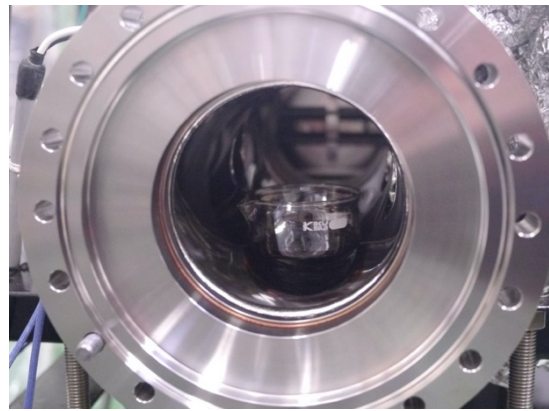

Fig. 38. Ferrofluid placed in an evaporation pan and installed inside the vacuum chamber.

| | |
|---|---|
| Total Pressure: | $9.27 \times 10^{-7}$ Pa |
| Upper Limit of Partial Pressure of Hydrocarbon Gases: | $7.36 \times 10^{-8}$ Pa |

From this value, a model was assumed, considering the conductance from around the target to the first acceleration tube downstream and the arrangement of vacuum pumps, to determine the upper limit of the partial pressure of hydrocarbon gases at the upstream end of the first acceleration tube.

Upper Limit of Partial Pressure of Hydrocarbon Gases at the upstream end of the first acceleration tube: $1.5 \times 10^{-10}$ Pa

This value is sufficiently small, indicating that the operation of the acceleration tube is not hindered.

Note: For detailed experimental and analytical procedures, refer to "Appendix 1".

## (V) Consideration of Target Heating, Cooling, and Stress

Simulation of heating, cooling, and stress was conducted in stages.

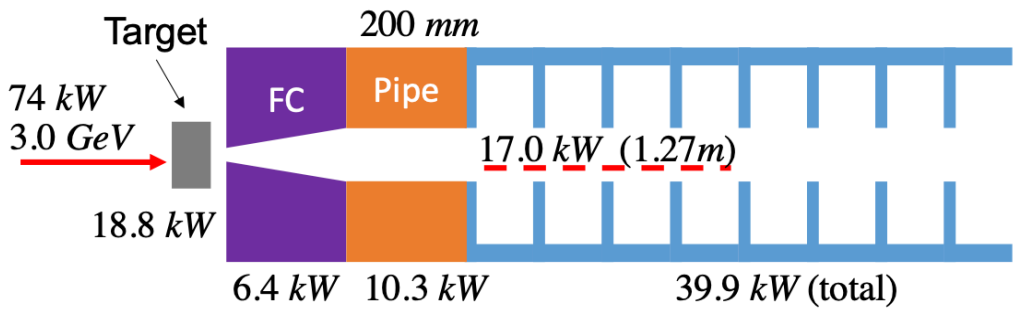

| | |
|---|---|
| **Target** | 18.8 *kW* |
| **Flux Concentrator** | 6.4 *kW* |
| **Pipe (absorber)** | 10.3 *kW* |
| **Capture Linac** | 39.9 *kW* (total), of which 17.0 *kW* goes 1$^{st}$ accelerator tube (1.27m) |

Fig. 39. Average heat load of the positron production target and its immediate downstream. Values during ILC250 GeV standard luminosity operation.

Figure 39 depicts the results obtained through simulation for the positron target and its downstream average heat load at the standard parameters of the ILC's TDR at 250 GeV, specifically with a positron yield of 2 x 10$^{10}$ per bunch and 1312 bunches per pulse in the main linac. The simulation was conducted using GEANT4. It is mandated as a design guideline for the ILC positron source to ensure a safety factor of 1.5 for the positron yield after the damping ring (DR) injection. Therefore, the positron source is required to generate 3 x 10$^{10}$ positrons per bunch within the DR acceptance. Currently, the design specifies an injection electron beam energy of 3 GeV, and the efficiency of generating positrons within the DR acceptance from the injection electron beam is determined from simulation to be e+/e- = 1.28 [3]. Thus, the required number of electrons per bunch in the injection electron beam is calculated to be 2.34 x 10$^{10}$. These are to be injected into the target at the rate of 1312 bunches x 5 times per second (Note: repetition frequency of the main linac is 5Hz). Therefore, the average power of the injection electron beam is 74 kW [3]. At this rate, the average energy deposition to the target (tungsten alloy, thickness 16 mm) amounts to 18.8 kW [3].

In terms of design considerations, the key points are to remove this average heat load through water cooling, to mitigate concentration of heat in one location through proper rotation, and to suppress stress within the target metal. With these three goals in mind, simulations were conducted based on the considerations outlined in (1) to (5) below.

### (1) Initial Trial and Error

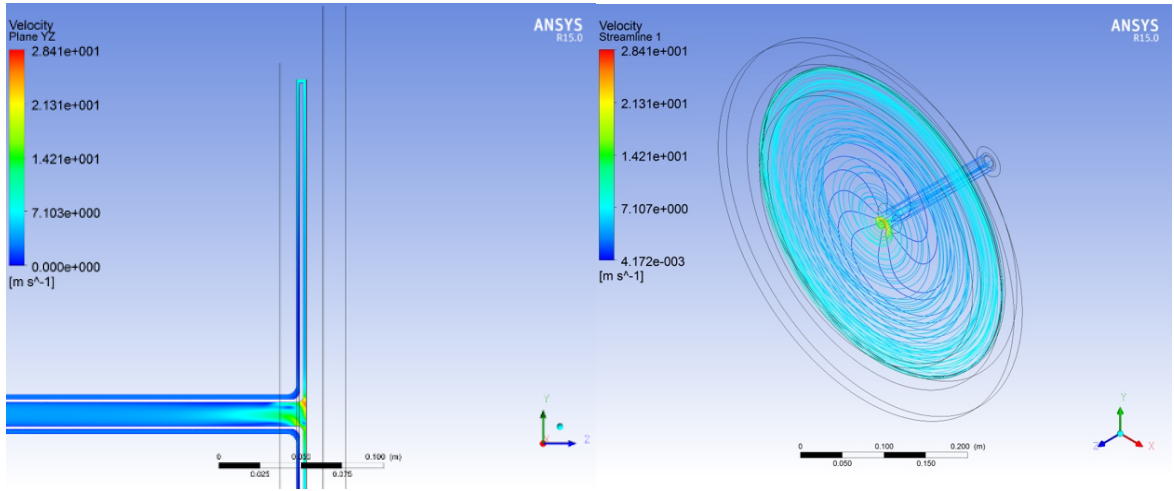

Fig. 40. Example of a simple water pathway with the separator disk

In the initial stage from 2013 to 2015, simulations were conducted assuming a relatively simple water pathway with a disc-shaped hollow within the copper portion of the target disk, into which a separator disk was inserted to separate the inlet and outlet water flows (Figure 40). Within this framework, the two cases were compared (Figure 41): one in which the cooling channel passes through the tungsten and copper joint and the cooling water directly touches the tungsten (direct cooling), and the other in which the channel passes only through the copper and the tungsten is cooled through heat conduction in the copper (indirect cooling). Since these initial calculations were conducted using simplified simulations that did not incorporate the actual pulse structure, quantitative data comparisons are omitted here. Instead, Figures 40 and 41, which illustrate the general structural differences, are presented, and the discussion is limited to conclusions. Direct cooling was expected to provide better cooling, but there were concerns about the risk of cooling water leakage into the vacuum. While details are omitted, there was no significant difference between direct and indirect cooling. Therefore, the conclusion was drawn that indirect cooling, which poses less concern for water leakage into the vacuum, is preferable. Additionally, a comparison was made between the fixed separator disk configuration and the configuration where the separator disk rotates along with the entire target (rotating separator disk). It was concluded that in the simple water pathway of the separator disk method, both fixed and rotating separator disks had insufficient cooling capacity. Hence, it was decided to adopt the capillary water pathway discussed in the next section (2).

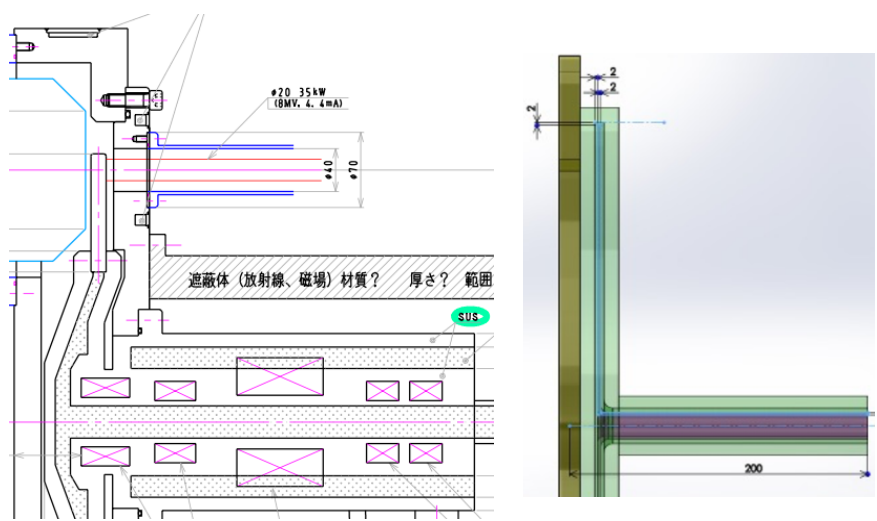

Fig. 41. Direct and Indirect Cooling. Left: Direct cooling, where water comes into contact with the tungsten component. Right: Indirect cooling, where water exists only within the copper. The tungsten component is cooled by contact with the copper.

## (2) Adoption of Capillary Water Pathways

In the final design, indirect cooling was adopted, and a configuration with multiple capillary water pathways running through the copper disk was implemented. These pathways include both inlet side pathways (multiple) and outlet side pathways (multiple). The inlet and outlet pathways are independently connected to the inlet and outlet passages within the rotating shaft, respectively. For the arrangement of water pathways within the rotating shaft, please refer to the description of Prototype 1 in section (III)(1). The tungsten ring was designed with a rather wider shape to allow for a larger contact area with the copper disk at the joint. Within the copper near the junction with the tungsten ring, the inlet and outlet pathways are connected. This area was specifically designed to narrow the pathways, increase flow velocity, and increase the contact area with the hot region by arranging a large number of pathways. These ingenuities were made to ensure efficient heat exchange. In designing these pathways, consideration was given to ensure a large water flow rate per unit time and to account for centrifugal forces due to target rotation. The methodology involved conducting simulations using computational fluid dynamics (CFD) to confirm that the heat transfer rates met the requirements. In designing a narrower water channel, it is necessary to balance the gains and losses between the velocity of the flow and the increased pressure loss due to the narrower channel. This point was also analyzed using CFD, and it was confirmed that the pressure of a typical chiller system is met. Additionally, placing multiple capillary pathways near the hot region results in variations in heat exchange efficiency on each face of each pathway. Some faces may efficiently exchange heat when facing the hot region, while others may experience reduced efficiency due to being shaded by neighboring pathways. This aspect was also analyzed using CFD to confirm that the overall cooling capacity of the cooling system was sufficient.

Specific diagrams illustrating the arrangement of water pathways are omitted from this paper to avoid infringement on proprietary knowledge held by Rigaku Corporation.

(3) Fine-Tuning of Rotation Speed

Based on the results of initial simulations (refer to Section (I)), the rotation speed was adjusted to achieve a tangential speed of approximately 5 m/s. For a rotating target with a diameter of 500 mm, this corresponds to a rotation speed of approximately 200 rpm. Figure 42 compares the temperature distribution on the target at rotation speeds of 200 rpm and 220 rpm. These snapshots were taken after a sufficient amount of time had passed following the start of electron beam irradiation. At 200 rpm, heat generation is localized to three specific areas, whereas at 220 rpm, heat generation appears to be more uniform (Note). The main linac of the ILC operates at a repetition rate of 5 Hz, so in the positron source, electron beam is directed onto the target for 63 ms to generate positrons, followed by a 137 ms pause to stop the electron beam and positron generation. It was found that the combination of this 5 Hz operation (operation interval 200 msec) and the time structure in which positrons are generated for only about one-third of 200 msec creates a thermal distribution in which heat is localized in three locations, as shown on the left side of Figure 42, in synchronization with the rotational speed of 200 rpm. Therefore, the rotation speed was increased by 10% to 220 rpm to introduce a slight shift in the heating pattern with each rotation, as shown on the right side of Figure 42. Simulations were also conducted at a reduced rotation speed of 180 rpm, but the results showed a similar, nearly uniform heating pattern as observed at 220 rpm. Since uniform heating is advantageous, the decision was made not to adopt the 200 rpm rotation speed. Ultimately, after fine-tuning based on simulation observations, a rotation speed of 225 rpm was determined.

Note: At 220 rpm, a region with higher temperature is observed moving from the zenith towards the left compared to other areas. This corresponds to the portion corresponding to the most recent beam injection.

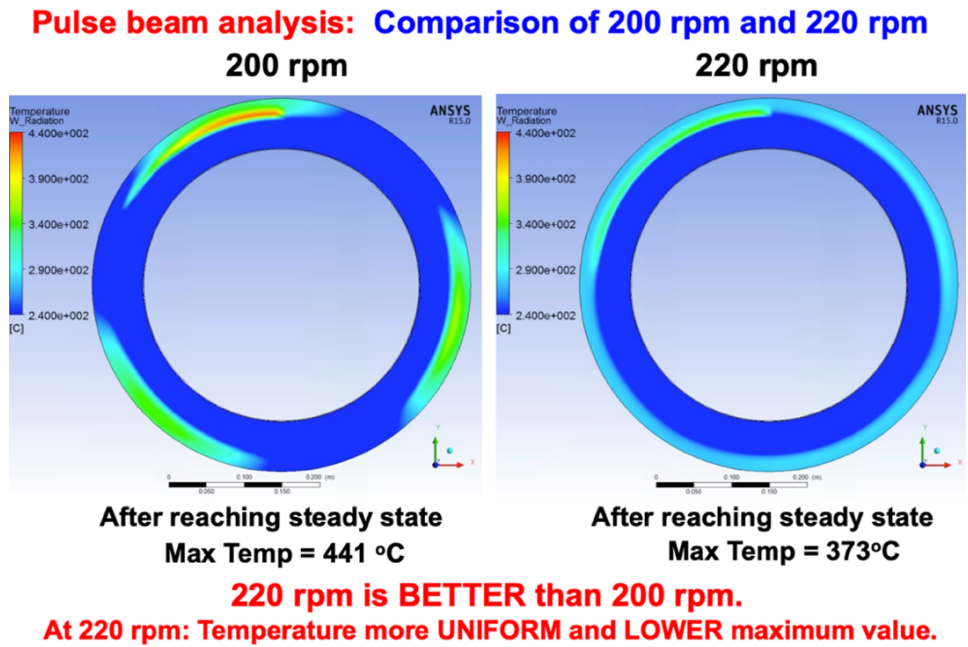

Fig.42. Example of significant differences caused by a slight difference of only about 10% in rotation speed. Left: 200 rpm, Right: 220 rpm. Finnaly, 225 rpm was adopted. Note: The beam conditions in this simulation are slightly different from those in Figure 39.

(4) Simulation Results of Heat Load and Stress on the Target

The simulations were conducted with parameters optimized for the standard operating mode of the ILC at 250 GeV.

| Electron Beam Parameters | |
|---|---|
| Electron Beam Energy: | 3 GeV |
| Spot Size of Electron Beam on Target: | $\sigma = 2$ mm |
| Number of Bunches (a) (Note 1): | 66 bunches/sub-pulse (some with 65 bunches) |
| Number of Sub-pulses (b) (Note 1): | 20 |
| Sub-pulse Spacing: | 3.3 ms (300 Hz) |
| Sub-pulse Duration (c) (Note 2): | 63 ms (20 pulses continuously) |
| Sub-pulse Pause Time (d) (Note 2): | 137 ms |
| Bunch Charge: | 3.75 nC (Note 3) |
| Bunch Spacing: | 6.15 ns (same as DR bunch spacing) |
| | |
| Target Parameters | |
| Target Material: | W75Re25 (alloy: 75% W, 25% Re, Note 4) |
| Target Diameter: | Approx. 500 mm (W75Re25 only covers the ring section) |
| Mass: Approximately | 52 kg (total of ring section and copper disk) |
| Tangential Velocity of Rotation: | Approximately 5 m/s |
| Rotation Speed: | 225 rpm |
| Target Thickness: | 16 mm |
| Cooling Water Flow Rate: | 60 L/min |
| | |
| Flux Concentrator (FC) Parameters | |
| Peak Magnetic Field: | 5 Tesla |
| Distance from Target Disk: | 5 mm (Note 5) |
| Target Heating due to Pulse Magnetic Field: | 1.1 kW (Note 5) |

Note 1: A total of 1312 bunches for the main linac are generated approximately as (a) x (b) = 66x20.

Note 2: The repetition rate of the main linac is 5 Hz, meaning the pulse interval of the main linac is 200 ms. (c) + (d) matches this duration.

Note 3: In the ILC, number of positrons required is $2 \times 10^{10}$ per bunch (3.2 nC). Therefore, taking into account a safety factor of 1.5 for the number of positrons entering the DR acceptance, as well as the results of positron capture simulations, the bunch charge of the electron beam was determined.

Note 4: The same material as the SLC target is assumed. This allows comparison with the beam tests of the target material conducted at the SLAC End Station [8], and also allows reference to the long-term operation track record of the SLC target [9]. SLC targets have been operated for long periods of time, and their damages have been confirmed by opening the vacuum vessel after the experiment is over [10]. W75Re25 has specific information on its durability against repeated large electromagnetic showers, which is a major advantage.

Note 5: The average heating due to the magnetic field is 1.1 kW when the distance between the target and FC is 5 mm. At this distance, the capture positron number per incident electron is e+/e- = 1.08. When the distance is 1 mm, e+/e- = 1.28. Further calculations of heating due to magnetic fields at a distance of 1 mm, optimization of the target-FC distance, and optimization of the capture field will be necessary in future.

Table 2: Parameters Used for Heat Load and Stress Analysis

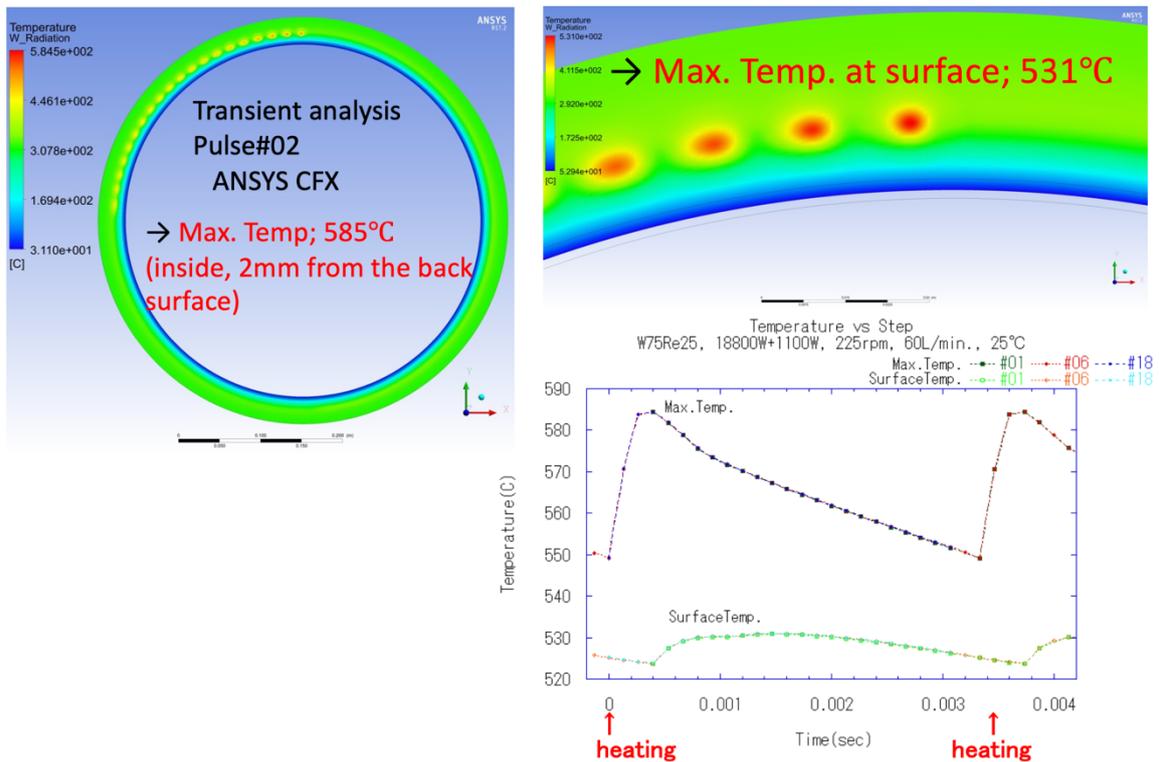

Fig. 43. Analysis Results of Heat and Stress. 66 bunches hit nearly the same spot on the target. This is reflected in the spatial distribution of temperature (top left and top right diagrams) and the temporal evolution of temperature (bottom right diagram). The spatial distribution is viewed from a coordinate system fixed to the rotating target. The bottom right diagram is presented in a slightly unconventional manner for visualizing temperature changes. The label "Heating Step" on the left side of the diagram indicates when the beam hits the target. The temperature change at that point is shown thereafter. The next beam hits the target at the label "Heating Step" on the right side of the diagram, approximately 16.5 mm away from the previous point on the target. The temperature change thereafter represents the temperature at this new point.

The analysis was conducted using the finite volume method, which includes cooling water and heat. Additionally, transient analysis was performed to obtain information on the heat distribution with time resolution. From these analyses, the distribution of heat and stress was determined. The software used for this analysis was ANSYS CFX from ANSYS Inc. To process the transient analysis efficiently, parallel computing was performed using 96 logical processors. The processing was done using a Precision 7920 workstation from Dell Technologies. In this analysis using such a processing system, transient analysis for one condition took more than one month.

The findings from this analysis can be summarized as follows, with the description provided from the coordinate system fixed to the rotating target:

(a) Upon immediate impact of the beam, the point where it hits becomes the hottest point within the entire target (refer to Figure 43). In Figure 43, multiple bright spots (yellow or with hints of red within yellow) represent the points where the beam hits. The brightest spot at the top represents the point immediately after the beam hits. The bright spot to the left of it corresponds to the point where the beam hit 3.3 ms earlier.

(a-1) Since the target has thickness, it is necessary to consider positions not only in the circumferential direction but also in the thickness direction. In the thickness direction, point about 2 mm inside from the downstream surface (the surface where the beam passes out), is the point of the maximum temperature. The temperature at this point is 585 ℃ (refer to Figure 44). At this point, all three principal stresses are negative, indicating compression. The maximum absolute value of the principal stress is 810 MPa, making it the highest among all principal stresses in the target. Since this is compression, the issue regarding the possibility of target failure is considered small. The von Mises stress at this point is 470 MPa. The Yield Strength of the W75Re25 tungsten alloy with 25% rhenium at 585 ℃ is 1230 MPa.

(a-2) The location of maximum tensile stress is at the downstream surface. The tensile stress at this point is 500 MPa, with a temperature of 531 ℃ (refer to Figure 44).

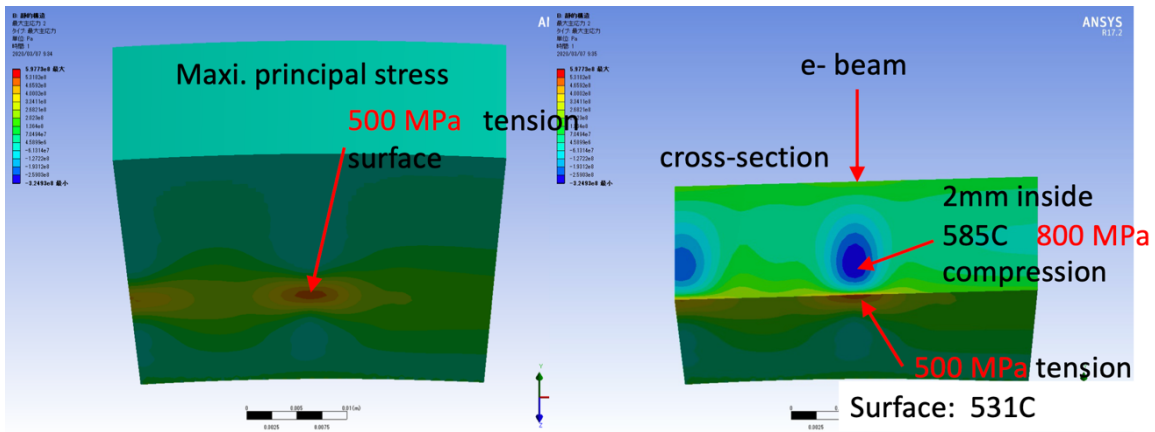

Fig.44. Analysis results of heat and stress. This figure illustrates the stress distribution immediately after the beam impact near the point of beam impact, represented by colors. Here, the maximum absolute value of the three principal stresses at each point is depicted. Blue represents compression (negative stress sign), while red represents tension (positive stress sign). The notation "5.0e+8 Pa" in the figure represents 5.0 x 10^8 Pa. Similar notation is used for other values.

(b) The point of maximum von Mises stress does not coincide with the point where the beam impacts. In other words, the point of maximum von Mises stress does not coincide with the point of maximum temperature. The location of the maximum von Mises stress is offset from the point where the beam strikes. The point of maximum von Mises stress is located roughly one-third of the distance between the two temperature local maxima points (the point hit by the beam most recently and the one hit previously). The temperature at the point of maximum von Mises stress is significantly lower than the temperature at the point of maximum temperature. The temperature at the point of maximum von Mises stress is 350 ℃ (refer to Figure 45).

The contents of Figures 43, 44, and 45 are summarized in Figure 46.

(c) Characteristics such as "the point of maximum von Mises stress does not coincide with the point where the beam impacts" and "the maximum absolute value of the principal stress is compressive" are observed in SLAC's analysis of the SLC target as well [12].

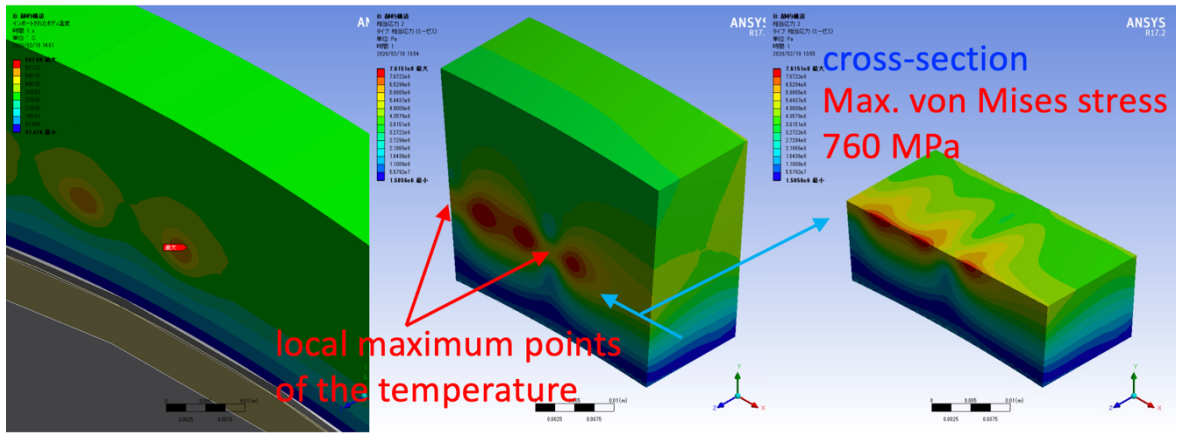

Fig. 45. Thermal and Stress Analysis Results. The figure illustrates the distribution of von Mises stress using colors, where color indicates the magnitude of the von Mises stress. Red represents high von Mises stress, decreasing to yellow, green, and finally blue, indicating the lowest value. In the figure, two points labeled as "Maximum Temperature Locations" with red arrows represent the points where the beam has most recently and previously impacted. The maximum von Mises stress point lies between these two points.

(d) The PEDD (Peak Energy Deposit Density) is 34 J/g. On the other hand, during the high luminosity period of the SLC (the last run), the PEDD is estimated to be around 31 J/g. The PEDD of the ILC target is slightly higher than that of the SLC during high luminosity periods. The target's destruction limit is estimated to be approximately 76 J/g [8], so the PEDD of the ILC target is less than half of that. A margin of approximately twice the limit has been secured.

(e) The instantaneous heat load and resulting stress on the target as shown in (a), (b), and (c) above are comparable to or slightly lower than those of the SLC [12][13]. (See Appendix 2 for details.)

(f) The number of repeated hits per year to the same point is 1/10 of that for SLC. This is because, although the effective repetition rate for SLC and ILC is approximately the same, the target circumference for ILC is approximately 10 times larger than that for SLC targets [14] (see Appendix 3). Therefore, ILC is approximately 10 times safer than SLC with respect to the stress fatigue limit.

    (Note) Effective repetition: In the ILC electron-driven positron source, 66 bunches of driving electron beams effectively hit a single point on the target. When this is counted as 1, the average repetition number is nearly identical to that of the SLC.

From the above considerations, it can be concluded that the ILC target offers equal or greater safety compared to the SLC target.

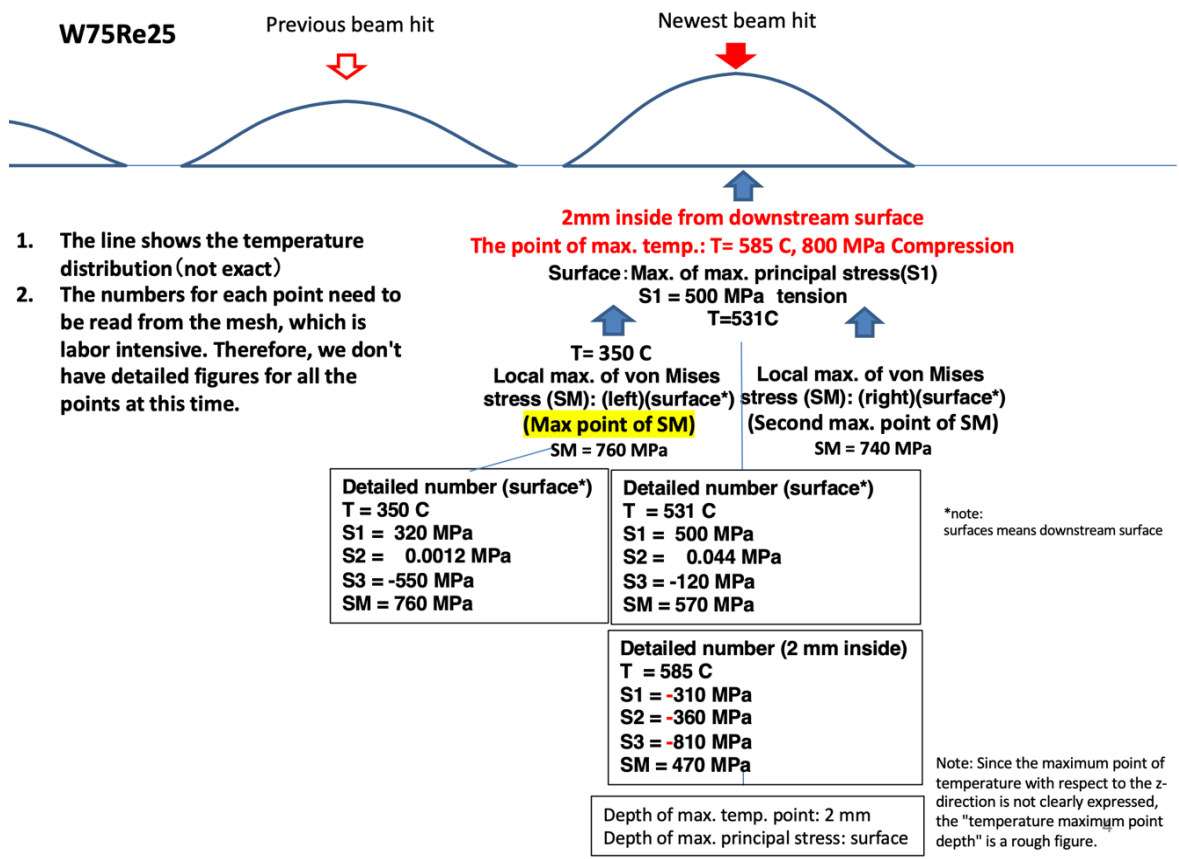

Fig. 46. Summary of Thermal and Stress Analysis Results. The "peaks" depicted at the top of the figure represent the temperature distribution (not to scale). The largest "peak" on the right side represents the temperature distribution near the point where the beam has most recently impacted. The slightly smaller "peak" to its left represents the temperature distribution near the point where the beam impacted previously. The point of maximum von Mises stress lies approximately one-third of the way from the peak on the right to the peak on the left.

It should be noted that some people often simply state that "the SLC target was destroyed," but that is not a fair statement. The SLC target was operated for several years. Subsequently, upon inspection, it was observed that the downstream surface was heavily worn [10]. Specifically, the SLC trolling target was in use for six years, but only during the last years (1997-1998), which is called "the last run," was the target injected with a high intensity electron beam as assumed at the time of design, and high luminosity was produced. [9]. Therefore, it seems reasonable to consider that "after being used under design conditions for one year, numerous cracks were discovered upon inspection of the downstream surface."

Considering the operational experience of the SLC and the fatigue limit of stress, the design of the ILC target is far more safe. Therefore, it is estimated that the ILC target can withstand operation for over one year.

## (5) Notes on Target Material Choices and Manufacturing Methods

Regarding the target material, this paper conducted an analysis assuming a tungsten alloy, W75Re25, containing 25% rhenium. On the other hand, pure tungsten (W) has a lower yield strength compared to W75Re25 but exhibits better thermal conductivity. It is difficult to determine the superiority of W75Re25 over W based solely on known material properties in handbooks. By simulating electromagnetic showers and analyzing the stress induced by heat using finite element methods, it is possible to compare the induced stress to the yield strengths of a material. Actually we did such comparison in this paper. However, effects such as fatigue due to repeated stress cannot be fully reflected in the analysis using finite element methods. Considering those points, W75Re25 appears to be a promising material choice at present, because for W75Re25 we can refere to the operating records at SLC and experimental results at the SLAC End Station. However, further consideration is necessary regarding whether it is feasible to manufacture large ring-shaped plates with a diameter of approximately 500 mm, as assumed in this paper.

On the other hand, is there absolutely no possibility of using pure tungsten (W) or entirely different materials not based on tungsten? The target is subjected to very severe stress, thus we think that continuous efforts to explore new possibilities are necessary. Evaluations through beam tests are essential when considering these new candidate materials. It is considered necessary to conduct beam tests, similar to those carried out at the SLAC End Station for W75Re25, for candidate materials such as pure tungsten. Additionally, if pure tungsten plates were used, it might be easier to manufacture larger ones than W75Re25, but further investigation is required to determine whether rings with a diameter of approximately 500 mm can be produced.

Regarding the target ring, this paper analyzed it as a single-piece component. However, there is room for considering splitting it into multiple parts and assembling them into a ring shape. In this case, the difficulty of manufacturing a single large ring can be avoided. However, if the individual components forming the ring have edges, there is a possibility of stress concentration or insufficient heat propagation at the edges, requiring further in-depth analysis if this structure is adopted.

The thickness of the target ring is 16 mm, but it is conceivable to employ pure tungsten with good thermal conductivity in the upstream side (e.g., about 5 mm) where the electromagnetic shower is not developed well, and to use W75Re25 on the downstream side (e.g., about 11 mm) where the electromagnetic shower develops and undergoes significant stress. In this case, research on joining methods for both materials would be necessary.

Furthermore, during the construction of the ILC, it is desirable to complete the injector parts such as the positron source and electron source earlier than the overall completion and conduct preliminary test runs. If there are materials that have performed well in beam tests, it is desirable to use targets made from them during the preliminary test runs of the positron source to verify their performance.

# Summary and Conclusions

We conducted the design, evaluation, prototyping, and test operation of the positron generation target for the ILC electron-driven positron source.

This target is a rotating target utilizing a ferrofluid rotation seal, with a diameter of 500 mm. The outer rim of the disk consists of a ring-shaped plate made of W75Re25 with a thickness of 16 mm. The electron beam is directed onto this ring-shaped plate to generate positrons via electromagnetic showers. Except for the ring-shaped plate, the target is made of copper and copper part contains water channels for cooling. By rotating the target disc at 225 rpm, the ring portion struck by the electron beam moves at a tangential speed of 5 m/s, dispersing the heat generated by the electron beam.

We designed and manufactured a full-scale prototype with a dummy disc identical in mass and rotational inertia to the actual target disc and conducted continuous operation tests. As a result, we confirmed the compatibility of rotation with the requirement of maintaining stable vacuum of $10^{-6}$ Pa, as demanded for the ILC target. Although the full-scale prototype lacks cooling channels in the disc due to budget constraints, it shares the same structure as the actual target, except for the disc part. The rotary seal of this prototype employs a two-stage ferrofluid seal with an intermediate point evacuation by an ion pump, demonstrating its effectiveness in maintaining stable high vacuum through experiments.

This design includes analysis based on simulations of thermal load, cooling, and stress. By comparing the results with simulations and performance of the SLC target, the lifetime of the ILC target ring, which is the part that the beam hits, due to beam-induced stress was estimated to be more than one year.

We conducted tests on the radiation resistance of ferrofluids, confirming their ability to withstand radiation exposure equivalent to six years of normal operation at the ILC without compromising vacuum performance. We also conducted tests on the overall target's radiation resistance, confirming its operability after exposure to radiation equivalent to two years of normal ILC operation.

Based on these findings, we conclude that we have successfully demonstrated and established the technology for the positron generation target of the ILC electron-driven positron source.

# Acknowledgments

The development of the ILC is being conducted under international cooperation. International collaboration is widely conducted through initiatives such as the Global Design Effort focused on ILC design and development, the Linear Collider Collaboration focusing on linear collider development, the International Development Team aimed at realizing ILC construction, and international conferences on polarized positron source development like the POSIPOL series. We express our gratitude to the following individuals who, throughout these collaborative efforts, provided advice, discussions, and constructive criticisms at various stages of this research:
Dr. Sabine Riemann (DESY), Dr. Gudrid Moortgat-Pick (University of Hamburg), Dr. Peter Sievers (CERN), Dr. Andriy Ushakov (University of Hamburg), Dr. Friedrich Staufenbiel (DESY), Dr. Felix Dietrich (DESY), Dr. Wei Gai (ANL), Dr. Wanming Lui (ANL), Dr. Ian Bailey (Lancaster University), Dr. Jeff Gronberg (LLNL), Dr. Pavel Martyshkin (BINP), Dr. Song Jin (IHEP), Dr. Sun X.J. (IHEP), Dr. Gao Jie (IHEP), Dr. Guoxi Pei (IHEP), Dr. Louis Rinolfi (CERN), Dr. Takuya Kamitani (KEK), Dr. Steffen Doebert (CERN), Dr. Joe Grames (JLab), Dr. Jim Clarke (STFC), Dr. Hitoshi Hayano (KEK), Dr. Akiya Miyamoto (KEK), Dr. Benno List (DESY)

The design of Prototype 1 and Prototype 2 of the rotating target was carried out by Mr. Manabu Noguchi of Rigaku Corporation. Both of these prototypes exhibited the specified performance without any trouble after completion, contributing significantly to the smooth progress of the overall development. This achievement is attributed to Mr. Noguchi's excellent design skills, for which we express our gratitude.

The $^{60}$Co gamma-ray irradiation tests at the QST Takasaki Institute for Advanced Quantum Science were conducted with the cooperation of many individuals. We would like to thank Dr. Yosuke Yuri and Dr. Naotsugu Nagasawa of the institute for their various assistance during the irradiation. We also express our gratitude to Dr. Akira Idesaki of the same institute for his preparation of irradiation and post-irradiation analysis. We are grateful to Dr. Koichiro Hirano of J-PARC Center/JAEA, and Dr. Masafumi Fukuda and Mr. Yu Morikawa of KEK for their assistance at the test site.

The development of the pendulum target prototype was conducted with the full cooperation of the Mechanical Engineering Center at KEK. In particular, we received cooperation from Dr. Masaru Yamanaka, the center's director at the time, from the conceptual stage, and design and fabrication were undertaken by Mr. Koichi Yoshida. We are grateful to both individuals.

The development of the pendulum target prototype was carried out with the full cooperation of the KEK Mechanical Engineering Center. We especially thank Dr. Masaru Yamanaka, the center director, for his collaboration from the conceptual stage. We also acknowledge Mr. Kouichi Yoshida for his work in design and fabrication. We express our gratitude to both individuals.

The tests of the rotating target prototype and pendulum target prototype were conducted at the ATF building at KEK. We extend our gratitude to Dr. Nobuhiro Terunuma, the ATF leader, for providing the facility. We appreciate the assistance provided by Mr. Sakae Araki and Dr. Alexander Aryshev of ATF and Mr. Hiroaki Edakawa of Assist Engineering Co., Ltd., in the on-site work.

Dr. Akira Yamamoto, head of the Linear Collider Promotion Office at KEK, and Dr. Shinichiro Michizono, head of the Innovation Center for Applied Superconducting Accelerators at KEK, understood the importance of positron source target development in the midst of tight budgetary conditions and allocated budget for this development. We are grateful for their support.

note: Affiliations etc. are as of the time

# Development Timeline

## Fiscal Year 2013: April 2013 - March 2014

1. Study on Required Tangential Speed (Determination of Basic Technology)

   Conducted a simplified simulation of the relationship between tangential speed and temperature rise to select the basic technology of the target, i.e., whether it should be a rotating target or a pendulum target.

2. Design, Fabrication, and Operation of Pendulum Target Prototype

   Due to concerns about the durability of bellows as the mechanism for pendulum motion and vacuum sealing, a test machine (s ort of prototype) was designed and fabricated to evaluate them. The test machine was completed in July 2013, and operation began immediately. But, vacuum pumping was not implemented yet. It experienced three failures during operation from July to November, each time undergoing repairs to continue operation. Particularly after the third failure, significant modifications and reinforcements were made.

3. Basic Data on Vacuum Seal Leakage Rate Using Commercial X-ray Generation Target

   Basic data on the vacuum seal leakage rate was collected using a rotating target of commercial X-ray generation equipment from December 2013 to February 2014, confirming that this target format is essentially durable for use.

4. Investigation of Radiation Resistance of Ferrofluid Used in Rotating Targets

   We investigated the radiation resistance of ferrofluid. Preliminary experiments were conducted using cobalt-based gamma-ray irradiation facilities at the QST Takasaki Research Institute from December 2013 to February 2014, revealing that the fluorine-based fluid endured up to 0.27 MGy but decomposed at 3.2 MGy. It was confirmed that the hydrocarbon-based fluid did not decompose even at 3.2 MGy.

## Fiscal Year 2014: April 2014 - March 2015

1. Study on Required Tangential Speed (Determination of Basic Technology)

   Based on simulations conducted by the German research group (DESY, University of Hamburg, University of Applied Sciences Wildau), it was concluded that a tangential speed of 5 m/s is required.

2. Design, Fabrication, and Operation of Pendulum Target Prototype

   Significant modifications and enhancements were made to address the previous year's damage, and operation resumed in the summer of 2014. The vacuum vessel was evacuated and tested. Long-term operation for approximately one year was conducted until the following summer, concluding operation without damage. As a result of these two years of testing, it was concluded that a tangential speed of approximately 1 m/s is the limit for this type of target.

3. Determination of Basic Technology

   Based on the above 1 and 2, a rotating target was selected.

4. Radiation Resistance of Ferrofluid Used in Rotating Targets

   Systematic irradiation with varying radiation doses was conducted for hydrocarbon-based fluids. The data at seven points ranging from 0 to 4.7 MGy was taken. It was found that viscosity increases almost proportionally with the irradiation dose.

5. Radiation Resistance of Ferrofluid (Put Irradiated Ferrofluid into Commercial Target)

   Irradiated hydrocarbon-based ferrofluid (4.7 MGy) was incorporated into a commercially available X-ray generator rotating target, then vacuum and rotation tests were conducted, confirming its suitability for use.

6. Radiation Resistance Test of Rotating Target (Whole Irradiation of Commercial Target)

   Radiation exposure was conducted on the entire commercially available target at Takasaki (March 2015). The dose was 0.6 MGy, equivalent to one year of radiation exposure in the ILC target motor section. It was confirmed that there were no issues with rotation and vacuum maintenance.

7. Heat, Stress, and Cooling Simulation in Rotating Targets

   Haet, stress, and cooling simulations were performed. The waterway had a simple shape using disc separators, with the separator discs not rotating.

8. Design of Rotating Target Prototype

   Design considerations for the prototype were made, and a design outline was drafted.

## Fiscal Year 2015: April 2015 - March 2016

1. Design and Fabrication of Prototype 1

   Prototype 1 was completed in March 2016. Test preparations were conducted throughout the year 2016, and testing commenced in February 2017.

2. Heat, Stress, and Cooling Simulation in Rotating Targets

   The waterway employed a simple design using disc separators, with the separator discs also rotating. The region near tungsten featured wide and thin waterways. Starting from March 2016, the waterway design transitioned to a capillary design. All subsequent analyses have been conducted using the capillary waterway configuration.

## Fiscal Year 2016: April 2016 - March 2017

1. Vacuum Test: Preparation for Prototype 1

   Preparation for testing Prototype 1, fabricated in FY 2015, was conducted. Due to the extensive preparation required, actual testing commenced in February 2017. The rotation speed was set at 225 rpm.

2. Heat, Stress, and Cooling Simulation in Rotating Targets

   Analysis using the capillary waterway configuration was continued.

## Fiscal Year 2017: April 2017 - March 2018

1. Vacuum Test: Prototype 1 Experiment

   Continuation of the Prototype 1 experiment (225 rpm). The experiment has been ongoing since February and reached a vacuum level of $3 \times 10^{-6}$ Pa around March. At this point, it was confirmed that the vacuum level requirements for the ILC TDR were basically satisfied. Around April, sporadic short-term (2-3 minutes) increases in pressure (spikes) were observed, occurring at intervals of approximately 2-3 hours. As there were no spikes until March, the possibility of degradation of the ferrofluid in some form was suspected. The vacuum level (base pressure) fluctuated between $3 \times 10^{-6}$ and $4 \times 10^{-6}$ Pa. Operation continued until July.

   On July 19, the experiment was halted, and the system was exposed to atmospheric pressure.

The ferrofluid unit was taken back to the Rigaku factory for disassembly and inspection. Subsequently, it was refilled with fresh fluid, returned to KEK, reinstalled, and the experiment resumed. Various tests were conducted intentionally, including high-speed rotation (900 rpm - 1150 rpm) and frequent alternation of rotation speeds (225 <-> 900 rpm) every 3 minutes, repeated 20 times, among others.

At the end of October, the target was put into steady-state operation at 225 rpm, with a vacuum level of 3 x 10$^{-6}$ Pa, and the occurrence of spikes resumed at intervals of approximately 2-3 hours, similar to the period from April to July. It was recognized that the occurrence of spikes was inherent to the system and not due to degradation of the ferrofluid. The progress from February 2017 to the end of October 2017 (as described above) is omitted from the main text of this paper.

In November 2017, preparations began for an experiment involving the flow of dry nitrogen gas on the atmospheric side. At the end of November, rotational and vacuum tests with the flow of dry nitrogen gas were started.

In the end of November to beginning of December, observations were made with various rotation speeds. For high-speed rotation at 1150 rpm, spikes occurred at a very high frequency (approximately every 20 minutes).

Then, also in December, tests were conducted at the rated rotation speed of 225 rpm. Significant improvement in vacuum was achieved, with a base pressure of 6 x 10$^{-7}$ Pa. The flow rate of dry nitrogen gas at this time was 100 mL/min. This is illustrated in Figure 20 of this paper. Tests were also conducted with different flow rates, and it was found that a lower flow rate of 10 mL/min resulted in a negligible improvement in vacuum.

From January to March 2018, experiments were conducted to evacuate the atmospheric side of the housing using a rotary pump. However, no improvement was observed. It should be noted that the atmospheric side of Prototype 1's housing was originally designed to be open, and evacuation with a vacuum pump was not intended. The holes were temporarily sealed with rubber sheets and putty, and the rotary pump was used for evacuation. Consequently, the atmospheric side only reached a level of "vacuum" where it could barely operate with a Pirani gauge. Although the experiment to evacuate the atmospheric side with a rotary pump did not yield positive results, this concept would be thoroughly tested later with Prototype 2.

2. Radiation Resistance Test (Ferrofluid Alone)

In February 2018, a radiation exposure test was conducted on the hydrocarbon-based ferrofluid to investigate changes in the molecular structure of the base oil due to radiation. Samples consisting of ferrofluid and base oil only were prepared, and they were exposed to gamma radiation up to a maximum of 4.7 MGy. After irradiation, viscosity measurements were conducted, and changes in molecular weight and structure were examined using gel permeation chromatography (GPC) and UV-Vis spectroscopy.

3. Potential Evaporation of Ferrofluid

Leakage rates were calculated based on the achieved vacuum level of Prototype 1 (conducted between February and April 2017). Subsequently, under the assumption that these leaks were not from the atmosphere but solely due to "evaporation" of the ferrofluid, the upper limit of the evaporation rate was estimated. It was confirmed that this evaporation rate was sufficiently small. This matter is omitted from the main text of this paper.

**Fiscal Year 2018: April 2018 - March 2019**

1. Radiation Resistance Test (Ferrofluid Alone)

   Regarding the radiation effects on ferrofluids, we conducted more detailed investigations compared to the 2017 fiscal year, utilizing not only GPC and UV-Vis but also gas chromatography to collect and analyze gases generated during radiation exposure. This experiment was conducted in February 2019. We ensured the consistency of the gas chromatography data with other data sets and deepened our understanding of the changes in molecular structure.

2. Continuation of Prototype 1 Testing

   In the 2018 fiscal year, we continued testing with Prototype 1. We operated it steadily for about one year and confirmed that there was no degradation or other issues.

3. Demonstration Experiment on the Potential Evaporation of Ferrofluid

   We conducted a verification experiment to investigate the potential evaporation of ferrofluids. By placing the ferrofluid in a 3 cm diameter evaporating pan, we increased the exposed surface area. The pan was then placed in a vacuum chamber, evacuated using an ion pump to create a vacuum, and analyzed using RGA (Residual Gas Analysis). We determined the upper limit of the evaporation rate of hydrocarbon-based gases and demonstrated that it was sufficiently small.

**Fiscal Year 2019: April 2019 - March 2020**

1. Design and Fabrication of Prototype 2

   In response to the results obtained from Prototype 1, an improved version was designed to achieve higher performance, particularly aiming at spike suppression. The improved version employed a two-stage ferrofluid seal, with an intermediate section evacuated by an ion pump. Design and fabrication of Prototype 2 were carried out, and it was completed in February 2020. Testing was conducted in the following fiscal year.

2. Radiation Resistance Test With Prototype 1 (Utilizing Ferrofluid after Radiation Exposure)

   Preparations to incorporate ferrofluid post-radiation exposure into Prototype 1 were initiated but interrupted due to the COVID-19 pandemic.

3. Simulation of Heat, Cooling, and Stress in Rotating Targets

   Similar simulations have been conducted several times in the past, all assuming operation with 2600 bunches. Specifically, simulations were carried out assuming a spot size of $\sigma = 4$ mm for the electron beam on the target, with 132 bunches hitting a single point (effectively considered the same point) on the target. Subsequently, in the International Linear Collider (ILC) Technical Design Report (TDR) released by the Linear Collider Collaboration in 2013, the initial parameters for ILC construction were formally determined as a center-of-mass energy of 250 GeV and 1300 bunches. In light of this, adjustments were made to the initial parameters for positron source construction, changing the electron beam spot size to $\sigma = 2$ mm, with 66 bunches hitting a single point on the target. Detailed simulations were then conducted based on these new parameters. This represents the latest version as of 2023.

**Fiscal Year 2020: April 2020 - March 2021**

    1. Vacuum Test: Prototype 2 Experiment

        Testing of Prototype 2 (with intermediate point pumping 30 L/s) was conducted. A long-term rotating vacuum test was performed from September 2020 to January 2021. The result was highly satisfactory, achieving a vacuum level of 1.1 x $10^{-6}$ Pa with no spikes. The experiment concluded in January 2021 to prepare for the next modification (adding a dummy disk). As the pressure continued to decrease at the end of the experiment, it is believed that the vacuum would have improved further if the test had continued.

    3. Addition of Dummy Disk to Prototype 2

        A dummy disk was added to Prototype 2. After completing the testing of Prototype 2 in January 2021, the main body of Prototype 2 was transferred to the RIGAKU factory. The dummy disk was installed, and the assembly was placed into a large vacuum chamber.

    3. Radiation Resistance Test With Prototype 1 (Utilizing Ferrofluid after Radiation Exposure)

        The testing, which had been interrupted due to the COVID-19 pandemic in the previous fiscal year, was restarted. The ferrofluid in Prototype 1 was replaced with one irradiated to 4.7 MGy. Rotating vacuum testing commenced at the end of September 2020 and continued into the next fiscal year.

**Fiscal Year 2021: April 2021 - March 2022**

    1. Vacuum Test: Testing of Prototype 2 with Dummy Disk

        Testing was conducted on Prototype 2 with a dummy disk. A long-term rotating vacuum test was carried out from April 2021 to July 2021. The result was favorable, with a base pressure of 7.4 x $10^{-7}$ Pa. While there were spikes, they were small and occurred infrequently.

    2. Radiation Resistance Test With Prototype 1 (Utilizing Ferrofluid after Radiation Exposure)

        A rotating vacuum test was conducted using Prototype 1 with ferrofluid irradiated to 4.7 MGy. This test continued from the previous fiscal year, spanning from the end of September 2020 to the end of July 2021. The achieved vacuum level was approximately 3.5 x 10-6 Pa. It was demonstrated to meet the requirements of the Technical Design Report (TDR) using irradiated ferrofluid. However, the target pressure of 6 x 10-7 Pa achieved in December 2017 could not be reproduced. For discussion on the reasons behind this, refer to section VI(3) of this paper.

**Fiscal Year 2022: April 2022 - March 2023**

    1. Vacuum Test: Testing of Prototype 2 with Dummy Disk

        Testing was conducted on Prototype 2 with an increased intermediate pumping of 50 L/s for the dummy disk configuration. A long-term rotating vacuum test was performed from September 2022 to early April 2023. The results were favorable, showing significant improvements. There was an enhancement in the achieved vacuum level, as well as an improvement in vacuum level during spikes. Additionally, there was a substantial reduction in the frequency of spikes.

# Appendix 1: Summary of the Evaporation Pan Experiment in November 2018

<＜Experiment＞

Ferrofluid was placed in an evaporation pan inside a vacuum chamber, which was then sealed with a black cover and subjected to rough evacuation. Baking was not performed (reason: to protect the ferrofluid from heat), and the ion pump was turned on.

The experiment commenced on the morning of November 8, 2018 (Thursday) and concluded on the morning of November 26, 2018 (Monday). The state just before the end on the morning of November 26, 2018 (Monday) was analyzed.。

## ＜Results and Discussion＞

Firstly, the results and discussion are presented. Subsequently, analysis and data are provided. The plot of the mass spectrum and a photographs taken just before the start of the experiment are presented at page 60.

Total pressure: 9.27e-7 Pa (9.27 x $10^{-7}$ is represented as 9.27e-7. Similarly for the following.)

Partial pressure:

| | | |
|---|---|---|
| $H_2$ | 6.41e-7 Pa | (6.21e-11／8.973e-11 = 0.692,  9.27e-7 x 0.692) |
| XCH | 7.36e-8 Pa | (7.12e-12／8.973e-11 = 0.0793, 9.27e-7 x 0.0793) |
| $H_2O$ | 9.25e-8 Pa | (8.81e-12／8.973e-11 = 0.0998, 9.27e-7 x 0.0998) |
| $N_2$+CO | 8.74e-8 Pa | (8.46e-12／8.973e-11 = 0.0943, 9.27e-7 x 0.9943) |
| $O_2$ | 4.01e-9 Pa | (3.88e-13／8.973e-11 = 0.00432, 9.27e-7 x 0.00432) |
| $CO_2$ | 1.01e-8 Pa | (9.81e-13／8.973e-11= 0.0109,  9.27e-7 x 0.0109) |

Note: XCH comprises components identified as $CH_4$ along with other hydrocarbons and unidentified components generated during analysis (including errors). For a detailed definition, please refer to page 57. The classification of XCH was established to conduct discussions on a more conservative basis. In other words, XCH represents an estimate of the upper limit of hydrocarbon-based molecules from a pragmatic standpoint. (Those identified as $CH_4$ account for 1/3.5 of XCH.)

Evaporation pan area:   7 x $10^{-4}$ $m^2$    (Pan diameter is approximately 3 cm)
Vacuum pump capacity: 100 L/s (=100x$10^{-3}$ $m^3$/sec) (ion pump)

"Leak" rate of XCH per unit area (1 m²) calculated from XCH partial pressure
(i.e., evaporation rate)

$$(7.36e\text{-}8 \text{ Pa}) \times (100 \times 10^{-3} \text{ m}^3/\text{sec}) \diagup (7 \times 10^{-4} \text{m}^2)$$

$$= (7.36e\text{-}9 \text{ Pa m}^3/\text{sec}) \diagup (7 \times 10^{-4} \text{m}^2)$$

$$= 1.1e\text{-}5 \text{ Pa m/sec} \quad \text{(Note: Rate per unit area of 1 m}^2\text{)}$$

In actual usage, the ferrofluid appears as a ring with a diameter of approximately 3 cm and a width of about 100 $\mu$ m when viewed from inside the vacuum chamber. This area is calculated as $(3.14 \times 0.03 \text{ m}) \times (100 \times 10^{-6} \text{ m}) = 1 \times 10^{-5} \text{ m}^2$

The "leak" rate (i.e., evaporation rate) of XCH under actual usage conditions is calculated to be 1.1e-10 Pa m³/sec

Based on this, the partial pressure of XCH at the upstream end of the downstream first acceleration tube, calculated using the following model and considering the pragmatic estimate of the upper limit of hydrocarbon molecular species, is 1.5e-10 Pa.

Based on calculations using the model below, the partial pressure of XCH at the uppermost part of the first acceleration tube immediately downstream from the target is estimated to be 1.5e-10 Pa.

This estimation is considered from a pragmatic perspective as the upper limit of the presence of hydrocarbon-based gases molecules at the upstream end of the first downstream acceleration tube, which is deemed to represent the most adverse conditions in terms of the impact of ferrofluid evaporation.

## Model

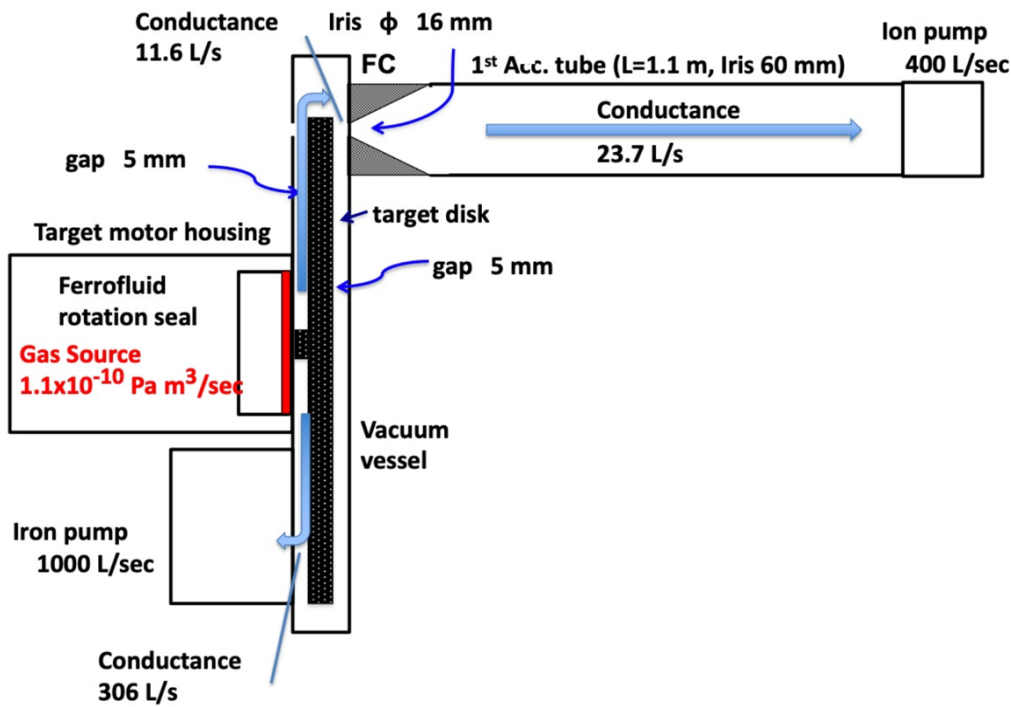

**Calculation results (partial pressure of XCH at each part)**

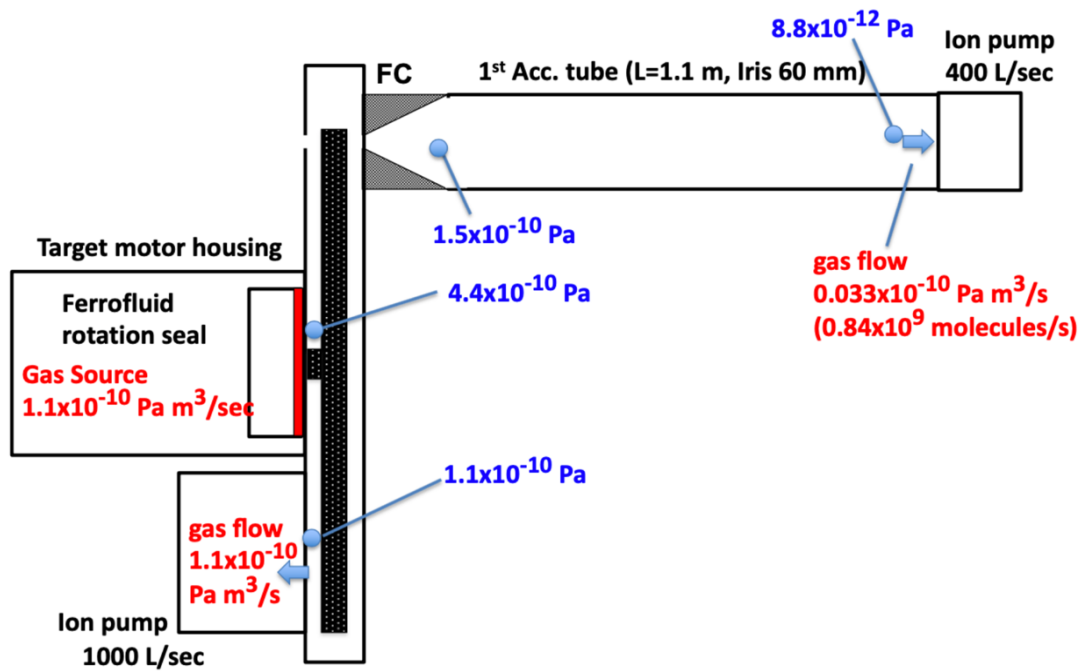

See below for details of calculation method:
"Asymptopic Coverage of the Solvent Molecule of the Ferro-fluid Vacuum Seal on the Accelerator Cavity Wall in ILC E-driven Positron Source", Masao KURIKI July 17, 2017

## ＜Basic Approach to Analysis＞

In conducting the analysis, a pragmatic approach was adopted as follows:

(a) In cases where there is uncertainty in the results, calculate the value of $CH_4$ plus the uncertain part and use it as the upper limit for hydrocarbon-based gases molecules. (Safety side)

(b) Each of $H_2O$, $N_2$, $CO$, $O_2$, and $CO_2$ is assumed to be acceptable even if not completely separated.
In the experiment conducted using the evaporation pan, no baking was performed to prevent damage to the ferrofluid due to heat. However, in actual target usage, baking is conducted before use. In the target, the ferrofluid part is protected from heat by cooling water, so baking is not a problem. Baking eliminates almost all $H_2O$, $N_2$, $CO$, $O_2$, and $CO_2$. Therefore, the breakdown of these components is not essential.

(c) The main concern is the partial pressure of $CH_4$ (harmful component), the partial pressure of $H_2$ (remaining due to incomplete exhaust by the pump) as a comparative reference, and the total pressure.

<Analysis>

(1) There are five locations in the cracking pattern table where only one number is written in the horizontal column. They are as follows: ME(1), ME(15), ME(18), ME(32), ME(44). Here, ME(1) represents m/e = 1. ME(15) represents m/e = 15, and so forth.

Cracking pattern table used：
(a) AIST bulletin of metrology, Vol.5,No.1, page 67, 「分圧真空標準に関する調査研究("Research on partial pressure vacuum standards"(written in Japanese)」 Hajime Yoshida(吉田 肇)、Table 2
(a) J. of the Mass Spectrometry Society of Japan, No 5, p. 34-43, 1955 「電子衝撃による気体の電離および解離の研究("Study of ionization and dissociation of gases by electron impact"(written in Japanese)」 Teruo Hayakawa(早川 晃雄), Toshio Sugiura(杉浦 俊男), Table 1
We use (a) for ME(1), ME(15), ME(18) and ME(44). We use (b) for ME(32).

| m/e | H₂ | CH₄ | H₂O | N₂ | CO | CO₂ |
|-----|-----|------|------|-----|-----|------|
| 1 | 1.9 | | | | | |
| 2 | 100 | 2.6 | | | | |
| 12 | | 8.6 | | | 4.7 | 8.7 |
| 14 | | 17.1 | | 4.0 | | |
| 15 | | 85.6 | | | | |
| 16 | | 100 | | | 1.7 | 9.6 |
| 17 | | 1.1 | 21.2 | | | |
| 18 | | | 100 | | | |
| 22 | | | | | | 1.9 |
| 28 | | | | 100 | 100 | 9.8 |
| 29 | | | | | 1.2 | |
| 44 | | | | | | 100 |
| 45 | | | | | | 1.2 |

(a)AIST bulletin of metrology, Vol.5,No.1, page 67
table 2

（電子加速電圧，100 V；全電子電流，200 μA；イオン加速電圧，1300V）

| 試 料 | M/e | イ オ ン | クラッキング・パターン（平均値） | 平均 偏差 | 間, % | 測定回数 |
|-------|-----|----------|-------------------------------|----------|--------|----------|
| A | 13.3 | A⁺⁺⁺ | 0.041 | | | |
| | 20 | A⁺⁺ | 12.54 | 0.23 | 1.8 | |
| | 36 | ³⁶A⁺ | 0.308 | | | 41 |
| | 38 | ³⁸A⁺ | 0.061 | | | |
| | 40 | A⁺ | 100 | — | — | |
| N₂ | 14 | N⁺(N₂⁺⁺) | 17.3 | 0.31 | 1.8 | |
| | 28 | N₂⁺ | 100 | — | — | 19 |
| | 29 | ¹⁵NN⁺ | 0.75 | | | |
| O₂ | 16 | O⁺ | 28.2 | 0.47 | 1.6 | |
| | 32 | O₂⁺ | 100 | — | — | 7 |
| | 34 | ¹⁸OO⁺ | 0.41 | | | |
| H₂ | 1 | H⁺ | 2.82 | 0.04 | 1.4 | 16 |
| | 2 | H₂⁺ | 100 | — | — | |
| NH₃ | 14 | N⁺ | 2.23 | 0.11 | 4.9 | |
| | 15 | NH⁺ | 5.81 | 0.23 | 3.9 | 9 |
| | 16 | NH₂⁺ | 78.0 | 1.0 | 1.3 | |
| | 17 | NH₃⁺ | 100 | — | — | |
| CO | 12 | C⁺ | 5.49 | 0.11 | 2.0 | |
| | 14 | CO⁺⁺ | 0.19 | | | |
| | 16 | O⁺ | 3.04 | 0.05 | 1.6 | 17 |
| | 28 | CO⁺ | 100 | — | — | |
| | 29 | ¹³CO⁺, ¹⁷CO⁺ | 1.13 | | | |
| CO₂ | 12 | C⁺ | 11.2 | 0.21 | 1.9 | |
| | 16 | O⁺ | 13.9 | 0.27 | 1.9 | |
| | 22 | CO⁺⁺ | 1.14 | | | |
| | 28 | CO⁺ | 16.25 | 0.34 | 2.0 | 14 |
| | 29 | ¹³CO⁺, ¹⁷OC⁺ | 0.22 | | | |
| | 32 | O₂⁺ | 1.84 | | | |
| | 44 | CO₂⁺ | 100 | — | — | |
| | 45 | ¹³CO₂⁺, ¹⁷OCO⁺ | 1.20 | | | |

(b) J. of the Mass Spectrometry Society of Japan, No 5, p. 34-43, 1955, Table 1

(2) The measured value for ME(1) is 9.51e-13, all of which is attributed to H2. However, this is not the main peak for $H_2$. The value in the cracking pattern table is 1.9.

(3) The measured value for ME(15) is 8.09e-13, all of which is attributed to CH4. However, this is not the main peak for $CH_4$. The value in the cracking pattern table is 85.6. However, due to its significant proportion, this value will be used as a reference to determine each cracking (as explained later).

(4) The measured value for ME(18) is 7.27e-12, all of which is attributed to H2O. This is the main peak for $H_2O$. The value in the cracking pattern table is 100.

(5) The measured value for ME(32) is 3.03e-13, all of which is attributed to O2. This is the main peak for $O_2$. The value in the cracking pattern table is 100.

(6) The measured value for ME(44) is 7.54e-13, all of which is attributed to CO2. This is the main peak for $CO_2$. The value in the cracking pattern table is 100.

(7) Cracking from the main peak of $H_2O$ (18) appears in ME(17). The ratio is 21.2/100, and the corresponding current value is calculated as 7.27e-12 x (21.2/100) = 1.54e-12.

(8) Regarding $H_2O$, all determinations are made based on (4) and (7). The total current of $H_2O$-derived current (partial pressure of H2O) is 8.81e-12.

(9) Cracking from the main peak of $CO_2$ (44) appears in ME(12), ME(16), ME(22), ME(28), and ME(45).
   ME(12): ratio 8.7/100,    current 7.54e-13 x(8.7/100) = 6.53e-14
   ME(16): ratio 9.6/100.    current 7.54e-13 x(9.6/100) = 7.24e-14
   ME(22): ratio 1.9/100,    current 7.54e-13 x(1.9/100) = 1.43e-14
   ME(28): ratio 9.8/100,    current 7.54e-13 x(9.8/100) = 7.39e-14
   ME(45): ratio 1.2/100,    current 7.54e-13 x(1.2/100) = 9.05e-15
   (The value of ME(45) above is greater than the measured value, so it is replaced by the measured value of 1.42e-15.)

(10) All determinations regarding $CO_2$ are made based on (6) and (9). The total current of $CO_2$-derived current (partial pressure of CO2) is 9.87e-13.

(11) From the measured value of ME(15) (all assumed to be derived from $CH_4$ as per (3)), the contribution of $CH_4$ to the current value of CH4's main peak, ME(16), is estimated to be 8.09e-13 x 100/85.6 = 9.45e-13.

(12) Similarly, the cracking of $CH_4$ from ME(15) to ME(2), ME(12), ME(14), and ME(17) is calculated based on their ratios and current values.
   ME(2):   ratio 2.6/85.6,    current 8.09e-13 x(2.6/85.6) = 2.46e-14
   ME(12): ratio 8.6/85.6    current 8.09e-13 x(8.6/85.6) = 8.12e-14
   ME(14): ratio 17.1/85.6,    current 8.09e-13 x(17.1/85.6) = 1.61e-13
   ME(17): ratio 1.1/85.6,    current 8.09e-13 x(1.1/85.6) = 1.04e-14
   The total current of $CH_4$-derived current can be obtained by sum of (3), (11), and(12); it is 2.03e-12.

(13) The estimated value (2.46e-14) of ME(2) from ME(15) (all assumed to be derived from $CH_4$) is significantly smaller than the measured value (6.11e-11) of ME(2), thus it is negligible. Therefore, the measured value of ME(2) is considered to be all derived from $H_2$.

(14) Cracking from the main peak of $H_2$ (2) appears in ME(1). The ratio is 1.9/100, and the current value is calculated as 6.11e-11 x (1.9/100) = 1.16e-12. This closely matches the measured value of ME(1), 0.951e-12. We use the measured value of $H_2$ (2) in delivering the conclusion.

(15) All determinations regarding $H_2$ are made based on (13) and (14). The total current of $H_2$-derived current (partial pressure of H2) is 6.21e-11.

(16) Cracking from the main peak of $O_2$ (32) appears in ME(16) and ME(34).
   ME(16): ratio 28.2/100,   current 3.03e-13x(28.2/100) = 8.54e-14
   ME(34): ratio 0.41/100,   current 3.03e-13x(0.41/100) = 1.24e-15 (Set to be zero. Measured value is also zero.)

(17) All determinations regarding $O_2$ are made based on (5) and (16). The total current of $O_2$-derived current (partial pressure of $O_2$) is 3.88e-13.

(18) Attention should be paid to the handling of data for ME(12), ME(14), and ME(16). The measured values for $CH_4$ are coupled with $N_2$, CO, and $CO_2$ in ME(14), ME(12), and ME(16), respectively.

(19) Both $N_2$ and CO have their main peaks at ME(28), making separation difficult.

(20a) The measured value for ME(14) is 6.69e-13. Assuming that 1.61e-13 is derived from $CH_4$ (referring to (12)), the remaining 5.08e-13 is assumed to be derived from $N_2$. (Note: Eventually, 5.08e-13 was classified as unknown in the conclusion.)

(20b) 100/4.0 of this 5.08e-13 should appear in ME(28). Its value is 5.08e-13 x (100/4.0) = 1.27e-1 . This is slightly above the measured value of 8.31e-12 for ME(28). Hence, all peaks in ME(28) are attributed to $N_2$ and there is no CO. (Hypothesis 1)

(21a) The measured value for ME(12) is 7.17e-13. Among this, 8.12e-14 is assumed to be derived from $CH_4$ (see (12)). Additionally, 6.53e-14 is assumed to be derived from CO2 (see (9)). The remaining 5.67e-13 is likely derived from CO (assumption). (Note: Eventually, 5.67e-13 is classified as unknown in the conclusion.)

(21b) 100/4.7 of this 5.67e-13 should appear in ME(28). Its value is 5.67e-13 x (100/4.7) = 1.21e-11. This is slightly above the measured value of 8.31e-12 for ME(28). Hence, all peaks in ME(28) are attributed to CO and there is no $N_2$. (Hypothesis 2). Note that hypothesis 2 contradicts hypothesis 1 from (20b).

(22a) The measured value for ME(16) is 2.48e-12. Among this, 0.945e-12 is assumed to be derived from $CH_4$ (refer to (12)). Additionally, 7.24e-14 is assumed to be derived from $CO_2$ (referenced in (9)). Furthermore, 8.54e-14 is assumed to be derived from $O_2$ (referenced in (16)). The remaining 1.38e-12 is likely derived from CO (assumption). (Note: Eventually, 1.38e-12 is classified as uknown.)

(22b) 100/1.7 of this 1.38e-12 should appear in ME(28). Its value is 1.38e-12 x (100/1.7) = 81.2e-12. It significantly exceeds the measured value of ME(28), 8.31e-12, by approximately ten times (contradiction). However, considering the substantial factor by which it is divided, the error is significant.

(23) Thus far, hypothesis 1 from (20b) contradicts hypothesis 2 from (21b). Additionally, there is a contradiction in (22b).

(24) Attention should be paid to handling the data for ME(17). The measured values for $CH_4$ are coupled with $H_2O$ in ME(17). The measured value for ME(17) is 3.33e-12. Among this, 1.54e-12 is assumed to be derived from $H_2O$ (referenced in (7)). The remaining 1.79e-12 is likely derived from $CH_4$ (hypothesis 3). However, this value significantly exceeds the contribution of $CH_4$ to ME(12) as determined from the measured value of ME(15), 8.12e-14, by a factor of 22, indicating a contradiction in hypothesis 3. (Note: Eventually, 1.79e-12 is classified as unknown.)

## ＜Summary of Analysis Results＞

| Mass Number | Results of Measurement (Amps.) | Molecule Assignment | | | | | | |
|---|---|---|---|---|---|---|---|---|
| | | H₂ | CH₄ | H₂O | N₂ + CO | O₂ | CO₂ | Unknown（＊） |
| 1 | 9.51e-13 | 9.51e-13 | 0 | 0 | 0 | 0 | 0 | 0 |
| 2 | 6.11e-11 | 6.11e-11 | 2.46e-14 | 0 | 0 | 0 | 0 | 0 |
| 12 | 7.17e-13 | 0 | 8.12e-14 | 0 | 0 | 0 | 6.53e-14 | 5.67e-13（＊） |
| 14 | 6.69e-13 | 0 | 1.61e-13 | 0 | 0 | 0 | 0 | 5.08e-13（＊） |
| 15 | 8.09e-13 | 0 | 8.09e-13 | 0 | 0 | 0 | 0 | 0 |
| 16 | 2.48e-12 | 0 | 9.45e-13 | 0 | 0 | 8.54e-14 | 7.24e-14 | 1.38e-12（＊） |
| 17 | 3.33e-12 | 0 | 1.04e-14 | 1.54e-12 | 0 | 0 | 0 | 1.79e-12（＊） |
| 18 | 7.27e-12 | 0 | 0 | 7.27e-12 | 0 | 0 | 0 | 0 |
| 22 | 1.50e-14 | 0 | 0 | 0 | 0 | 0 | 1.43e-14 | 0 |
| 28 | 8.31e-12 | 0 | 0 | 0 | 8.24e-12 | 0 | 7.39e-14 | 0 |
| 29 | 2.22e-13 | 0 | 0 | 0 | 2.22e-13 | 0 | 0 | 0 |
| 32 | 3.03e-13 | 0 | 0 | 0 | 0 | 3.03e-13 | 0 | 0 |
| 34 | 0 | 0 | 0 | 0 | 0 | 0 | 0 | 0 |
| 44 | 7.54e-13 | 0 | 0 | 0 | 0 | 0 | 7.54e-13 | 0 |
| 45 | 1.42e-15 | 0 | 0 | 0 | | 0 | 1.42e-15 | 0 |
| Amps. of each molecule | – | 6.21e-11 | 2.03e-12 | 8.81e-12 | 8.46e-12 | 3.88e-13 | 9.81e-13 | 4.25e-12 |
| Sum of others | 8.46e-13（＊＊） | – | – | – | – | – | – | – |

・Current of Each Molecule
  H₂          6.21e-11
  CH₄         2.03e-12
  H₂O  8.81e-12
  N₂+CO       8.46e-12
  O₂          3.88e-13
  CO₂         9.81e-13

Note: To be on the safer side when seeking the final conclusion, XCH is defined as follows

XCH = identified as CH₄ + unknowns in ME(12)(14)(16)(17)(*) + all other unknowns (**)
        ((*) and (**) refer to the inside of the table above)

XCH 7.12e-12

・Total Current
  8.973e-11

・Total pressure: (measured by an ionization vacuum gauge)
  9.27e-7   Pa

# Unified table of measurement data and cracking patterns

Measurement #：RGA 2018, Nov 26, 07:25　　scan#1825
Cracking pattern table used：
    (a) AIST bulletin of metrology, Vol.5, No.1, page 67, Hajime Yoshida(吉田 肇), Table 2
    (b) J. of the Mass Spectrometry Society of Japan, No 5, p. 34-43, 1955, Teruo Hayakawa(早川 晃雄), Toshio Sugiura(杉浦 俊男), Table 1
    We use (a) for ME(1), ME(15), ME(18) and ME(44). We use (b) for ME(32).

|    | data | $H_2$ | $CH_4$ | $H_2O$ | $N_2$ | CO | $O_2$ | $CO_2$ |
|----|------|------|-------|-------|------|-----|------|-------|
| 1  | 9.51e-13 | 1.9 | – | – | – | – | – | – |
| 2  | 6.11e-11 | 100 | 2.6 | – | – | – | – | – |
| 3  | 9.71e-15 | – | – | – | – | – | – | – |
| 4  | – | – | – | – | – | – | – | – |
| 5  | – | – | – | – | – | – | – | – |
| 6  | 3.72e-15 | – | – | – | – | – | – | – |
| 7  | 6.99e-15 | – | – | – | – | – | – | – |
| 8  | 6.13e-15 | – | – | – | – | – | – | – |
| 9  | – | – | – | – | – | – | – | – |
| 10 | – | – | – | – | – | – | – | – |
| 11 | – | – | – | – | – | – | – | – |
| 12 | 7.17e-13 | – | 8.6 | – | – | 4.7 | – | 8.7 |
| 13 | 1.58e-13 | – | – | – | – | – | – | – |
| 14 | 6.69e-13 | – | 17.1 | – | 4.0 | – | – | – |
| 15 | 8.09e-13 | – | 85.6 | – | – | – | – | – |
| 16 | 2.48e-12 | – | 100 | | | 1.7 | 28.2 | 9.6 |
| 17 | 3.33e-12 | – | 1.1 | 21.2 | – | – | – | – |
| 18 | 7.27e-13 | – | – | 100 | – | – | – | – |
| 19 | 3.04e-14 | – | – | – | – | – | – | – |
| 20 | 1.02e-13 | – | – | – | – | – | – | – |
| 21 | – | – | – | – | – | – | – | – |
| 22 | 1.50e-14 | – | – | – | – | – | – | 1.9 |
| 23 | – | – | – | – | – | – | – | – |
| 24 | 3.72e-14 | – | – | – | – | – | – | – |
| 25 | 1.24e-13 | – | – | – | – | – | – | – |
| 26 | 5.70e-13 | – | – | – | – | – | – | – |
| 27 | 5.83e-13 | – | – | – | – | – | – | – |
| 28 | 8.31e-12 | – | – | – | 100 | 100 | – | 9.8 |
| 29 | 2.22e-13 | – | – | – | – | 1.2 | – | – |
| 30 | 2.83e-13 | – | – | – | – | – | – | – |
| 31 | 1.79e-14 | – | – | – | – | – | – | – |
| 32 | 3.03e-13 | – | – | – | – | – | 100 | – |
| 33 | – | – | – | – | – | – | – | – |
| 34 | – | – | – | – | – | – | 0.41 | – |
| 35 | – | – | – | – | – | – | – | – |
| 36 | 3.82e-15 | – | – | – | – | – | – | – |
| 37 | 1.81e-14 | – | – | – | – | – | – | – |
| 38 | 2.75e-14 | – | – | – | – | – | – | – |
| 39 | 8.79e-14 | – | – | – | – | – | – | – |
| 40 | 4.61e-13 | – | – | – | – | – | – | – |
| 41 | 1.00e-13 | – | – | – | – | – | – | – |
| 42 | 4.46e-14 | – | – | – | – | – | – | – |
| 43 | 3.80e-14 | – | – | – | – | – | – | – |
| 44 | 7.54e-13 | – | – | – | – | – | – | 100 |
| 45 | 1.42e-15 | – | – | – | – | – | – | 1.2 |
| 46 | 4.57e-15 | – | – | – | – | – | – | – |
| 47 | – | – | – | – | – | – | – | – |
| 48 | – | – | – | – | – | – | – | – |
| 49 | – | – | – | – | – | – | – | – |
| 50 | 8.15e-15 | – | – | – | – | – | – | – |
| 51 | 6.81e-15 | – | – | – | – | – | – | – |
| 52 | 4.81e-15 | – | – | – | – | – | – | – |
| 53 | 7.41e-15 | – | – | – | – | – | – | – |
| 54 | 5.69e-15 | – | – | – | – | – | – | – |
| 55 | 1.51e-14 | – | – | – | – | – | – | – |
| 56 | 9.03e-15 | – | – | – | – | – | – | – |
| 57 | 7.15e-15 | – | – | – | – | – | – | – |
| 58 | – | – | – | – | – | – | – | – |
| 59 | – | – | – | – | – | – | – | – |
| 60 | – | – | – | – | – | – | – | – |
| 61 | – | – | – | – | – | – | – | – |
| 62 | – | – | – | – | – | – | – | – |
| 63 | – | – | – | – | – | – | – | – |
| 64 | – | – | – | – | – | – | – | – |
| 65 | – | – | – | – | – | – | – | – |
| 66 | – | – | – | – | – | – | – | – |
| 67 | 5.24e-15 | – | – | – | – | – | – | – |
| 68 | 1.63e-15 | – | – | – | – | – | – | – |
| 69 | 4.00e-15 | – | – | – | – | – | – | – |
| 70 | 2.64e-15 | – | – | – | – | – | – | – |
| 71 | – | – | – | – | – | – | – | – |
| 72 | – | – | – | – | – | – | – | – |
| 73 | – | – | – | – | – | – | – | – |
| 74 | – | – | – | – | – | – | – | – |
| 75 | – | – | – | – | – | – | – | – |
| 76 | – | – | – | – | – | – | – | – |
| 77 | – | – | – | – | – | – | – | – |
| 78 | – | – | – | – | – | – | – | – |
| 79 | 1.90e-15 | – | – | – | – | – | – | – |
| 80 | – | – | – | – | – | – | – | – |
| 81 | 2.36e-15 | – | – | – | – | – | – | – |
| 82 | – | – | – | – | – | – | – | – |

Morning of November 26, 2018        07:26, Scan #1825

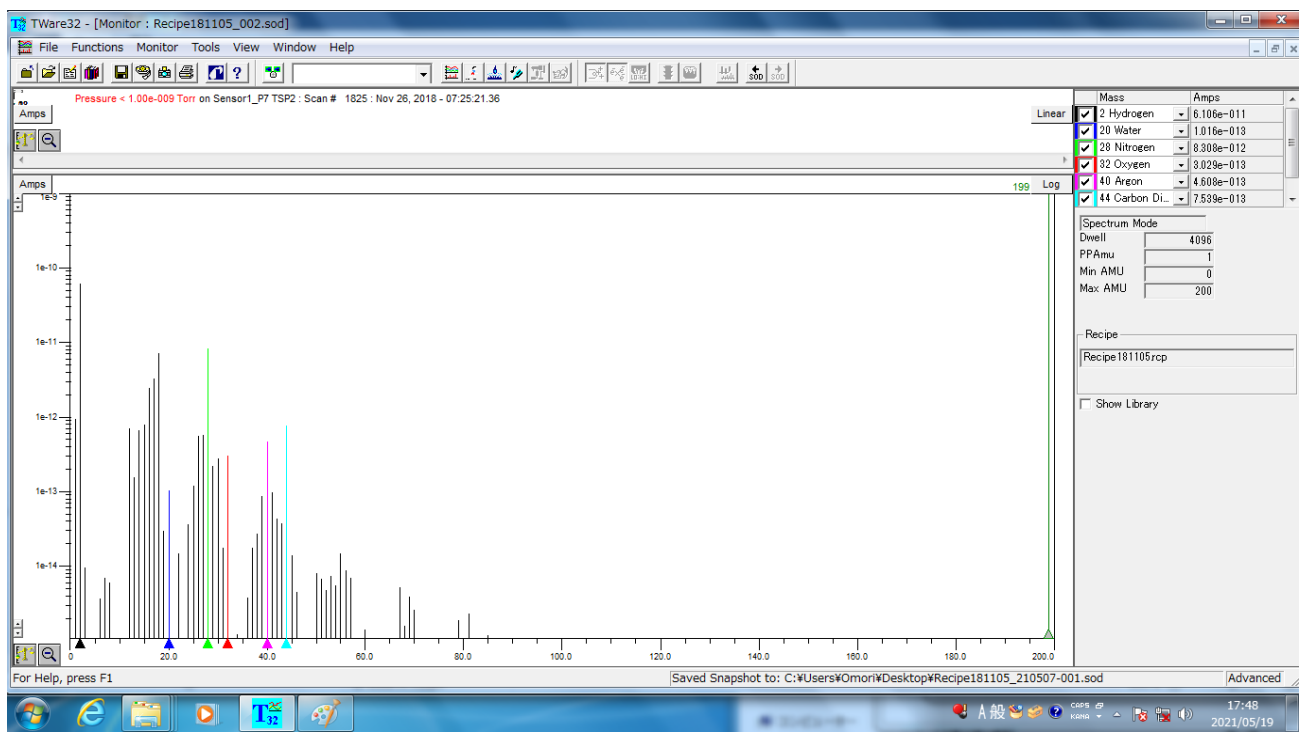

Morning of November 8, 2018        Photos of just before the start of the experiment.

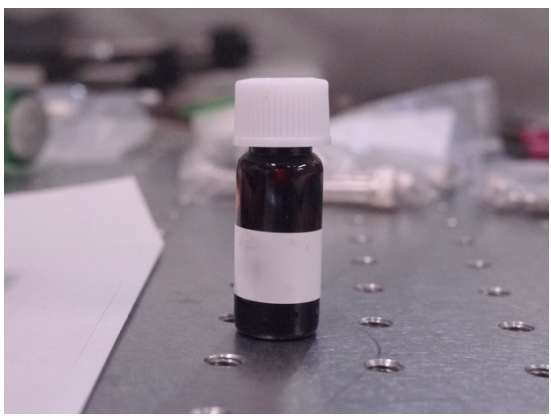

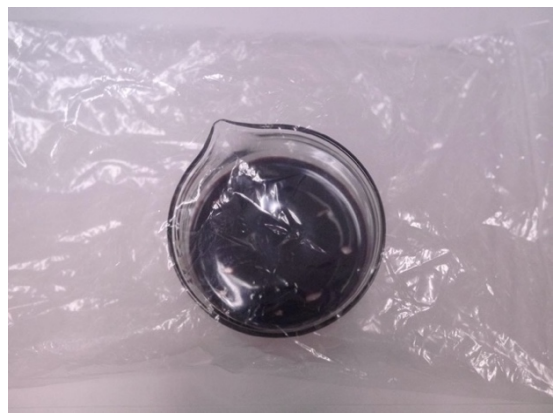

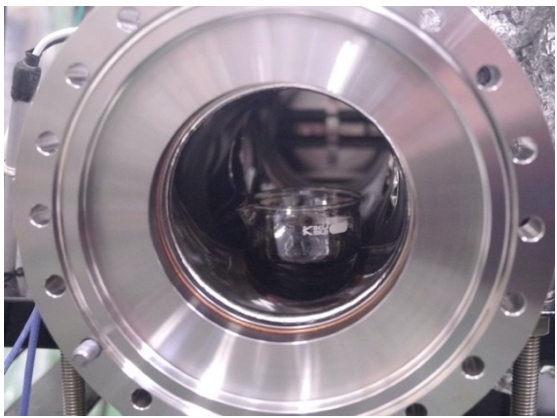

# Appendix 2: Stress Analysis of ILC Positron Source Target

In Chapter V(4), an analysis of the heat and stress induced by the electromagnetic shower created by the electron beam was conducted, concluding that the ILC target possesses safety equal to or greater than that of the SLC target. Furthermore, a comparison with the stress of the SLC target was conducted in the process of reaching this conclusion. In this appendix, we will examine the comparison between the ILC target and the SLC target in more detail.

When comparing the ILC target and the SLC target, it can be said that, in general, the stress on the ILC target is similar to that on the SLC target. Broadly speaking, they are in the same physical condition.

Common Features of Stress between ILC Target and SLC Target 1

＊ ILC Target: Near the point of impact of the beam
The maximum principal stress (tensile stress in this case) is maximum on the downstream surface (exit-side surface) Slightly deeper at this location, it is in compression, with the maximum absolute value of the principal stress being 800 MPa, greater than the tensile stress of 500 MPa on the surface. Furthermore, all three principal stresses—maximum (S1), intermediate (S2), and minimum (S3)—are negative, indicating a compressive state being tightly squeezed from the surroundings.

＊SLC Target: Near the point of impact of the beam (Note: presumed point)
The maximum stress is compressive at 180 ksi (1240 MPa), at a temperature of 650°C, within a compression environment surrounded by solid material (See [12]SLAC-PUB-5370).

> (Note) Although "point of impact of the beam" is not explicitly stated in
>         SLAC-PUB-5370, it is interpreted as such from the context.

| The point of beam hit<br>The point of max. temp. | SLC<br>(Inside) | ILC<br>(Inside: 2mm from downstream surface) |
|---|---|---|
| Compressive stress | 1240 MPa (absolute vale) | 810 MPa (absolute vale = ABS(S3)) |
| von Mises stress (SM) | 600 MPa | 470 MPa |
| Temperature | 650 C | 585 C |
| Yield Strength | 1030 MPa (650C) | 1230 MPa (585 C) |
| Comment 1 | The paper says "the material is captured" so we assume not surface. | All S1, S2, S3 are negative. Captured by compressive stresses in all directions. |
| Comment 2 | The principal stress with the greatest absolute value. It's 120% of the yield strength. But OK, because compressive. | |

| The point of max von Mises stress<br>(between two hits) | SLC<br>(downstream surface) | ILC<br>(downstream surface) |
|---|---|---|
| Tensile stress | 650 MPa | 320 MPa |
| Compressive stress | 434 MPa (absolute vale) | 550 MPa (absolute vale) |
| Von Mises stress (SM) | 965 MPa | 760 MPa |
| Temperature | 370 C | 350 C |
| Yield Strength | 1380 MPa (370C) | 1400 MPa (350 C) |
| Comment | SM is 70% of the yield strength, it is OK. No comparison with fatigue stress. | SM is 54% of the yield strength. SM is 1.36 of the fatigue stress. |

Table. Comparison of SLC target and ILC target: The stress of the ILC target is almost the same as that of the SLC target, but the ILC target is slightly lower. The temperature is also slightly lower for the ILC target.

Common Features of Stress between ILC and SLC Targets 2

Let's also examine the downstream surface. However, since the stress analysis of the SLC target described in SLAC-PUB-5370 does not include a description of the downstream, some parts are speculative.

The point of maximum von Mises stress is shifted from the beam impact point and lies between the two impacts.

If tensile stress is the primary concern (from the perspective of target destruction), then the downstream surface near the point of beam impact, where tensile stress is maximum, becomes the primary concern. Here, the tensile stress of the ILC target is 500 MPa, which is 39% of the Yield Strength. The tensile stress at this point for the SLC target has not been described in SLAC-PUB-5370. However, the compression at the depth of this location is 1240 MPa, whereas for the ILC, it is 810 MPa. Thus, the compression at this location is 1.5 times greater for the SLC than the ILC. Presumably, the tensile stress on the surface for the SLC is also larger. It can be inferred that the situation is slightly more favorable for the ILC compared to the SLC.

Supplement: In the analysis of SLAC-PUB-5370, von Mises stress is compared with the Yield Strength at the point where the von Mises stress is maximum (offset from the point of beam impact). Tensile stress on the downstream surface of the point of beam impact has not been described.

| The point of beam hit | SLC (downstream surface) | ILC (downstream surface) |
|---|---|---|
| Tensile stress | | 500 MPa (S1) |
| Compressive stress | | 110 MPa (absolute value, ABS(S3)) |
| von Mises stress (SM) | | 570 MPa |
| Temperature | | 531 C |
| Yield Strength | | 1280 MPa (531 C) |
| Comment 1 | In SLAC-PUB-5370, there is no description of downstream surface. | the point where tensile stress is maximum |
| Comment 2 | | Tensile stress is 39% of yield strength. Tensile stress is about equals to fatigue limit. |

Table. Status of the downstream surface at the point of beam hit

Lastly, we discuss the value of 35 J/g, which is considered the limit for the Peak Energy Deposit Density (PEDD) of W75Re25 and is frequently cited. This value was obtained from experiments conducted at the SLAC end station (see [8] SLAC-CN-128). Converting this experimentally obtained limit to units of J/g yields 76 J/g. It is important to note that the frequently cited 35 J/g includes an approximate doubling as a safety margin. Since this experiment was conducted under conditions similar to ours, it holds significant relevance.

Note: Brief Summary on Stress

maximum principal stress :             S1
middle principal stress :              S2
minimum principal stress :             S3

von Mises stress:    $$SM = \frac{\sqrt{(S1-S2)^2+(S2-S3)^2+(S3-S1)^2}}{2}$$

Note:    "maximum" does not mean maximum of the absolute value.
         When referring to "maximum principal stress," the term "maximum"
         is determined including the sign. Among S1, S2, and S3, the largest
         positive value is S1.

sign  +:  tensile stress
      - :  compressive stress

The von Mises stress is commonly used as a criterion for failure in ductile materials and as a criterion for the onset of plastic deformation [∗]. In an isotropic stress state (hydrostatic stress), the von Mises stress is zero. Additionally, in uniaxial tension, it is equal to the tensile stress [∗].

[∗]『よくわかる連続体力学ノート("Easy-to-Understand Notes on Continuum Mechanics")(written in Japanese)』written by Takashi KYOYA（京谷孝史）, compiled by Japan Association on Non Linear CAE, Morikita Publishing Co., Ltd.,

# Appendix 3: Consideration from the Fatigue Perspective: Annual Repetition Hits to the Same Point

From the fatigue perspective, a comparison is made between the ILC target and the SLC target.

In the design of the ILC electron-driven positron source, the parameters are largely set based on the SLC positron source. Specifically, the parameters and materials of the target are selected to utilize the design and operational experience of the SLC trolling target, as well as experimental results from the End-station at SLAC, particularly concerning the heat and stress loads on the target. Specifically, parameters are chosen to make the energy deposition per hit and the resulting stress approximately the same as those in the SLC. Additionally, the material selected as the primary candidate is W75Re25, the same as that used in the SLC.

When considering the limit of target material failure, not only the stress per hit but also the effect of repeated stress, i.e., fatigue effect, becomes important. Therefore, let us compare the annual repetition hits to the same point on the target between the ILC and SLC.

ILC Electron-Driven Positron Source Target:
     Circumference:        Approximately 1500 mm
     Repetition:           5 Hz (*)

SLC Trolling Target [14]:
     Circumference:        Approximately 180 mm
     Repetition:           120 Hz

(*) The repetition of the electron beam in the ILC electron-driven positron source's linac is 300 Hz. During the 63-millisecond positron generation period (300 Hz), 1312 bunches are generated in 20 pulses of 66 bunches each. This is followed by a 137-millisecond pause. This generation-pause cycle corresponds to the main linac repetition of 5 Hz. Within the 5 Hz repetition, there are 20 pulses, so the effective repetition is 100 Hz, which is nearly the same as that of the SLC.

Below, the hit counts per unit circumference (year · mm) for the ILC and SLC are compared:

ILC:
   1300/66 x 5 Hz x 3600 s/h x 24 h/d x 365 d/y / 1500 mm= $2.1 \times 10^6$ hits/year.mm

SLC:
   120 Hz x 3600 s/h x 24 h/d x 365 d/y / 180 mm = $2.1 \times 10^7$  hits/(year · mm)

Definition of Hits:
     ILC:   Each hit corresponds to 66 bunches of the driving electron beam hitting the same point on the target.
     SLC:   Each hit corresponds to one bunch of the driving electron beam hitting the target.

As shown above, the annual repetition hits to the same point are approximately 1/10 for the ILC compared to the SLC.

For this reason, the ILC target is designed to operate under conditions approximately 10 times easier than those of the SLC target in terms of fatigue.